\documentclass[aps,nofootinbib,prd]{revtex4}

\usepackage{lscape}
\usepackage{epsfig,epsf}
\usepackage{amsmath}
\usepackage{amsthm}
\usepackage{amsfonts}
\usepackage{amssymb}
\usepackage{dsfont}
\usepackage{multirow}
\usepackage{appendix}
\usepackage{slashed}
\usepackage[active]{srcltx}
\usepackage{psfrag}
\usepackage{subfigure}
\usepackage[colorlinks,citecolor=blue,urlcolor=red,linkcolor=red]{hyperref}
\usepackage{multirow}

\begin{document}

\title{{\Large{\bf Form factors of $B_{(s)}$ to light scalar mesons with the $B$-meson LCSR}}}

\author{\small
R. Khosravi\footnote {e-mail: rezakhosravi @saadi.shirazu.ac.ir }}

\affiliation{Department of Physics, School of Science, Shiraz
University, Shiraz 71946-84795, Iran}

\begin{abstract}
In this work, the transition form factors of the semileptonic decays
of $B_{(s)}$ to the light scalar mesons with masses close to
$1.5~\rm{GeV}$ such as $K_0^*(1430), a_0(1450)$, and $f_0(1500)$ are
calculated in the framework of the light-cone sum rules (LCSR). For
this purpose, the two- and three-particle $B$-meson distribution
amplitudes (DA's) are used. Note that it is possible to use the
$B$-meson DA's for $B_{s}$-meson in the $\rm SU(3)_F$ symmetry
limit. The transition form factors are obtained in terms of the
two-particle DA's up to twist-five accuracy, and the three-particle
up to twist-six level. We apply two classes of the phenomenological
models for the DA's of $B$-meson. The longitudinal lepton
polarization asymmetries and branching fractions for the
semileptonic decays of $B_{(s)}$ to the light scalar mesons are
estimated with the help of these form factors.
\end{abstract}


\maketitle

\section{Introduction}

Although scalar states with $J^P=0^+$ have been observed for more
than half a century, their inner structure is still controversial
both experimentally and theoretically. In order to discover their
underlying structure, many different theoretical and
phenomenological descriptions are presented including considering
the scalar mesons as the conventional mesons $q\bar{q}$ or
non-conventional mesons such as: tetraquark \cite{Ja}, molecule
\cite{Meissner}, hybrid \cite{HorMan}, and glueballs \cite{FriGell}.
A tetraquark is a complex structure made up of one diquark and one
antidiquark $qq-\bar{q}\bar{q}$. A molecule or a meson-meson bound
state is composed of two quark-antiquark couples
$q\bar{q}-q\bar{q}$. In addition, a hybrid is an object consisting
of a $q\bar{q}$ pair with at least one extra gluon $q\bar{q}-g$, and
the glueballs are made only of gluons (for instance see
\cite{Piotrowska}). It is very likely that some scalar mesons are
not made of one simple component but are the superpositions of these
contents. For example, it is suggested that $a_0(980)$ is a
superposition of $q\bar{q}$ and tetraquark \cite{Klempt}. The
dominant component of the scalar mesons can be found from the decay
and production of them.

A number of the scalar mesons have been discovered in the
spectroscopic studies. Due to the large decay widths of the scalar
mesons, the identification of them is more difficult in contrast to
the pseudoscalars and vector mesons. Among them, there are nine the
light scalar mesons together below or near $1~\rm{GeV}$, including
the isoscalars $f_0(500)$(also denoted as $\sigma$) and $f_0(980)$,
isodoublets $[K^{*+}_0(700)$(also refereed as $\kappa$),
$K^{*0}_0(700)$] and [$\bar{K}^{*0}_0(700)$, $\bar{K}^{*-}_0(700)$],
and isovector [$a^+_0(980)$, $a^0_0(980)$, $a^-_0(980)$] which can
form an $\rm{SU(3)}$ nonet, while the scalar mesons around $1.5~
\rm{GeV}$, consisting isoscalars $f_0(1370)$ and
$f_0(1500)/f_0(1700)$, isodoublets $[K^{*+}_0(1430)$,
$K^{*0}_0(1430)$] and [$\bar{K}^{*0}_0(1430)$,
$\bar{K}^{*-}_0(1430)$] and isovector [$a^+_0(1450)$, $a^0_0(1450)$,
$a^-_0(1450)$] can be members of another nonet. From a survey of the
accumulated experimental data, two scenarios can be suggested to
describe these two groups of nine scalar mesons in the quark model
\cite{CheChuYa}. In the first one, scenario 1(S1), it is supposed
that the light scalar mesons are composed from two quarks. The nonet
mesons below $1\, \mbox{GeV}$ are treated as the lowest lying
states, and the nonet mesons near $1.5\, \mbox{GeV}$ are the excited
states corresponding to the lowest lying states. In scenario 2 (S2),
the scalar states below $1\, \mbox{GeV}$ are considerd as the
members of a tetraquark nonet, while the nonet mesons near
$1.5~\rm{GeV}$ are viewed as the lowest lying states, with the
corresponding first excited states between $2.0\sim 2.3~\rm {GeV}$.
In the both scenarios, it is suggested that the heavier nonet near
$1.5~\rm{GeV}$ consists of the scalar mesons with two quarks in the
quark model. However in S1, those are regarded as the excited
states, and in S2, they are seen as the ground states. Therefore,
the calculation of the decay constant values and DA's for the scalar
mesons near $1.5~\rm{GeV}$ are different via the two scenarios.

In recent years, some experimental efforts have been devoted to
measuring the decay modes involving the light scalar mesons in final
state. BESIII collaboration has recently measured the nonleptonic
and semileptonic decays of $D_{s}$ to the light scalar mesons
\cite{BESIII,Ablikim1,Ablikim2}. In the same context, the
nonleptonic two-body $B$ meson decays involving a scalar final state
have been observed by Belle \cite{Belle Collaboration}, BABAR
\cite{BABARCollaboration}, and LHCb \cite{LHCbCollaboration}. These
observations provide a efficient way to investigate the features and
the possible inner structures of the scalar mesons.

In the particle physics an accurate calculation of the transition
form factors for the semileptonic $B_{(s)}$ decays  to the light
scalar mesons is important in two folds. First, to study the
quantities related to the semileptonic and nonleptonic decays
involving $B_{(s)}$ to the scalar mesons, it is necessary to know
the appropriate behavior of the transition form factors. Second, for
the indirect search of new physics beyond the standard model (SM),
these form factors are the essential ingredients. Since the heavy to
light transition form factors are nonperturbative in nature,
therefore the nonperturbative QCD approaches are applied to evaluate
them. Usually, the Lattice QCD (LQCD) works well to calculate the
form factors in these cases.

So far, the transition form factors of $B_{(s)} \to S$,
($S=K_0^*(1430), a_0(1450), f_0(1500)$), have been not estimated
through the LQCD, although they have been calculated from other
methods such as the perturbative QCD (pQCD) \cite{LiLuWaWa},
covariant light-front (CLF) \cite{ChChuHwa}, QCD sum rules (QCDSR)
\cite{AliAzSav,MZYang,Ghahramany}, light-front quark model (LFQM)
\cite{CheGenLiLi}, minimal supersymmetric standard model (MSSM)
\cite{AsluWang}, and also the light cone sum rules (LCSR).

The LCSR is a proper approach to evaluate the transition form
factors of the heavy to light meson decays. The conventional LCSR
starts with a two-point correlation function inserting the operators
between vacuum and light meson. Then, it develops in terms of the
nonlocal operators by using the operator product expansion (OPE)
near the light-cone region $x^2 = 0$. The matrix elements of the
nonlocal operators are parameterized as the light meson DA's of the
increasing twist, i.e twist-$2, 3, 4$ and so on. These DA's offer
valuable insights into the nonperturbative makeup of hadrons and the
distribution of partons in relation to their momentum fractions
within these particles. More researches on the $B_{(s)} \to S$
transition form factors have been performed in the framework of the
LCSR with the light scalar meson twist-$2$, $3$ DA's
\cite{WanAslLu,ZGWang1,ZGWang2}, only scalar meson twist-$2$ DA
\cite{SunLiHu,HuaZhoTon} or twist-$3$ DA's \cite{HanWuFu},
respectively. In this method, a reliable estimation of the form
factors depends on an accurate knowledge of the internal structure
of the light meson and its DA's. Since the intrinsic nature of the
light scalar mesons is still not completely clear, therefore the
DA's attributed to them can also be doubted.  Therefore, it is
important to use a new LCSR method to calculate the form factors
that is independent of the light scalar meson DA's, and then compare
its results with the conventional method.

In Ref. \cite{Offen}, the authors proposed a new method based on the
LCSR technique that relates the $B$-meson DA's to the $B \to \pi$
form factor. This model was independently suggested in the framework
of the soft-collinear effective theory (SCET) in Ref.
\cite{FazFeldHur}. In this new approach which is sometimes called
the $B$-meson LCSR, the main idea is to invert the correlation
function compared to the conventional LCSR, so that the light meson
interpolates with an appropriate light-quark current, and the
non-local operators between an on-shell $B$-meson state and the
hadronic vacuum are expressed as convolutions of hard scattering
kernels with light-cone distribution amplitudes (LCDA's) of
$B$-meson. Recently, considering the next-to-leading order QCD
corrections to the correlation function in order to extract the hard
and jet functions, the form factors of semileptonic decays $B$ to
scalar mesons have been calculated in terms of leading twist
function of the $B$-meson DA \cite{HanLuShi}. Also, considering the
$\rm SU(3)_F$ symmetry limit and using the two-particle $B$-meson
DA's up to twist-three, and three-particle DA's up to twist-four,
the transition form factors of the semileptonic $B_{s}\to
K_{0}^*(1430)$ decays have been calculated in the framework of the
$B$-meson LCSR \cite{Khosravi2022}.

In this work, we focus on the three light scalar mesons
$K_0^*(1430), a_0(1450)$, and $f_0(1500)$, with the mass of about
$1.5~\rm{GeV}$. The production of the light scalar mesons $K_0^*,
a_0$, and $f_0$ can provide a different unique insight to the
mysterious structure of them. Our main goal is to calculate the form
factors of the $B_{(s)} \to (K_0^*, a_0, f_0)$ decays via the
$B$-meson LCSR applying the two-particle DA's up to twist-five, and
considering the new results for the complete set of the
three-particle DA's up to twist-six. The four-particle $B$-meson
DA's are not taken into account in this work due to the negligible
effects. Note that in the $\rm SU(3)_F$ symmetry limit, it is
possible to use the $B$-meson DA's for $B_{s}$-meson. The functional
form of the higher-twist $B$-meson DA's involve contributions of
multiparton states. To calculate the form factors, we use two
classes of the phenomenological models for the two- and
three-particle DA's of $B$-meson which contain a minimum number of
free parameters and satisfy equation of motion (EOM) constraints in
tree-level \cite{JiMan,ShenWangWei}. Utilizing these form factors,
the semileptonic $B_{(s)} \to (K_0^*, a_0)\, l \bar{\nu_{l}}$ and
$B_{(s)} \to (K_0^*, f_0)\, l \bar{l}/\nu \bar{\nu}$, $l=e, \mu,
\tau$ decays are analyzed. In the SM, the rare semileptonic $B_{(s)}
\to (K_0^*, f_0)\, l \bar{l}/\nu \bar{\nu}$ decays occur at loop
level instead of tree level, by electroweak penguin and weak box
diagrams via the flavor changing neutral current (FCNC) transitions
of $b\to s l^+ l^-$ at quark level.

The content of paper is as follows: In Sec II, the form factors of
the semileptonic $B_{(s)}\to S, ~(S=K_0^*, a_0, f_0)$ decays are
calculated with the $B$-meson LCSR approach using the two- and
three-particle DA's of $B$-meson up to twist-five, and twist-six,
respectively. These form factors are basic parameters to study the
other quantities such as forward-backward asymmetry, longitudinal
lepton polarization asymmetry and branching fraction of semileptonic
decays. In Sec III, the two phenomenological models for the DA's of
$B$-meson are presented. This section is also devoted to the
numerical and analytical results for the semileptonic $B_{(s)} \to
S$ decays.
\\
\section{Form Factors of $B_{(s)} \to S$ with higher-twist corrections}

Considering parity and using Lorentz invariance, the transition
matrix elements involved in $B_{(s)}\to S$ transitions can be
parameterized as:
\begin{eqnarray}\label{eq201}
\langle S(p^\prime)|J^{A}_{\mu}
|B_{(s)}(p)\rangle& =&-i\left[f_+(q^2) P_\mu +f_-(q^2)q_\mu\right], \nonumber \\
\langle S(p^\prime) |J^{T}_{\mu}|
B_{(s)}(p)\rangle&=&-\frac{f_T(q^2)}{m_{B_{(s)}}+m_{S}}\left[q^2P_\mu
-(m_{B_{(s)}}^2-m_{S}^2)q_\mu\right].
\end{eqnarray}
In these phrases $P_{\mu}=(p^\prime +p)_\mu$, $q_\mu = (p-
p^\prime)_\mu$. The transition currents $J^{A}_{\mu}=\bar
q_{_1}\gamma_\mu\gamma_5 b$, and $J_{\mu}^{T}=\bar
q_{_1}\sigma_{\mu\nu}\gamma_5q^\nu b$  ($q_{_1}=u, s$) are used to
calculate the transition form factors $f_+(q^2)$, $f_-(q^2)$ and
$f_T(q^2)$, respectively. $q^2$ is the momentum transfer squared.
The calculation of these form factors using the LCSR method are
described in this section.

To investigate the form factors in the frame work of the $B$-meson
LCSR, the two-point correlation function $\Pi(p^\prime,q)$ is
constructed from two currents inserted between vacuum and
$B_{(s)}$-meson as follows:
\begin{eqnarray} \label{eq202}
\Pi(p^\prime,q)=i\int {d^{4}x}~ e^{ip^\prime.x} \langle0| T\left\{
J^{S}(x)J(0)\right\}|B_{(s)}(p)\rangle,
\end{eqnarray}
where $T$ is the time ordering operator,  $J^{S}(x)$ is the
interpolating current of the scalar meson $S$, so that
$J^{K_0^{*+}}(x)=\bar{s}(x) u(x)$, $J^{K_0^{*0}}(x)=\bar{d}(x)
s(x)$, $J^{a_0}(x)=\bar{d}(x) u(x)$, and $J^{f_0}(x)=\bar{s}(x)
s(x)$. The matrix element of $J^{S}$ between the vacuum and scalar
meson $S$ is given in terms of the decay constant $f_{S}$, and mass
of the the scalar meson as $\langle 0|J^{S}|S\rangle =f_{S}\,
m_{S}$. In the correlation function, $J(0)$ is the transition
current; $J=J_\mu^{A}$ or $J_\mu^{T}$. In addition, $p^\prime$ and
$q$ are the momenta of the interpolating and transition current,
respectively. The relation between these two quantities is: $p^2=
(p^\prime + q)^2 =m^2_{B_{(s)}}$.

The correlation function in Eq. (\ref{eq202}) can be investigated
from two aspects; hadronic representation and QCD calculations. The
form factors are obtained via the LCSR method in terms of the DA's
of $B$-meson by equating the two sides. As mentioned before, it is
possible to use the $B$-meson DA's for $B_{s}$-meson in the $\rm
SU(3)_F$ symmetry limit.
\\\\
$\bullet$ \textbf{Hadronic representation}

Inserting a complete set of the intermediate states with the same
quantum number as the interpolating current $J^{S}$, in Eq.
(\ref{eq202}), and isolating the pole term of the lowest scalar
meson $S$, and then applying the Fourier transformation, the
hadronic representation of the correlation function is obtained.
Defining the spectral density function of the higher resonances and
continuum states as $\rho (s)\equiv \sum_h\,\langle 0 |
J^{S}(p^\prime) |h(p^\prime) \rangle \langle h(p^\prime) | J|
B_{(s)}(p) \rangle \, \delta(s-m^{2}_{h})$, the correlation function
can be written in terms of the scalar meson state and the higher
resonance contributions as
\begin{eqnarray}\label{eq203}
\Pi^{\rm{HAD}}(p^\prime,q) &=& \frac{\langle 0 | J^{S} |S(p^\prime)
\rangle \langle S(p^\prime) | J | B_{(s)}(p)
\rangle}{m_{S}^2-p^{\prime2}}+ \int_{s_0}^\infty ds
\frac{\rho(s)}{s-p'^2},
\end{eqnarray}
where $s_{0}$ is the continuum threshold of the scalar meson $S$.
\\\\
$\bullet$ \textbf{QCD calculations}

At the quark level in QCD, the correlation function can be evaluated
in the deep Euclidean region as a complex function. Using the
dispersion relation, the correlation function can be written as
\begin{eqnarray}\label{eq204}
\Pi^{\rm{QCD}}(p^\prime, q)=\frac{1}{\pi}\int_{0}^{\infty} ds
\,\frac{{\rm{Im}}{\Pi}^{\rm {QCD}}(s)}{s-p'^2}.
\end{eqnarray}

Applying Borel transformation in Eq. (\ref{eq203}) and Eq.
(\ref{eq204}) with respect to the variable $p'^{2}$ as:
$\textbf{\emph{B}}_{p'^2}(\frac{1}{x^2-p'^2})=\frac{e^{-x^2/M^2}}{M^2}$,
where $M^2$ is  the Borel parameter, guarantees that the
contributions of the higher states and continuum in the hadronic
representation are effectively suppressed. In addition, it assures
that the contributions of higher dimensional operators in the QCD
side are small. Equating the both sides of the correlation function,
and using the quark-hadron duality approximation at large spacelike
$p'^2$ as $\rho(s)\simeq \frac{1}{\pi}{\rm Im} \Pi^{\rm QCD} (s)$,
the following equality is determined
\begin{eqnarray}\label{eq205}
\langle 0 | J^{S} |S(p^\prime) \rangle \langle S(p^\prime) | J |
B_{(s)}(p) \rangle\, e^{-m_{S}^2/M^2}= \int_{0}^{s_0}ds \,\,
\rho(s)\,e^{-s/M^2}.
\end{eqnarray}
Investigation the spectral density at the quark level allows to
derive the form factors $f_+$, $f_-$ and $f_T$ in Eq. (\ref{eq205}).

Based on the heavy quark effective theory (HQET), the
$B_{(s)}$-meson state in the limit of large $m_b$ can be estimated
by the relativistic normalization of it $|B_{(s)}(p)\rangle =
|B_{(s)}(v)\rangle$, where $v$ is four-velocity of $B_{(s)}$-meson.
Up to $1/m_b$ corrections, the correlation function of the $B_{(s)}
\to S$ transition can be approximated as:
$\Pi^{\rm{QCD}}(p^\prime,q)=\widetilde{\Pi}^{\rm{QCD}}(p^\prime,\widetilde{q})+\mathcal{O}(1/m_b)$,
where $\widetilde{q}=q - m_b v$ is called static part of $q$.
Replacing the $b$-quark field by the HQET field $h_{v}$, the
correlation function (Eq. (\ref{eq202})) in the heavy quark limit,
($m_b \to \infty$), become \cite{KMO}:
\begin{eqnarray}\label{eq206}
\widetilde{\Pi}^{\rm{QCD}}(p^\prime,\widetilde{q})&=& i\int d^4x~
e^{ip^\prime.x}\langle 0| T \{\bar{q}(x) S_{q_{_1}}(x)\, \Gamma\,
h_v(0)\} |B(v)\rangle,
\end{eqnarray}
where quark field $\bar{q}$ stands for $\bar{s}$ or $\bar{d}$, and a
matrix composition $\Gamma$ is $\gamma_{\mu}\gamma_{5}$ or
$\sigma_{\mu\nu} \gamma_5 q^\nu $ corresponding to the transition
current $J_{\mu}^{A}$ or $J_{\mu}^{T}$, respectively. The full-quark
propagator $S_{q_{_1}}(x)$ of a massless quark $q_{_1}$ ($u$ or $s$)
in the external gluon field in the Fock-Schwinger gauge is as
follows \cite{BalBra}:
\begin{eqnarray}\label{eq207}
S_{q_{_1}}(x) &=& i \int \frac{d^{4}k}{(2\pi)^{4}}
e^{-ik.x}\left\{\frac{\not\! k}{k^{2}}+\int^{1}_{0}du~
G_{\lambda\rho}(u x)\left[\frac{1}{k^2}
ux^{\lambda}\gamma^{\rho}-\frac{1}{2k^{4}}\not\! k
\sigma^{\lambda\rho}\right] \right\}.
\end{eqnarray}
If the full-quark propagator $S_{q_{_1}}(x)$ in Eq. (\ref{eq207}) is
replaced in Eq. (\ref{eq206}), operators between vacuum mode and
$B(v)$-state create the non-zero matrix elements as $\langle 0|\bar
{q}{_{\alpha}(x){h_{v\beta}(0)}}|B(v)\rangle$ and $\langle
0|\bar{q}_\alpha(x) G_{\lambda\rho}(ux) h_{v\beta}(0)|B(v)\rangle$.
These matrix elements of the non-local heavy-light currents are
parametrized in terms of the $B$-meson DA's. The two-particle
higher-twist DA's of $B$-meson arise in the expansion of the
relevant nonlocal quark-antiquark operator close to the light-cone
as \cite{JiMan,ShenWangWei,GubKokDyk}:
\begin{eqnarray}\label{eq208}
\langle 0|\bar {q}{_{\alpha}(x)\,{h_{v\beta}(0)}}|B(v)\rangle &=&
-\frac{i f_{B}\, m_{B}}{4} \int^{\infty}_{0} d\omega~ e^{-i\,\omega
\,v \cdot x} \Big\{(1+\not\! v) \Big [ \left
(\varphi_{_+}(\omega)+x^2 g_{_+}(\omega)\right )- \frac{\not\!
x}{2\, v \cdot x}\left [ \left
(\varphi_{_+}(\omega)-\varphi_{_-}(\omega)\right )
\right.\nonumber\\
&&+\left. x^2 \left (g_{_+}(\omega)-g_{_-}(\omega)\right )\right
]\Big ] \gamma_{5} \Big\}_{\beta\alpha},
\end{eqnarray}
where $\varphi_{_+}(\omega)$, $\varphi_{_-}(\omega)$,
$g_{_+}(\omega)$ and $g_{_-}(\omega)$ are of leading-twist,
twist-three, twist-four, and twist-five, respectively. The
three-particle contributions involve a further gluon field, so that
the matrix elements of the non-local operator ${q}_\alpha(x)
G_{\lambda\rho}(ux) h_{v\beta}(0)$, are parametrized in terms of the
$B$-meson DA's of increasing twist as
\cite{JiMan,ShenWangWei,GubKokDyk}:
\begin{eqnarray}\label{eq209}
\langle 0|\bar{q}_\alpha(x)\, G_{\lambda\rho}(ux)\,
h_{v\beta}(0)|B(v)\rangle &=& \frac{f_B\, m_B}{4}\int_0^\infty
d\omega \int_0^\infty d\xi\,  e^{-i(\omega+u\,\xi)\, v \cdot x}
\bigg\{ (1 +
\not\!v)\Big[(v_\lambda\gamma_\rho-v_\rho\gamma_\lambda)
\left(\Psi_{_A}(\omega, \xi)-\Psi_{_V}(\omega, \xi)\right)\nonumber\\
&&-i\sigma_{\lambda\rho}\Psi_{_V}(\omega, \xi) -\frac{x_\lambda
v_\rho-x_\rho v_\lambda}{v \cdot x}\,X_{_A}(\omega,
\xi)+\frac{x_\lambda \gamma_\rho-x_\rho \gamma_\lambda}{v \cdot
x}\,\left(W(\omega, \xi)+Y_{_A}(\omega,
\xi)\right)\nonumber\\
&&-i\,\epsilon_{\lambda \rho \delta \eta}\,\frac{x^{\delta}
v^{\eta}\gamma_{5}}{v \cdot x}\, \overline{X}_{_A}(\omega,
\xi)+i\,\epsilon_{\lambda \rho \delta \eta}\,\frac{x^{\delta}
\gamma^{\eta} \gamma_{5}}{v \cdot x}\, \overline{Y}_{_A}(\omega,
\xi)-u\,\frac{x_\lambda v_\rho-x_\rho v_\lambda}{(v \cdot
x)^2}\not\! x\,\,W(\omega,
\xi)\nonumber\\
&&+u\,\frac{x_\lambda \gamma_\rho-x_\rho \gamma_\lambda}{(v \cdot
x)^2}\not\! x\,Z(\omega, \xi)\Big]\gamma_5\bigg\}_{\beta\alpha}.
\end{eqnarray}
There exist eight independent Lorentz structures and therefore eight
invariant functions, $\Psi_{_A}$, $\Psi_{_V}$, $X_{_A}$, $Y_{_A}$,
$\overline{X}_{_A}$, $\overline{Y}_{A}$, $W$, and $Z$ are the eight
independent three-particle DA's of $B$-meson. The three-particle
DA's are related to a basis of DA's such as $\phi_{i}\,
(i=3,...,6)$, $\psi_{j}$ and $\overline{\psi}_{j}$\, $(j=4,5)$  with
definite twist ($i$ and $j$ indicate the twist level), as follows
\begin{equation}\label{eq210}
\begin{aligned}
\Psi_{_A}(\omega, \xi)&= \frac{1}{2}\, [ \phi_{3} + \phi_{4}]\,, &
\Psi_{_V}(\omega, \xi)&=\frac{1}{2}\, [-\phi_{3} + \phi_{4}]\,, \\
X_{_A}(\omega, \xi)&=\frac{1}{2}\, [-\phi_{3}-\phi_{4}+
2\,\psi_{4}]\,, & Y_{_A}(\omega,\xi)&=\frac{1}{2}\, [-\phi_{3}-
\phi_{4}+ \psi_{4}- \psi_{5}]\,, \\
\overline{X}_{_A}(\omega, \xi)&=\frac{1}{2}\, [-\phi_{3}+ \phi_{4}-
2\,\overline{\psi}_{4}]\,, & \overline{Y}_{_A}(\omega, \xi)&=
\frac{1}{2}\, [-\phi_{3}+ \phi_{4}-\overline{\psi}_{4}+
\overline{\psi}_{5}]\,, \\
W(\omega, \xi)&=\frac{1}{2}\, [ \phi_{3}- \psi_{4}-
\overline{\psi}_{4}+{\phi}_{5}+\psi_{5}+ \overline{\psi}_{5}]\,, &
Z(\omega, \xi)&=\frac{1}{4}\, [-\phi_{3}+\phi_{4}-
2\,\overline{\psi}_{4}+{\phi}_{5}+ 2\, \overline{\psi}_{5}-
\phi_{6}]\,.
\end{aligned}
\end{equation}

Substituting the appropriate expressions of Eqs. (\ref{eq208}) and
(\ref{eq209}) instead of the matrix elements $\langle 0|\bar
{q_{_1}}{_{\alpha}(x){h_{v\beta}(0)}}|B(v)\rangle$ and $\langle
0|\bar{q_{_1}}_\alpha(x) G_{\lambda\rho}(ux)
h_{v\beta}(0)|B(v)\rangle$ that appear in the correlation function
in Eq. ({\ref{eq206}}), and integrating over the variables $x$ and
$k$, and then separating the solutions according to the Lorentz
structures $P_{\mu}$ and $q_{\mu}$, the general form of the
correlation function,
$\widetilde{\Pi}^{\rm{QCD}}(p^\prime,\widetilde{q})$ can be written
as $i\,\left[
\widetilde{\Pi}_{+}^{A}(p^\prime,\widetilde{q})\,P_{\mu}+\widetilde{\Pi}_{-}^{A}(p^\prime,\widetilde{q})\,q_{\mu}\right]$
and $\widetilde{\Pi}^{T}(p^\prime,\widetilde{q})\,P_{\mu} $
corresponding to $J^{A}_{\mu}$ and $J^{T}_{\mu}$, respectively.

Finally, inserting Eq. (\ref{eq201}) in Eq. (\ref{eq203}), and then
equating the coefficients of the same Lorentz structures on the both
sides of the correlation function, the form factors are obtained via
the LCSR in terms of the two- and three-particle DA's of $B$-meson.
Our results for $f_{+}(q^2)$, $f_{-}(q^2)$,  and $f_{T}(q^2)$ are
obtained as:
\begin{eqnarray}\label{eq211}
f_{+}(q^{2})&=&\frac{f_{B_{(s)}}m_{B_{(s)}}^2}{2
f_{S}m_{S}}e^{\frac{m_{S}^2}{M^2}}
\int_{0}^{\sigma_{0}}d\sigma\,e^{-\frac{{\textbf{s}}(\sigma)}{M^2}}
\left\{\varphi_{_+}(\sigma
m_{_{B_{(s)}}})-\frac{\widetilde{\overline{\varphi}}(\sigma
m_{_{B_{(s)}}})}{ \bar{\sigma} m_{_{B_{(s)}}} }-\frac{4g_{_+}(\sigma
m_{_{B_{(s)}}})}{\bar{\sigma} M^2}\right.\nonumber\\
&+&\left. \int_{0}^{\sigma m_{B_{(s)}}}d\omega \int_{\sigma
m_{B_{(s)}}-\omega}^{\infty }\frac{d\xi}{\xi}
\left\{\left[(2u+2)\left(\frac{\overline{q}^2}{\bar{\sigma}^3
M^2}+\frac{1}{\bar{\sigma}^2}\right)+\frac{(2u+1)m_{B_{(s)}}^2}{\bar{\sigma}
M^2}\right] \frac{\overline{\Psi}(\omega, \xi)}{m_{B_{(s)}}^2}\right.\right.\nonumber\\
&+&\left.\left.\frac{6u}{\bar{\sigma}M^2}\,\Psi_{_V}(\omega,
\xi)+\left[(2u-1)\left(\frac{\overline{q}^2}{\bar{\sigma}^3
M^4}+\frac{1}{{\bar{\sigma}}^2 M^2}\right)+\frac{3}{{\bar{\sigma}^2}
M^2}\right]\frac{\widetilde{X}_{_A}(\omega,\xi)}{m_{B_{(s)}}}-\frac{4(u+3)}{\bar{\sigma}^2
M^2}\, \frac{\widetilde{Y}_{_A}(\omega,\xi)}{m_{B_{(s)}}}
\right.\right.\nonumber\\
&-&\left.\left. \left(\frac{\overline{q}^2}{\bar{\sigma}^3
M^4}+\frac{1}{{\bar{\sigma}}^2
}\right)\frac{\widetilde{\overline{X}}_{_A}(\omega,\xi)}{m^2_{B_{(s)}}}
-\frac{4u}{m^2_{B_{(s)}}} \left(\frac{\overline{q}^2}{\bar{\sigma}^4
M^4}-\frac{m^2_{B_{(s)}}}{{\bar{\sigma}}^2
M^4}-\frac{2}{{\bar{\sigma}}^2
M^2}\right)\widetilde{\widetilde{W}}(\omega,\xi)-
\frac{48u}{\bar{\sigma}^2 M^4}\,
\widetilde{\widetilde{Z}}(\omega,\xi) \right\}\right\}, \nonumber\\
f_{-}(q^{2})&=&-\frac{f_{B_{(s)}}m_{B_{(s)}}^2}{2
f_{S}m_{S}}e^{\frac{m_{S}^2}{M^2}}
\int_{0}^{\sigma_{0}}d\sigma\,e^{-\frac{{\textbf{s}}(\sigma)}{M^2}}
\left\{\frac{(1+\sigma)}{\bar{\sigma}} \varphi_{_+}(\sigma
m_{_{B_{(s)}}})+\frac{\widetilde{\overline{\varphi}}(\sigma
m_{_{B_{(s)}}})}{ \bar{\sigma} m_{_{B_{(s)}}}
}-\frac{4(1+\sigma)}{\bar{\sigma}^2 M^2} g_{_+}(\sigma
m_{_{B_{(s)}}})\right.\nonumber\\
&-&\left. \int_{0}^{\sigma m_{B_{(s)}}}d\omega \int_{\sigma
m_{B_{(s)}}-\omega}^{\infty }\frac{d\xi}{\xi}
\left\{\left[(2u+2)\left(\frac{\overline{q}^2}{\bar{\sigma}^3
M^2}+\frac{1}{\bar{\sigma}^2}\right)-\frac{(2u+1) (1+\sigma)
m_{B_{(s)}}^2}{\bar{\sigma}^2 M^2}\right]
\frac{\overline{\Psi}(\omega,
\xi)}{m_{B_{(s)}}^2}\right.\right.\nonumber\\
&+&\left.\left.\frac{6u(1+\sigma)}{\bar{\sigma}^2
M^2}\,\Psi_{_V}(\omega,
\xi)+\left[\frac{(2u-1)(1+\sigma)}{\bar{\sigma}}\left(\frac{\overline{q}^2}{\bar{\sigma}^3
M^4}+\frac{1}{{\bar{\sigma}}^2
M^2}\right)+\frac{4(u+\bar{\sigma})}{{\bar{\sigma}^3}
M^2}\right] \frac{\widetilde{X}_{_A}(\omega,\xi)}{m_{B_{(s)}}}\right.\right.\nonumber\\
&-&\left.\left.\frac{4(u+3)}{\bar{\sigma}^2 M^2}\,
\frac{\widetilde{Y}_{_A}(\omega,\xi)}{m_{B_{(s)}}}-\left[\frac{(1+\sigma)(2q^2-\bar{\sigma}^2
)}{\bar{\sigma}^4 M^4}+\frac{3+\sigma}{{\bar{\sigma}}^3
}\right]\frac{\widetilde{\overline{X}}_{_A}(\omega,\xi)}{m^2_{B_{(s)}}}
- \frac{48u(1+\sigma)}{\bar{\sigma}^3 M^4}\,
\widetilde{\widetilde{Z}}(\omega,\xi)
\right.\right.\nonumber\\
&+&\left.\left. \frac{4u}{m^2_{B_{(s)}}}
\left(\frac{\overline{q}^2}{\bar{\sigma}^4
M^4}+\frac{(1+\sigma)m^2_{B_{(s)}}}{{\bar{\sigma}}^3
M^4}-\frac{2}{{\bar{\sigma}}^2
M^2}\right)\widetilde{\widetilde{W}}(\omega,\xi) \right\}\right\},
\nonumber\\ \nonumber\\
f_{T}(q^{2})&=&\frac{f_{B_{(s)}}m_{B_{(s)}}}{2f_{S}m_{S}}(m_{B_{(s)}}
+{m_{S}})\,e^\frac{m_{S}^2}{M^2}\int_{0}^{\sigma_{0}}d\sigma\,e^{-\frac{{\textbf{s}}(\sigma)}{M^2}}
\left\{\frac{\varphi_{_+}(\sigma m_{_{B_{(s)}}})}{\bar{\sigma}}\,
+\int_{0}^{\sigma m_{B_{(s)}}}d\omega \int_{\sigma
m_{B_{(s)}}-\omega}^{\infty }\frac{d\xi}{\xi}
\left\{\left[\frac{6u}{\bar{\sigma}^2 M^2}\right] \Psi_{_V}(\omega,
\xi)\right.\right.\nonumber\\
&+&\left.\left. \left[\frac{2u+1}{\bar{\sigma}^2 M^2}\right]
\overline{\Psi}(\omega,
\xi)-\left[\frac{\overline{q}^2}{\bar{\sigma}^4 M^4}+2\right]
\frac{\widetilde{X}_{_A} }{m_{B_{(s)}}}-\left[\frac{\overline{q}^2
}{2\bar{\sigma}^4 M^4}-\frac{1}{{\bar{\sigma}}^2 M^2
}\right]\frac{\widetilde{\overline{X}}_{_A}(\omega,\xi)}{m_{B_{(s)}}}
-\frac{u\, \sigma m_{B_{(s)}}(1+\bar{\sigma})}{\bar{\sigma}^4 M^4}\,
\widetilde{\widetilde{W}}(\omega,\xi)\right.\right.\nonumber\\
&-&\left.\left.\frac{6u\, \sigma (1+\bar{\sigma})}{\bar{\sigma}^4
M^4}\, \widetilde{\widetilde{Z}}(\omega,\xi)\right\}\right\},
\end{eqnarray}
where $\sigma=\omega/m_{B_{(s)}}\,$, $u = (\sigma
m_{B_{(s)}}-\omega)/\xi\,$, ${\textbf{s}}(\sigma)=\sigma\,
m^2_{B_{(s)}} -\frac{\sigma}{\bar{\sigma}}\,q^2\,$,
$\bar{\sigma}=1-\sigma$\,,
$\overline{\varphi}={\varphi}_{_+}-{\varphi}_{_-}$\,,
$\overline{\Psi}=\Psi_{_A}-\Psi_{_V}$\,,
$\overline{q}^2=q^2-\overline{\sigma}^2\,m_{B_{(s)}}^2$\,,
$\sigma_0=\frac{s_0+m_{B_{(s)}}^2-q^2-\sqrt{(s_0+m_{B_{(s)}}^2-q^2)^2-4s_0m_{B_{(s)}}^2}}{2m_{B_{(s)}}^2}$\,,
and:
\begin{eqnarray}\label{eq212}
\begin{aligned}
\widetilde{\varphi}_{_{\pm}}(\sigma m_{B_{(s)}}) &= \int_0^{\sigma
m_{B_{(s)}}} d\tau\, {\varphi}_{_{\pm}}(\tau)\,, &
\widetilde{Y}_{_A}(\omega,\xi) &= \int_0^\omega d\tau\,
Y_{_A}(\tau,\xi)\,, \\
\widetilde{X}_{_A}(\omega,\xi) &= \int_0^\omega d\tau\,
X_{_A}(\tau,\xi)\,,& \widetilde{\overline{X}}_{_A}(\omega,\xi) &=
\int_0^\omega d\tau\, \overline{X}_{_A}(\tau,\xi) \,,\\
\qquad \widetilde{\widetilde{W}}(\omega,\xi) &=\int_0^\xi
d\zeta\,\int_0^\omega d\tau\, W(\tau,\zeta)\,,&
\widetilde{\widetilde{Z}}(\omega,\xi) &= \int_0^\xi
d\zeta\,\int_0^\omega d\tau\, Z(\tau,\zeta)\,.
\end{aligned}
\end{eqnarray}

\newpage
\section{Numerical Analysis}
This section encompasses our numerical analysis of the form factors
$f_{+}$, $f_{-}$, and $f_{T}$ of the semileptonic decays $B_{(s)}\to
S, (S=K^*_0(1430), a_0(1450), f_0(1500))$, branching fractions,
longitudinal lepton polarization asymmetries, and discussion. The
$B$-meson LCSR expressions of the form factors in Eq. (\ref{eq211})
depict that the main input values are the masses and leptonic decay
constants of mesons. In addition, the two- and three-particle DA's
of $B$-meson are the effective terms that must be specified in
calculations of the form factors. The expressions of the form
factors contain also two auxiliary parameters; Borel mass square
$M^2$, and the continuum threshold $s_0$ of the scalar mesons.

The values for the masses and leptonic decay constants of $B, B_s,
K_0^*(1430), a_0(1450)$ and $f_0(1500)$ are given in Table
\ref{T301}.
\begin{table}[th]
\caption{ Masses and leptonic decay constants of mesons in \rm{GeV}
\cite{PDG,Aoki,CheChuYa}.} \label{T301}
\begin{ruledtabular}
\begin{tabular}{cccccc}
\rm{Meson}          & $B$    & $B_s$  & $K_0^*(1430)$  &
$a_0(1450)$& $f_0(1500)$
\\ \hline
\rm{Mass}           & $5.28$ & $5.37$ & $1.43\pm 0.05$ & $1.47\pm 0.02$ & $1.50\pm 0.00$ \\[2mm]
\rm{Decay constant} & $0.19$ & $0.23$ & $0.45\pm 0.05$ & $0.46\pm 0.05$ & $0.49\pm 0.05$\\
\end{tabular}
\end{ruledtabular}
\end{table}

To continue, we need to specify appropriate functions for the DA's
of $B$-meson. The $B$-meson light-cone DA's are the main
nonperturbative input to the QCD description of weak decays
involving light hadrons in the framework of QCD factorization. The
knowledge about the behavior of the higher-twist $B$-meson DA's is
still rather limited due to infrared divergences which appear in
power-suppressed contributions. To overcome the divergences in these
cases, some efforts have been made, including the calculation of
non-perturbative contributions of $B$-meson decays in terms of
increasing twist based on the LCSR approach
\cite{KMO,BraKhod,WangShen,WaWeShLu,BrManOff}. One of the problems
on this way is that the higher-twist $B$-meson DA's involve
contributions of multiparton states and are practically unknown.

In Ref. \cite{JiMan}, authors present a systematic study of the
higher-twist DA's of $B$-meson which give rise to power-suppressed
$1/m_B$ contributions to $B$-decays in final states with energetic
light particles in the framework of QCD factorization. As the main
result, they find that the renormalization group equations for the
three-particle distributions are completely integrable in the large
$N_c$ limit and can be solved exactly. Finally, they study the
general properties of the solutions and suggest two simple models
including the exponential-model (Exp-model), and local duality model
(LD-model) for the higher-twist DA's of $B$-meson with a minimum
number of free parameters which satisfy all tree-level EOM
constraints and can be used in phenomenological studies. Authors in
Ref. \cite{ShenWangWei} construct the LD-model for the twist-five
and -six $B$-meson DA's, in agreement with the corresponding
asymptotic behaviors at small quark and gluon momenta.

The Exp-model is the simplest model based on combining the regime of
low momentum of quarks and gluons with an exponential suppression at
large momentum, whereas the LD-model is based on the duality
assumption to match the $B$-meson state with the perturbative
spectral density integrated over the duality region. In this work,
we apply these two models, Exp and LD models, for estimation of the
semileptonic form factors of $B_{(s)}$ to the light scalar mesons
$K_0^*(1430), a_0(1450)$, and $f_0(1500)$ via the $B$-meson LCSR.
Note that in the $\rm SU(3)_F$ symmetry limit, it is possible to use
the $B$-meson DA's for $B_{s}$-meson.\\

{\textbf{{$\bullet$ Exp-model}}}

Combining the known low momentum behavior with an exponential
fall-off at large quark and gluon momenta, and considering the
normalization conditions, the Exp-model can be obtained
\cite{JiMan}. The shapes of the two-particle DA's
$\varphi_{_+}(\omega), \varphi_{_-}(\omega), g_{_+}(\omega)$ and
$g_{_-}(\omega)$ are presented as:
\begin{equation}\label{eq313}
\begin{aligned}
\varphi_{_+}(\omega)&=\frac{\omega}{\omega_0^2}\,e^{-\omega/\omega_0}\,,
& \varphi_{_-}(\omega)&=\bigg\{\frac{1}{\omega_0}-
\frac{\lambda_E^2-\lambda_H^2}{9\,\omega_0^3}\left[1-2\,\left(\frac{\omega}{\omega_0}\right)
+\frac{1}{2}\left(\frac{\omega}{\omega_0}\right)^2\right]\bigg\}\,e^{-\omega/\omega_0}\,,
\\
g_{_+}(\omega)&=\frac{15}{32\,\omega_0}\omega^2\,e^{-\omega/\omega_0}\,,
& g_{_-}(\omega)&=\bigg\{\frac{3}{4}-
\frac{\lambda_E^2-\lambda_H^2}{12\,\omega_0^3}\left[1-\,\left(\frac{\omega}{\omega_0}\right)
+\frac{1}{3}\left(\frac{\omega}{\omega_0}\right)^2\right]\bigg\}\,\omega\,e^{-\omega/\omega_0}\,.
\end{aligned}
\end{equation}
The values of the parameters $\lambda _{E}^{2}$ and $\lambda
_{H}^{2}$  of the $B$-meson DA's are chosen as: $\lambda
_{E}^{2}=(0.01 \pm 0.01) \,\mbox{GeV}^2$ and $\lambda _{H}^{2}=(0.15
\pm 0.05) \,\mbox{GeV}^2$\cite{Rahimi}. Implementing the EOM
constraint for the Exp-model leads to $\omega_{0}=\lambda_{B_{(s)}}$
\cite{BOLange}. Prediction of the $\lambda_{B}$ and $\lambda_{B_s}$
values are varied in different models \cite{Offen,Beneke,Heller}. In
this work, we use the values for $\lambda_{B}$ and $\lambda_{B_s}$
based on the recent researches. Analyzing the $\bar{B}_{u} \to
\gamma {l}^{-} \bar{\nu}$ decay by the LCSR leads to $\lambda_{B} =
(360\pm 110)\,~ \mbox{MeV}$ \cite{Janowski}. On the other hand, the
inverse moment of the $B_s$-meson DA is predicted from the QCDSR as
$\lambda_{B_s}=(438 \pm 150)~\rm{MeV}$ \cite{KhMaMa}. Therefore
\begin{eqnarray} \label{eq314}
\omega_0= \left\{
\begin{array}{l}
\lambda_{B}=0.360\pm 0.110 \,\, {\rm GeV}   \, \quad \hspace{1.6 cm}
({\rm For}\, B{\rm-meson}), \vspace{0.25 cm} \\
\lambda_{B_{s}}=0.438\pm 0.150 \,\, {\rm GeV}  \, \quad \hspace{1.5
cm} ({\rm For}\, B_s{\rm-meson}).
\end{array}
\hspace{0.5 cm} \right.
\end{eqnarray}

\begin{figure}[th]
\begin{center}
\includegraphics[width=4cm,height=4cm]{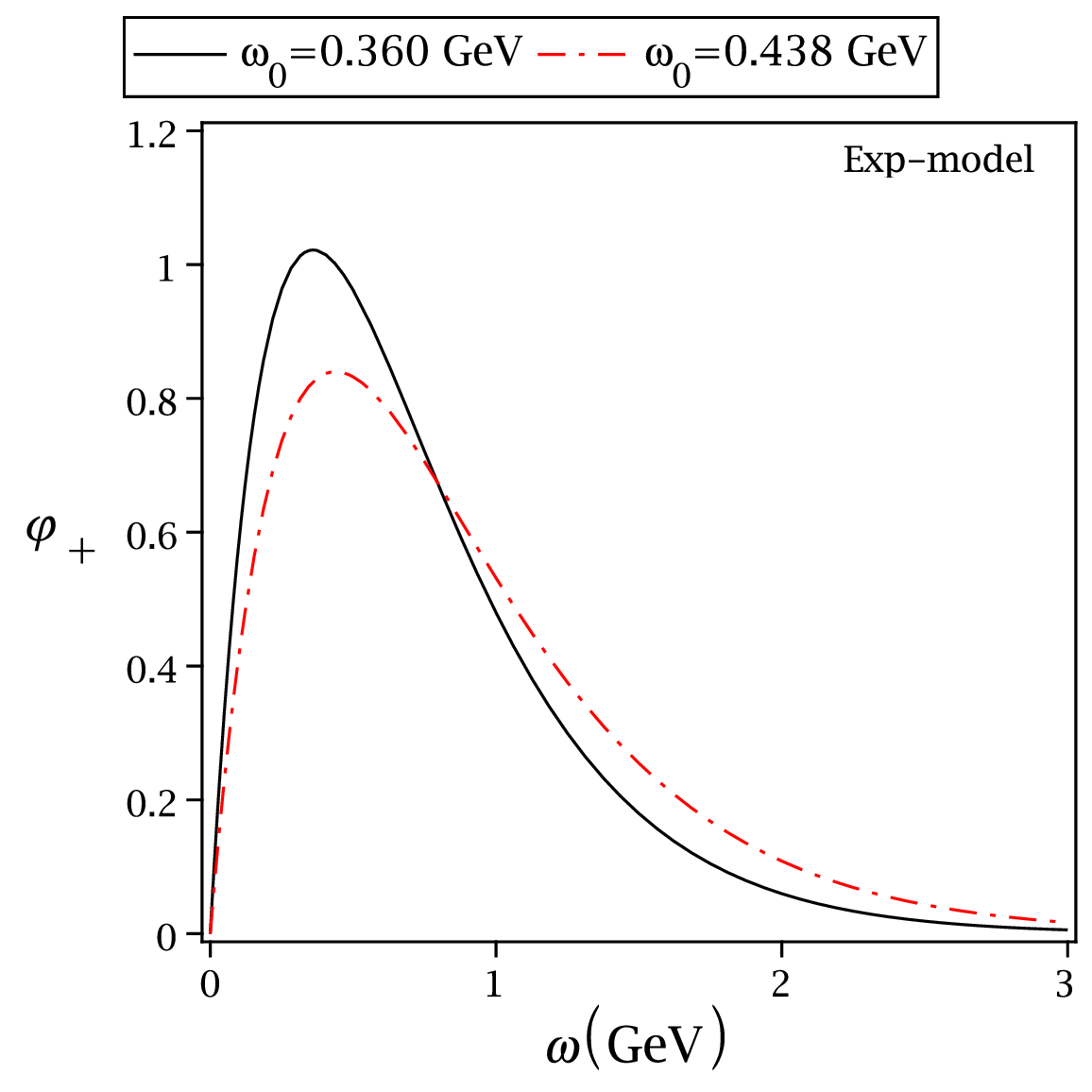}
\includegraphics[width=4cm,height=4cm]{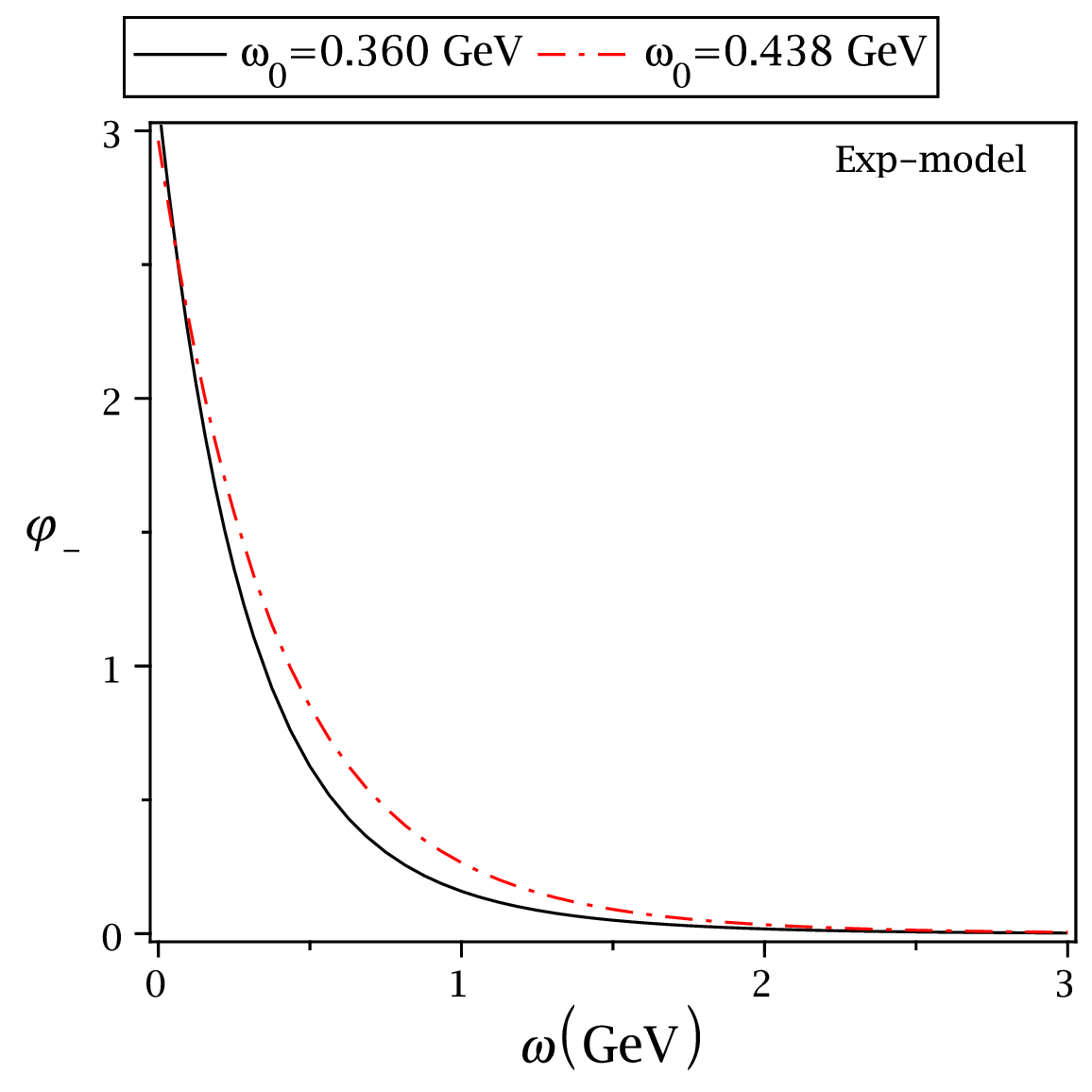}
\includegraphics[width=4cm,height=4cm]{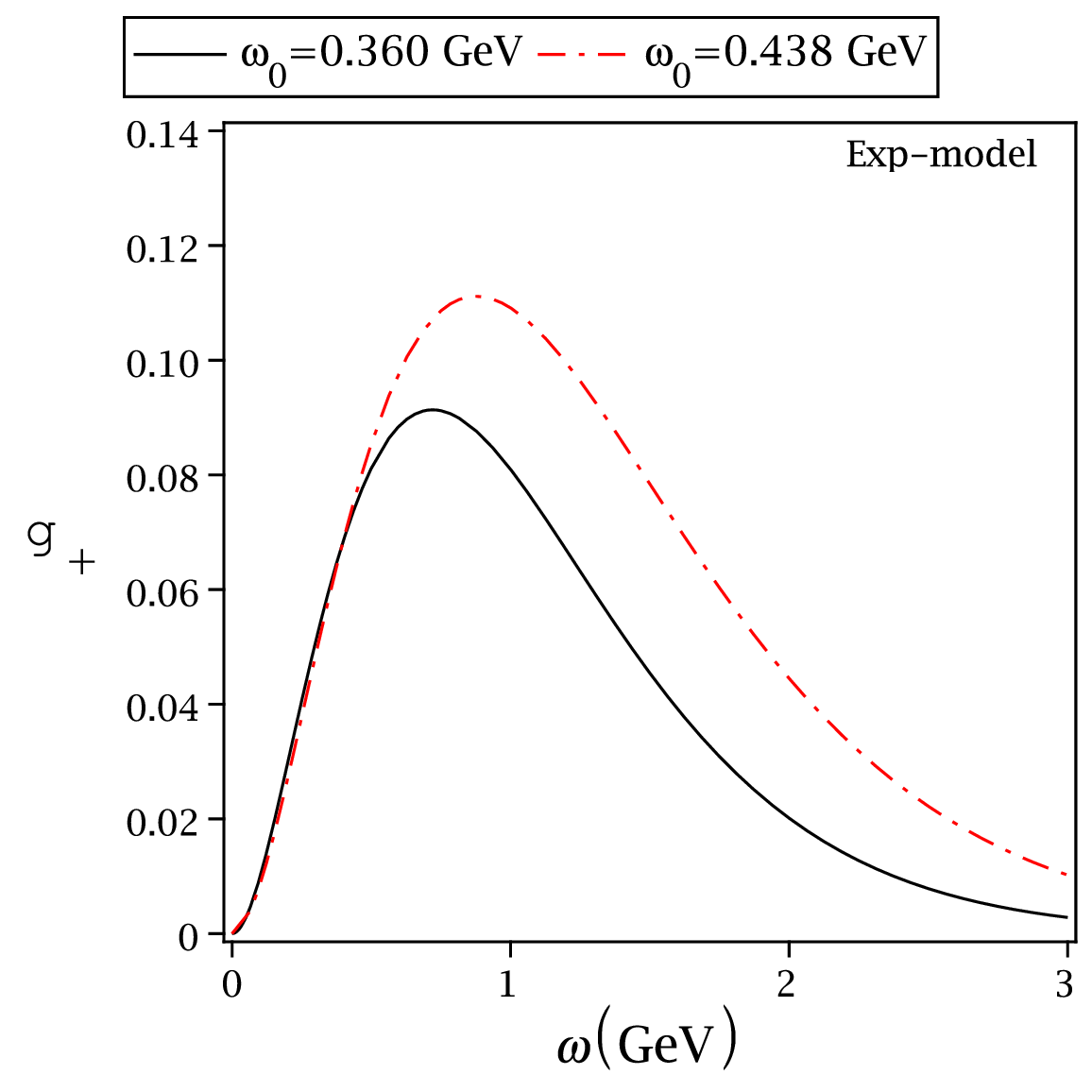}
\includegraphics[width=4cm,height=4cm]{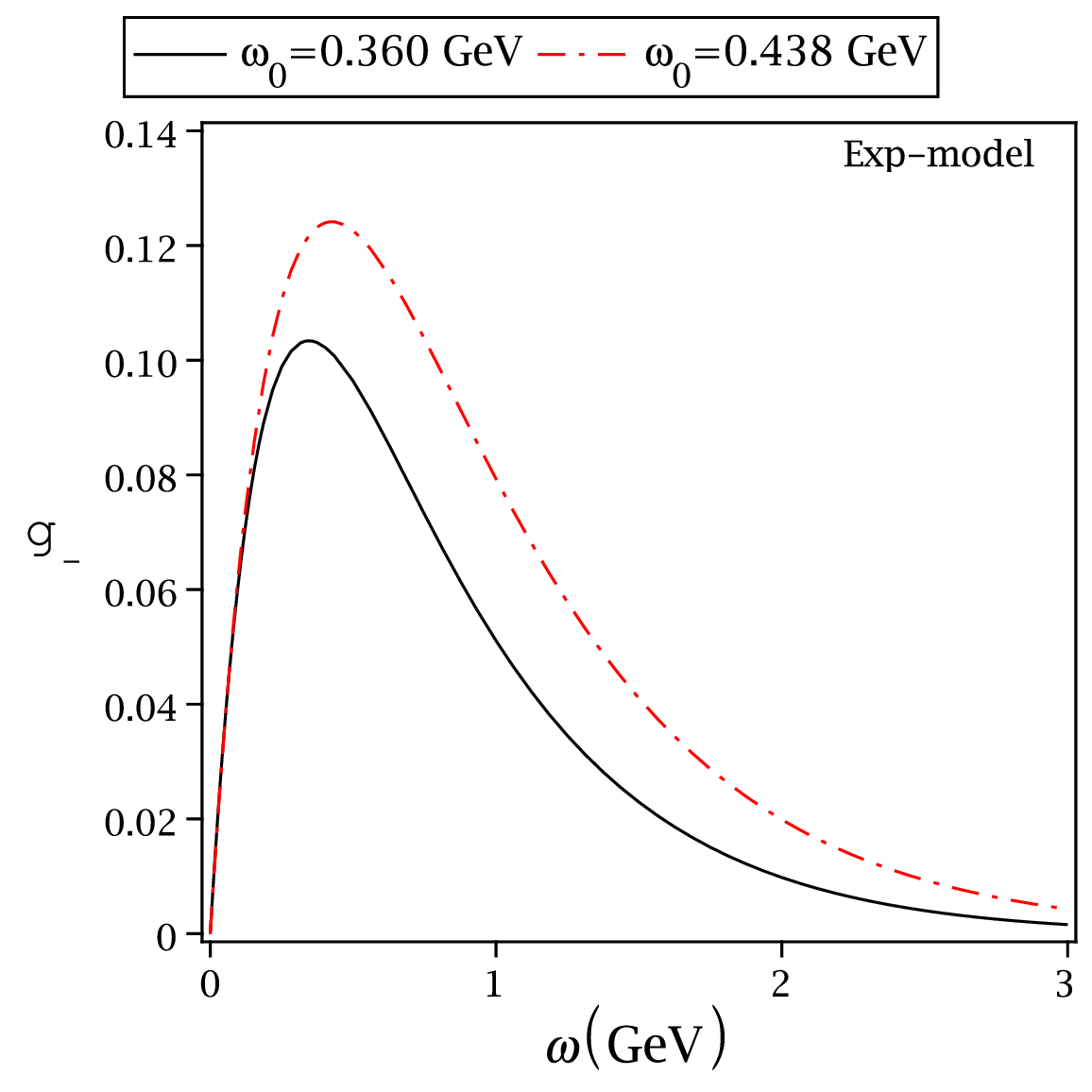}
\caption{The dependence of the two-particle DA's,
$\varphi_{_+}(\omega), \varphi_{_-}(\omega), g_{_+}(\omega)$, and
$g_{_-}(\omega)$ on $\omega$ for the Exp-model.} \label{F301}
\end{center}
\end{figure}

The dependence of the two-particle DA's in Eq. (\ref{eq313}) with
respect to $\omega$ is shown in Fig. \ref{F301} for the Exp-model.

The three-particle DA's of $B$-meson up to twist-six in the
Exp-model can be constructed as
\begin{equation}\label{eq315}
\begin{aligned}
\phi_3(\omega, \xi)&={\lambda_E^2 - \lambda_H^2 \over 6 \,
\omega_0^5} \, \omega \, \xi^2 \, e^{-(\omega + \xi)/\omega_0} \,, &
\phi_4(\omega, \xi)&={\lambda_E^2 + \lambda_H^2 \over 6 \,
\omega_0^4} \, \xi^2 \,
e^{-(\omega + \xi)/\omega_0} \,,\\
\psi_4(\omega, \xi)&={\lambda_E^2 \over 3 \, \omega_0^4} \, \omega
\, \xi \, e^{-(\omega + \xi)/\omega_0} \,,&
\overline{\psi}_4(\omega, \xi) &= {\lambda_H^2 \over 3 \,
\omega_0^4} \, \omega \, \xi \, e^{-(\omega +
\xi)/\omega_0} \,, \\
\psi_5(\omega, \xi)&=- {\lambda_E^2 \over 3 \, \omega_0^3} \, \xi \,
e^{-(\omega + \xi)/\omega_0} \,, & \overline{\psi}_5(\omega, \xi)&=
- {\lambda_H^2 \over 3 \, \omega_0^3} \, \xi \, e^{-(\omega +
\xi)/\omega_0} \,, \\
\phi_5(\omega, \xi)&={\lambda_E^2 +
\lambda_H^2 \over 3 \, \omega_0^3} \, \omega \, e^{-(\omega +
\xi)/\omega_0} \,, & \phi_6(\omega, \xi)&={\lambda_E^2 - \lambda_H^2
\over 3 \, \omega_0^2} \, e^{-(\omega + \xi)/\omega_0} \,.
\end{aligned}
\end{equation}

{\textbf{{$\bullet$ LD-model }}}

\begin{figure}[th]
\begin{center}
\includegraphics[width=4cm,height=4cm]{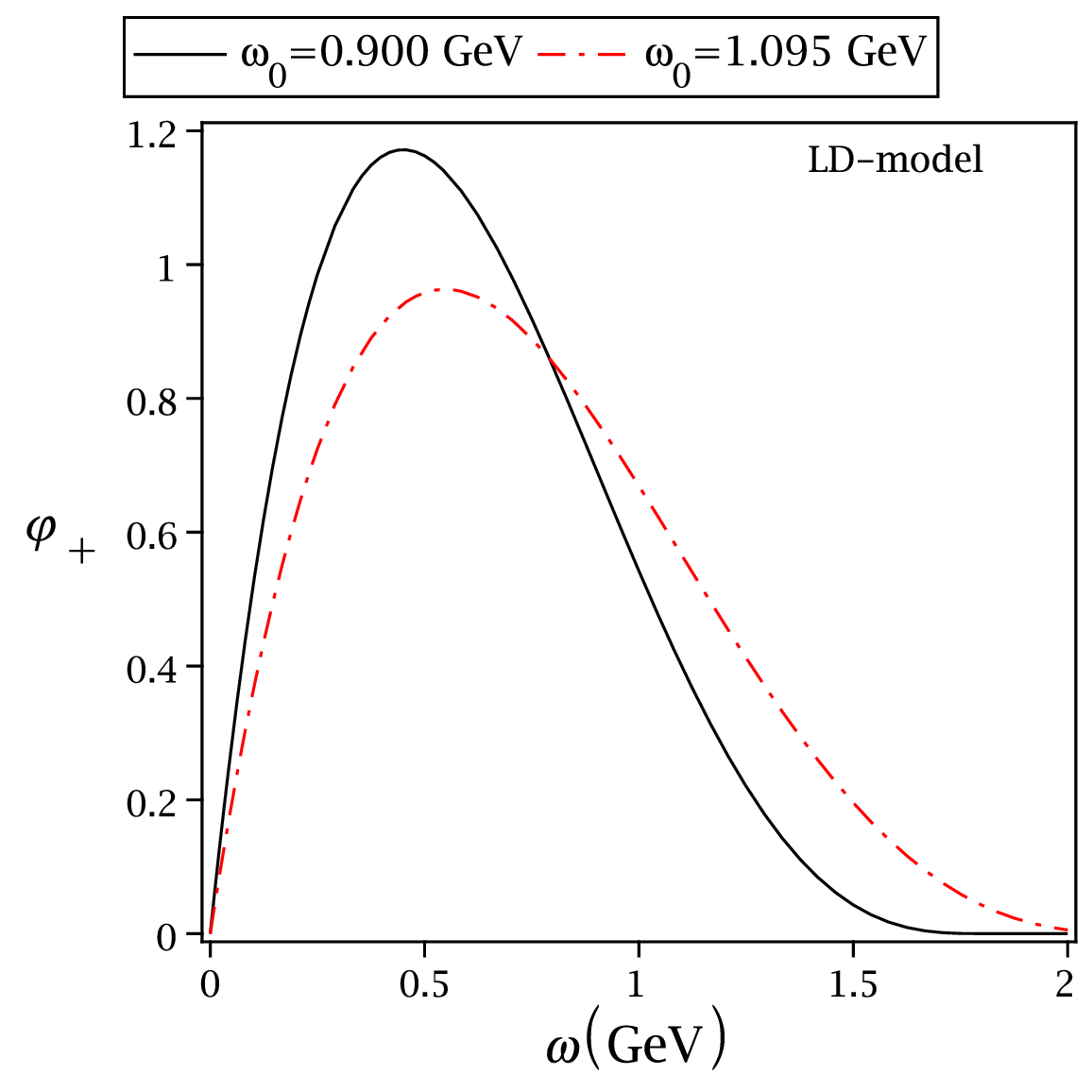}
\includegraphics[width=4cm,height=4cm]{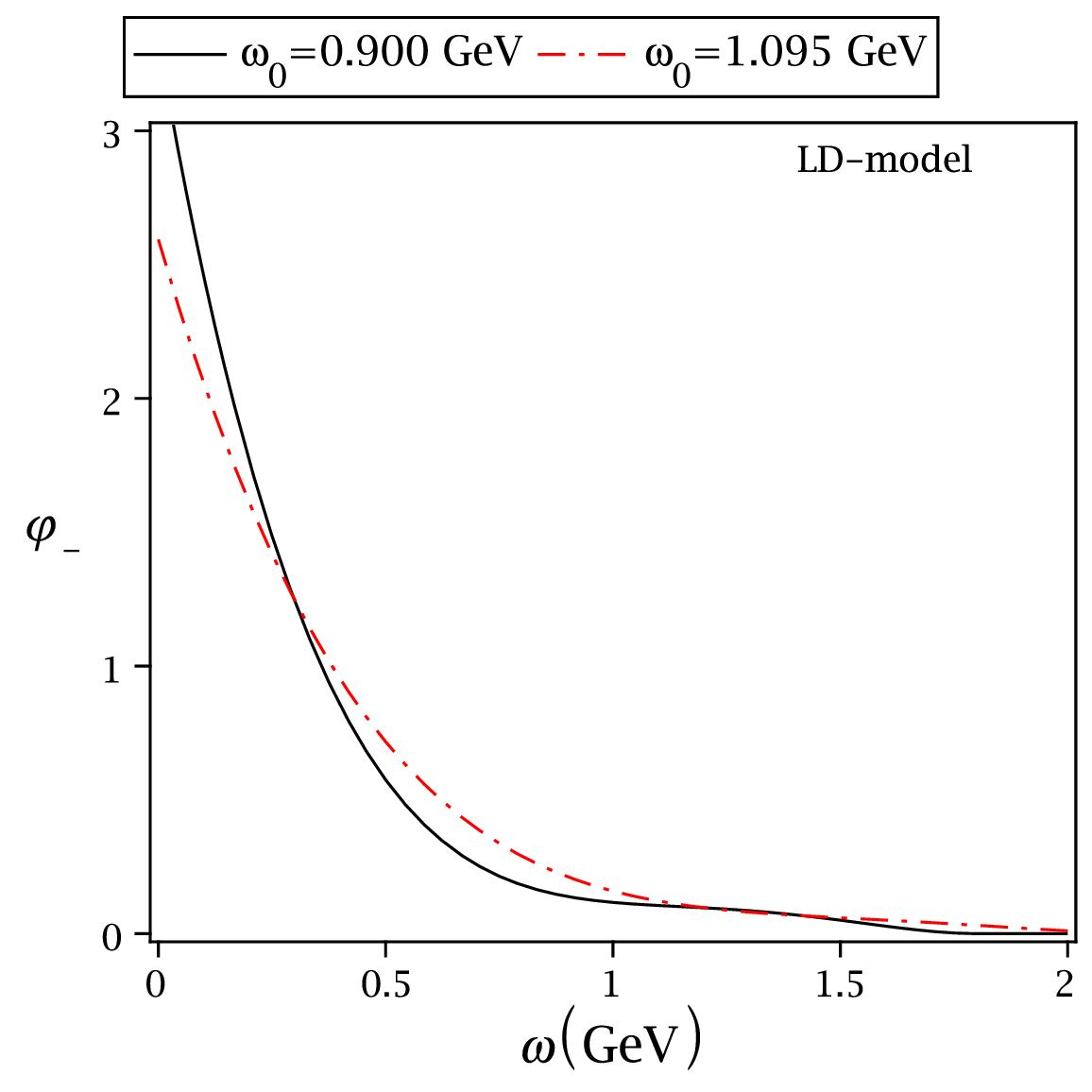}
\includegraphics[width=4cm,height=4cm]{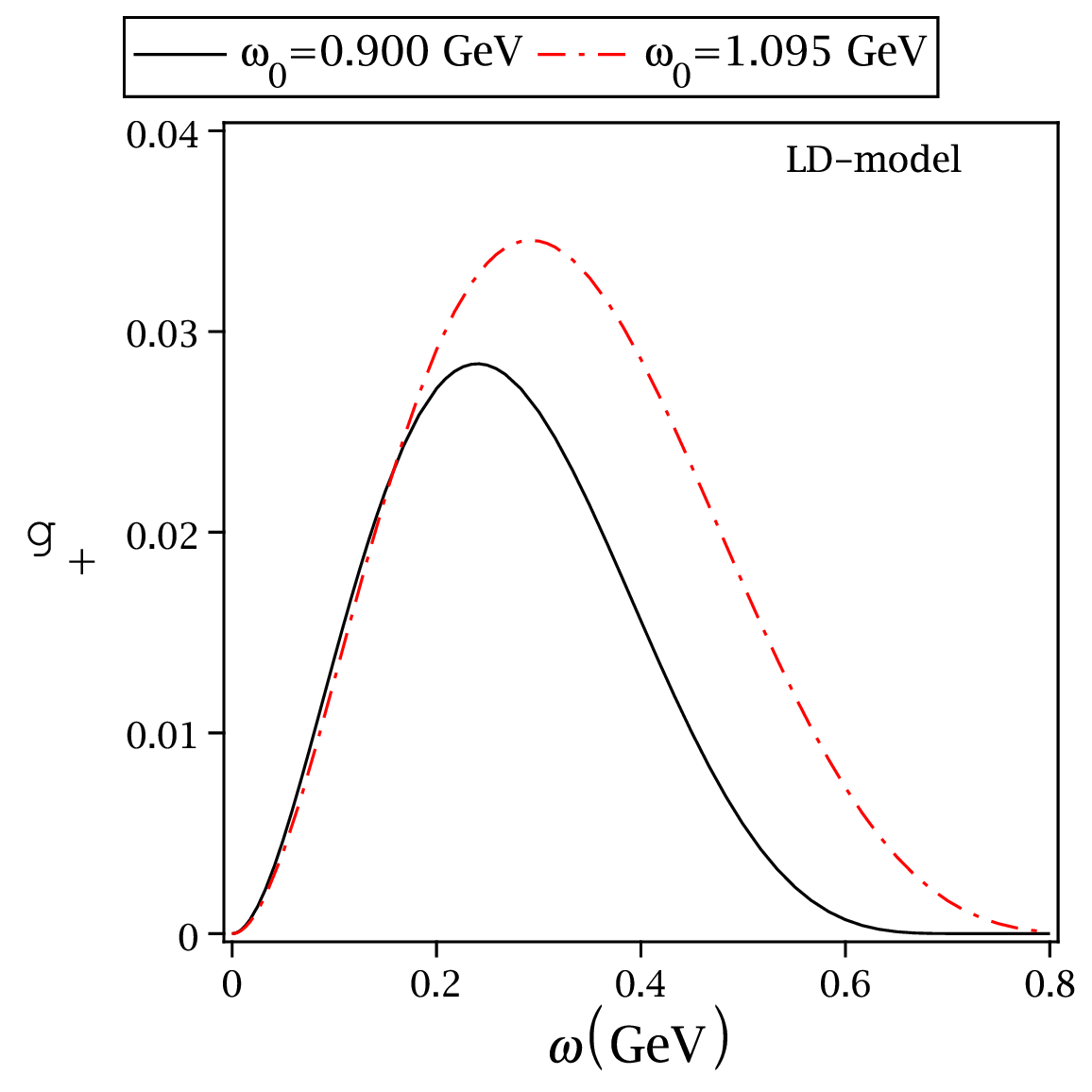}
\includegraphics[width=4cm,height=4cm]{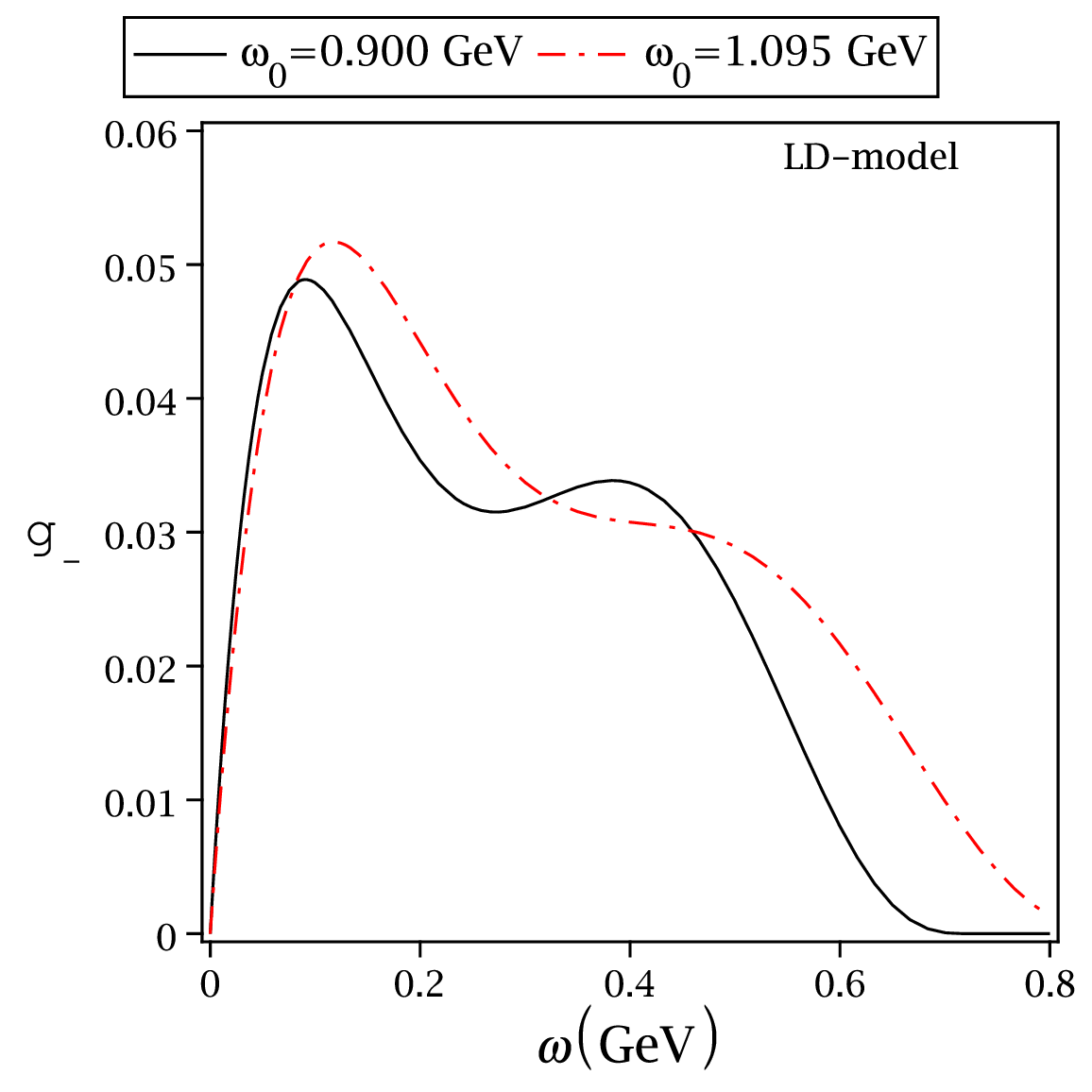}
\caption{The same as Fig. \ref {F301} but for the LD-model.}
\label{F302}
\end{center}
\end{figure}

Another class of the phenomenological models for the $B$-meson DA's
is the LD-model. In this model, the structures of the two-particle
DA's $\varphi_{_+}(\omega), \varphi_{_-}(\omega), g_{_+}(\omega)$
and $g_{_-}(\omega)$ are presented as \cite{JiMan}:
\begin{eqnarray}\label{eq316}
\varphi_{_+}(\omega) &=& {5 \over 8 \, \omega_0^5} \, \omega(2 \,
\omega_0 - \omega)^3 \,
\theta(2 \, \omega_0 - \omega)\,,  \nonumber  \\
\varphi_{_-}(\omega) &=& {5 (2 \, \omega_0 - \omega)^2 \over 192\,
\omega_0^5}  \, \bigg \{6 \, (2 \, \omega_0 - \omega)^2 - {7 \,
(\lambda_E^2 - \lambda_H^2) \over \omega_0^2} \, (15 \, \omega^2 -
20 \, \omega \, \omega_0 + 4 \, \omega_0^2) \bigg
\} \,\theta(2 \, \omega_0 - \omega)\,,  \nonumber  \\
g_{_+}(\omega)&=&\frac{115}{2048\, \omega_{0}^5}\,\omega^2\,
(2\,\omega_0-\omega)^4\,\theta(2\omega_0-\omega)\,,  \nonumber  \\
g_{_-}(\omega) &=& {\omega \, (2 \, \omega_0 - \omega)^3 \over
\omega_0^5} \, \left \{ {5 \over 256} \, (2 \, \omega_0 - \omega)^2
- {35 \, (\lambda_E^2 - \lambda_H^2) \over 1536} \, \left [ 4 - 12
\, \left ( {\omega \over \omega_0} \right ) + 11 \, \left ( {\omega
\over \omega_0} \right )^2  \right ]\right \}\, \theta(2 \, \omega_0
- \omega) \,.
\end{eqnarray}
Applying the EOM constraint between the leading-twist and the
higher-twist $B$-meson DA's, the HQET parameters entering the
LD-model for the $B_{(s)}$-meson DA's must satisfy the relation
$\omega_{0}= \frac{5}{2} \lambda_{B_{(s)}}$ \cite{JiMan}. So
according to Eq. (\ref{eq314}), we have
\begin{eqnarray} \label{eq317}
\omega_0= \left\{
\begin{array}{l}
\lambda_{B}=0.900\pm 0.275 \,\, {\rm GeV} \, \qquad \hspace{1.6 cm}
({\rm For}\, B{\rm-meson}), \vspace{0.25 cm} \\
\lambda_{B_{s}}=1.095\pm 0.375 \,\, {\rm GeV} \, \qquad \hspace{1.5
cm} ({\rm For}\, B_s{\rm-meson}).
\end{array}
 \hspace{0.5 cm} \right.\,.
\end{eqnarray}
The dependence of the two-particle DA's in Eq. (\ref{eq316}) with
respect to $\omega$ is shown in Fig. \ref{F302} for the LD-model.

\begin{figure}[th]
\begin{center}
\includegraphics[width=6cm,height=5cm]{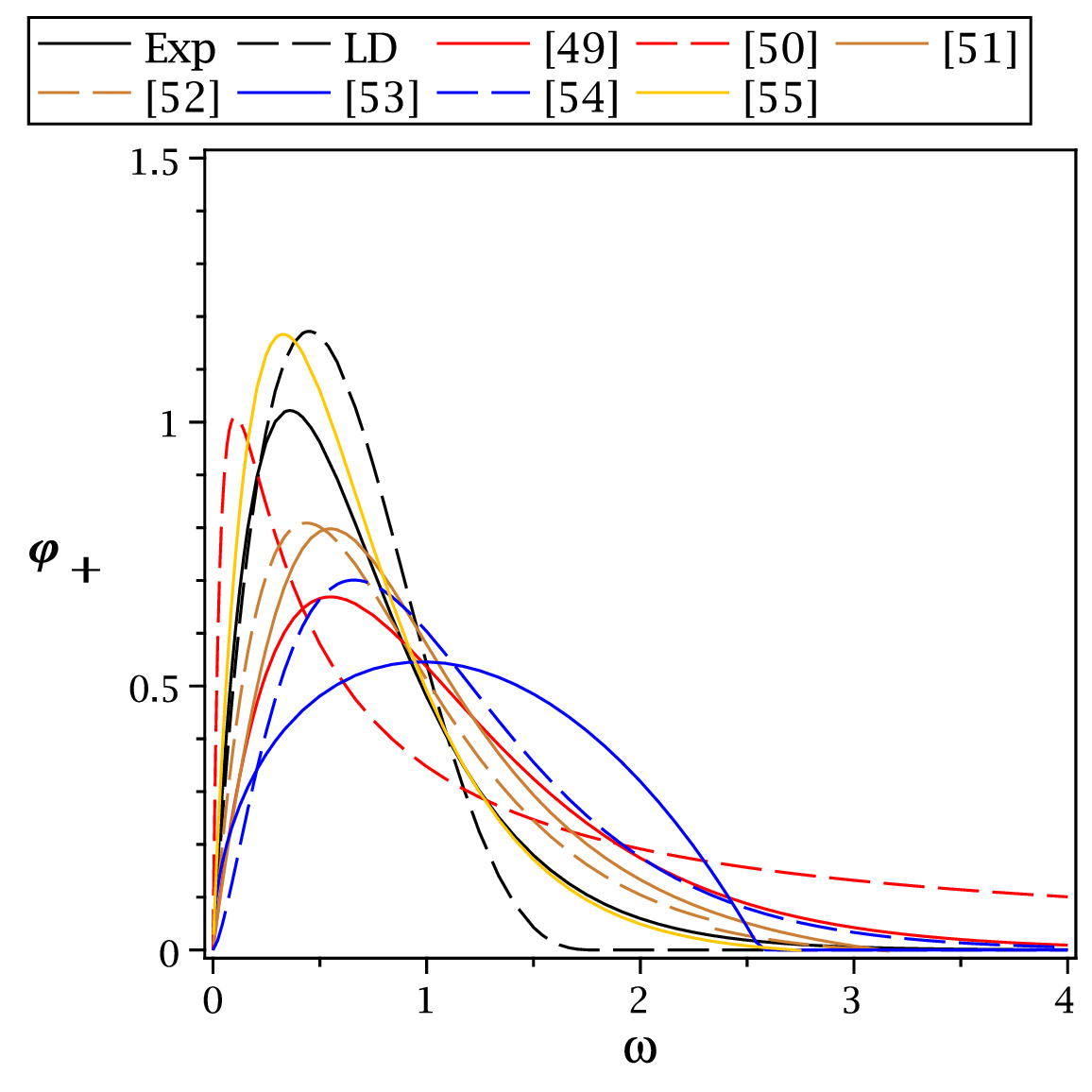}
\caption{The shape of $\varphi_{_{+}}(\omega)$ for $B$-meson in the
Exp and LD and other models. } \label{F303}
\end{center}
\end{figure}

The three-particle DA's of $B$-meson are derived in the LD-model up
to twist-six as \cite{JiMan,ShenWangWei}
\begin{eqnarray}\label{eq318}
\phi_3(\omega, \xi) &=&  {105 \, (\lambda_E^2 - \lambda_H^2) \over 8
\, \omega_0^7} \, \omega \, \xi^2 \, \left (\omega_0 - {\omega + \xi
\over 2} \right )^2  \,
\theta(2 \, \omega_0 - \omega - \xi)  \,,  \nonumber  \\
\phi_4(\omega, \xi) &=&  {35 \, (\lambda_E^2 + \lambda_H^2) \over 4
\, \omega_0^7} \, \xi^2 \, \left (\omega_0 - {\omega + \xi \over 2}
\right )^3  \,
\theta(2 \, \omega_0 - \omega - \xi)  \,,  \nonumber  \\
\psi_4(\omega, \xi) &=&  {35 \, \lambda_E^2 \over 2 \, \omega_0^7}
\, \omega \, \xi \, \left (\omega_0 - {\omega + \xi \over 2} \right
)^3  \,
\theta(2 \, \omega_0 - \omega - \xi)  \,,  \nonumber  \\
\overline{\psi}_4(\omega, \xi) &=&  {35 \, \lambda_H^2 \over 2 \,
\omega_0^7} \, \omega \, \xi \, \left (\omega_0 - {\omega + \xi
\over 2} \right )^3  \, \theta(2 \,
\omega_0 - \omega - \xi)  \,, \nonumber \\
\phi_5(\omega, \xi)  &=& \frac{35\, (\lambda_E^2 +
\lambda_H^2)}{64\, \omega_0^7}\, \omega  \, (2 \, \omega_0 -
\omega-\xi)^4 \,
\theta(2 \, \omega_0 - \omega - \xi)  \,, \nonumber \\
\psi_5(\omega, \xi)  &=& - \frac{35\, \lambda_E^2}{ 64\,\omega_0^7}
\, \xi\, (2 \, \omega_0 - \omega-\xi)^4 \,
\theta(2 \, \omega_0 - \omega - \xi)  \,, \nonumber \\
\overline{\psi}_5(\omega, \xi)  &=& - \frac{35\, \lambda_H^2}{
64\,\omega_0^7} \, \xi\, (2 \, \omega_0 - \omega-\xi)^4 \, \theta(2
\, \omega_0 - \omega - \xi) \,,
\nonumber \\
\phi_6(\omega, \xi)  &=& \frac{7\, (\lambda_E^2 - \lambda_H^2)}{64\,
\omega_0^7}\, (2 \, \omega_0 - \omega-\xi)^5 \, \theta(2 \, \omega_0
- \omega - \xi)  \,.
\end{eqnarray}
The two-particle leading-twist DA of $B$-meson,
$\varphi_{_{+}}(\omega)$ has the most important contribution in
estimation of the form factors. The evolution effects show that the
DA $\varphi_{_{+}}(\omega)$ satisfies the condition
$\varphi_{_+}(\omega) \sim \omega$ as $\omega \rightarrow 0$ and
falls off slower than $1/\omega$ for $\omega \rightarrow \infty$
\cite{BraManash}. In Fig. {\ref{F303}}, the shape of
$\varphi_{_{+}}(\omega)$ in the Exp and LD models is compared with
those proposed by other models in Refs.
\cite{Grozin,GenonSachrajda,Braun,LeeNeubert,FaFeHu,BeneBraun,Galda}.

\begin{figure}[th]
\begin{center}
\includegraphics[width=4cm,height=4cm]{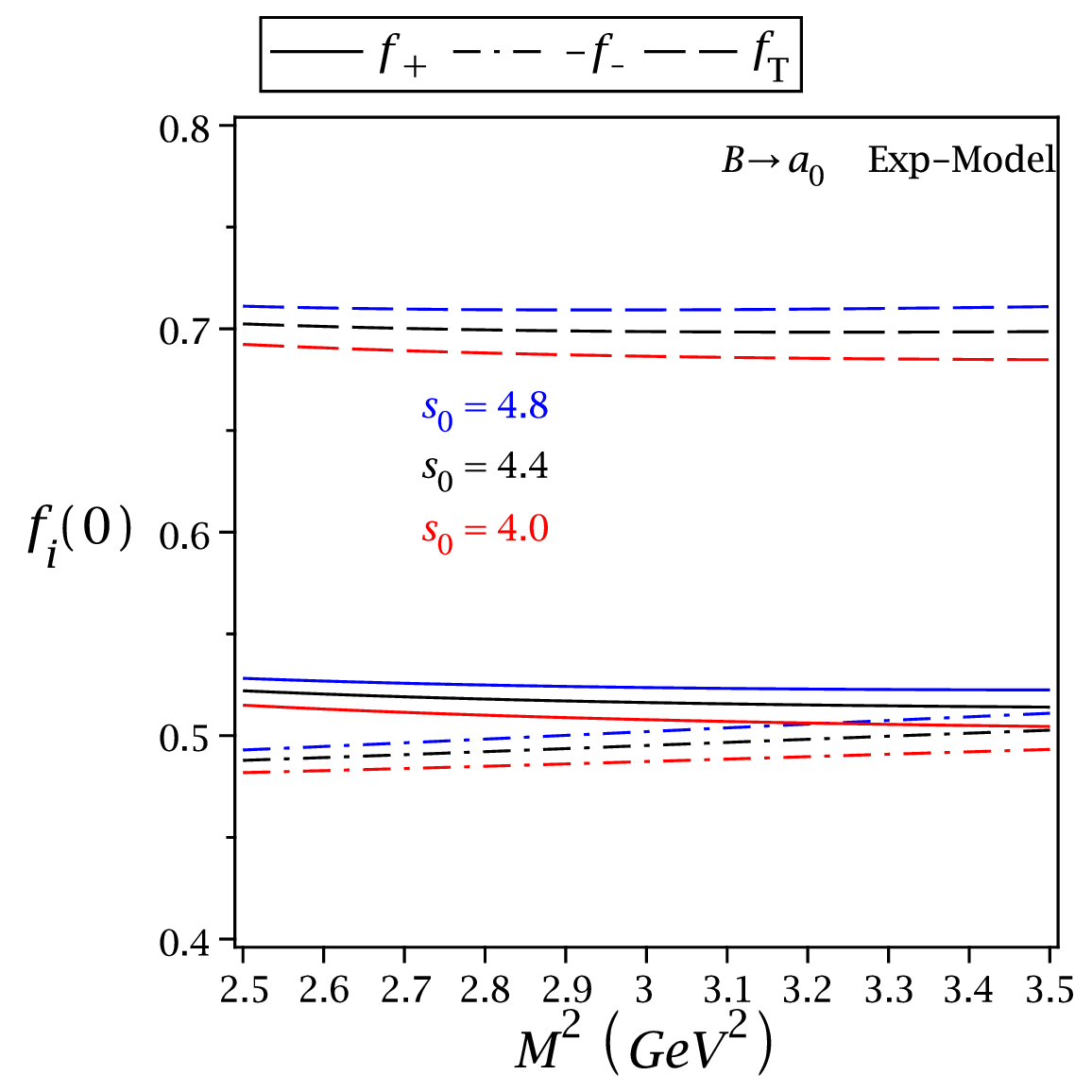}
\includegraphics[width=4cm,height=4cm]{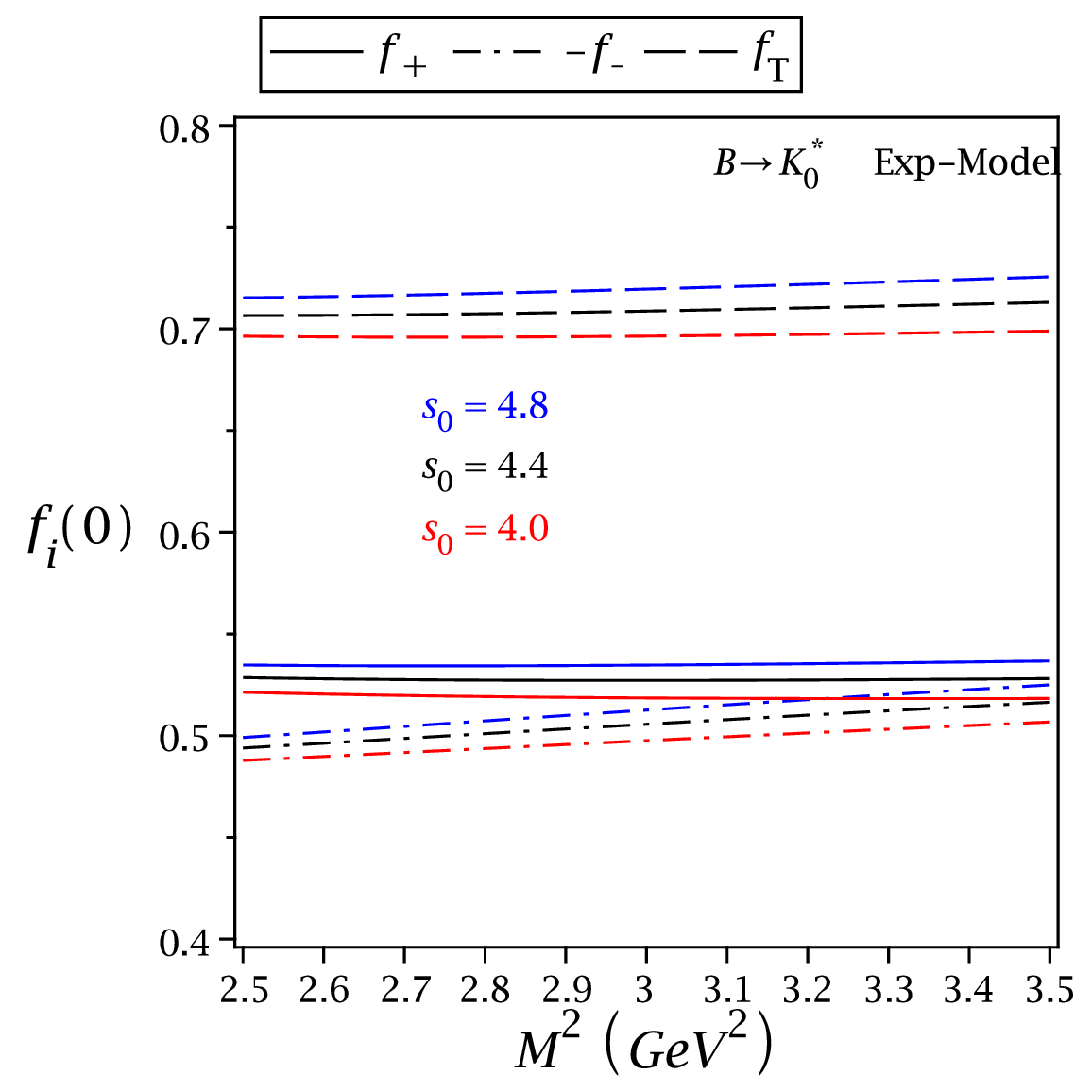}
\includegraphics[width=4cm,height=4cm]{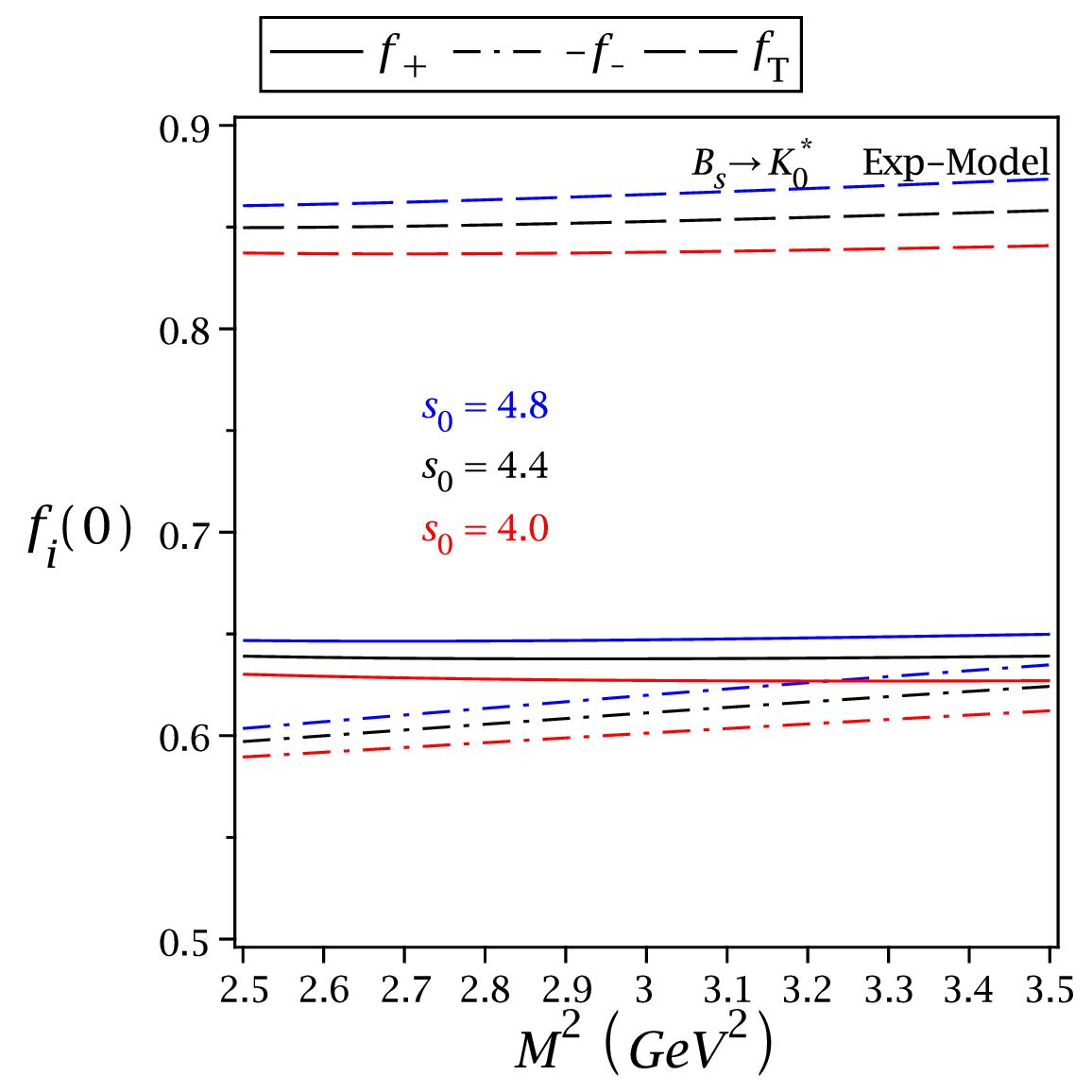}
\includegraphics[width=4cm,height=4cm]{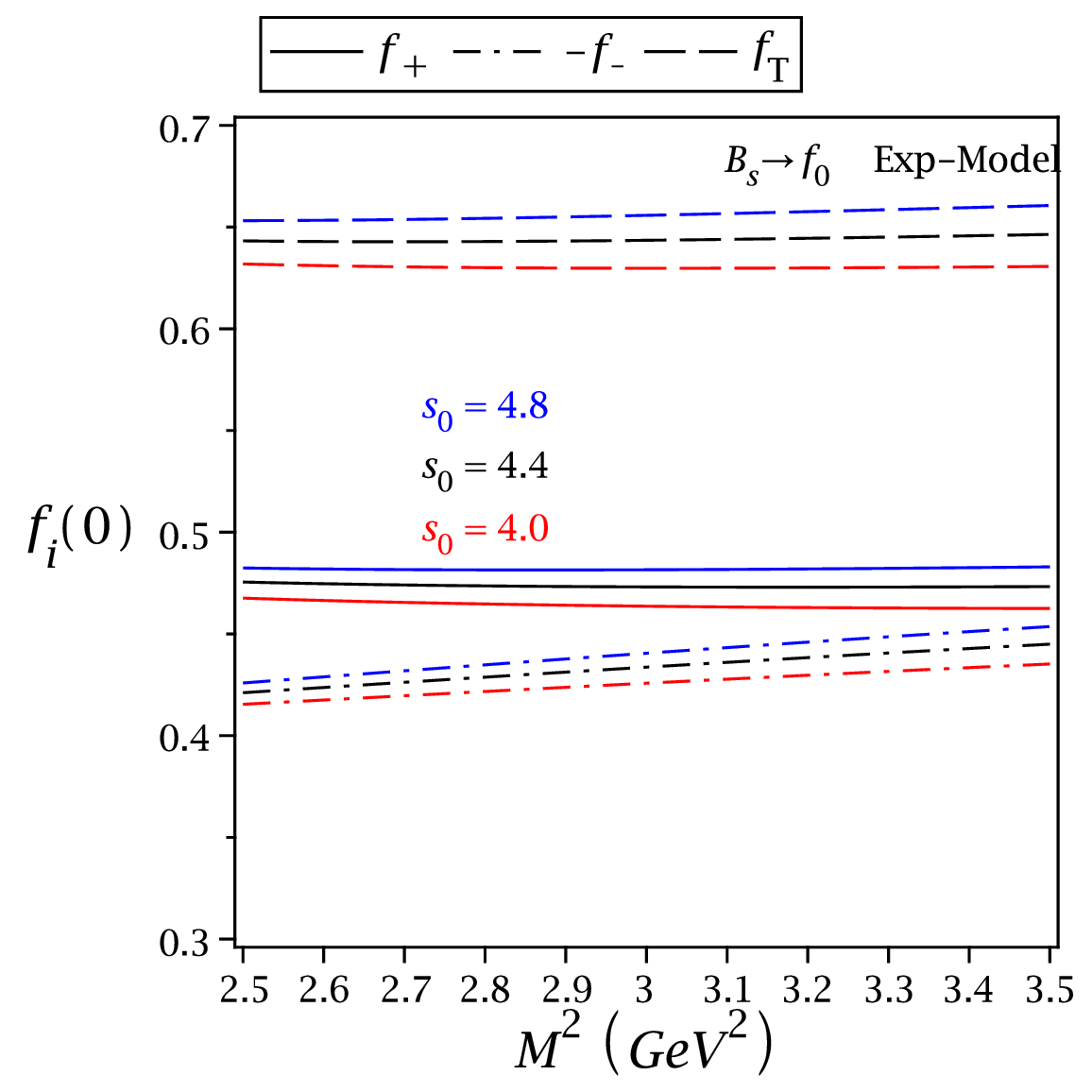}
\caption{The dependence of the form factors $f_+$, $-f_{-}$ and
$f_{T}$ on the Borel parameter $M^2$ at three fixed values $s_0=4.0,
~4.4$, and $4.8$ for $B\to (a_0, K_0^*)$ and $B_{s}\to (K_0^*, f_0)$
at $q^2=0$ in the Exp-model.} \label{F304}
\end{center}
\end{figure}

\begin{figure}[th]
\begin{center}
\includegraphics[width=4cm,height=4cm]{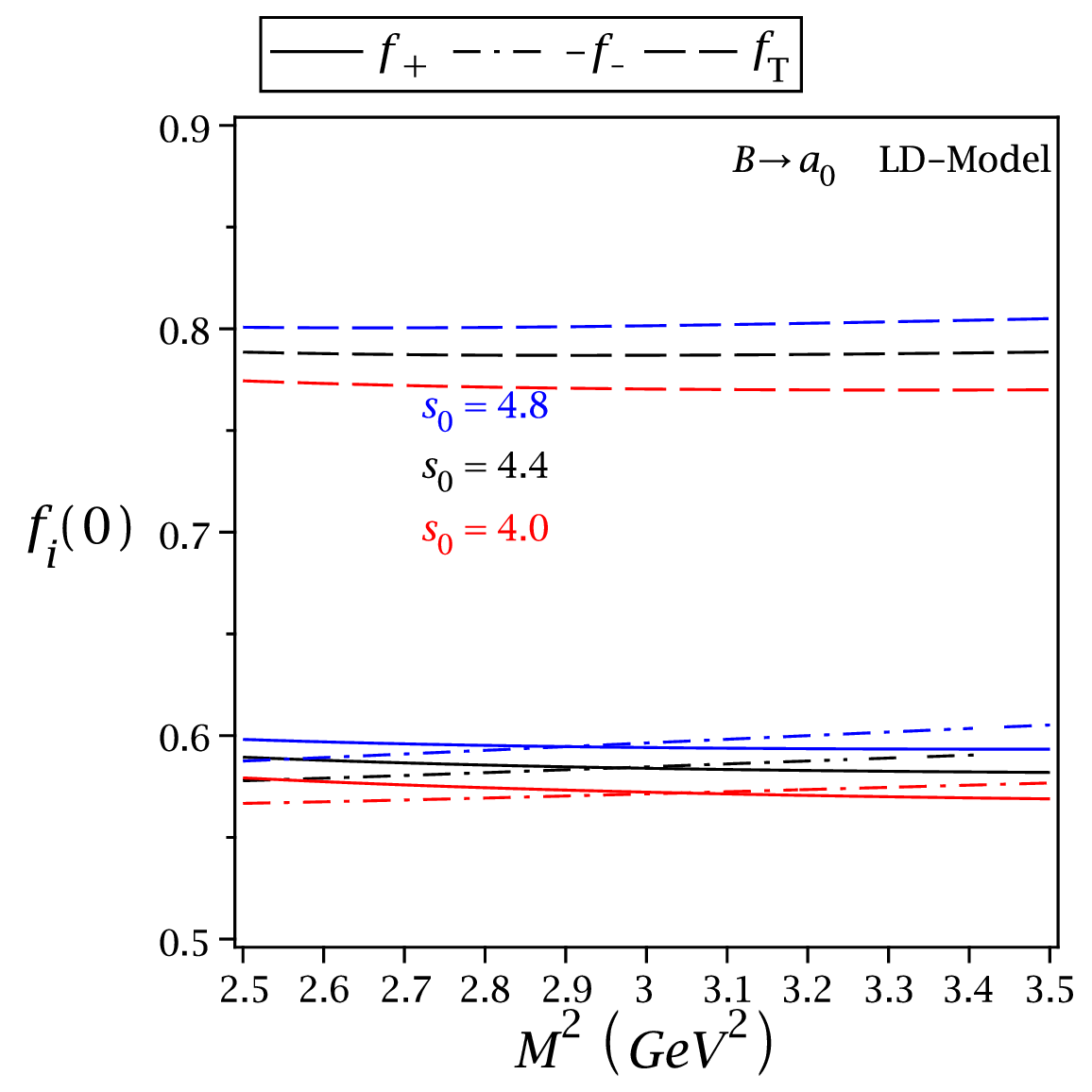}
\includegraphics[width=4cm,height=4cm]{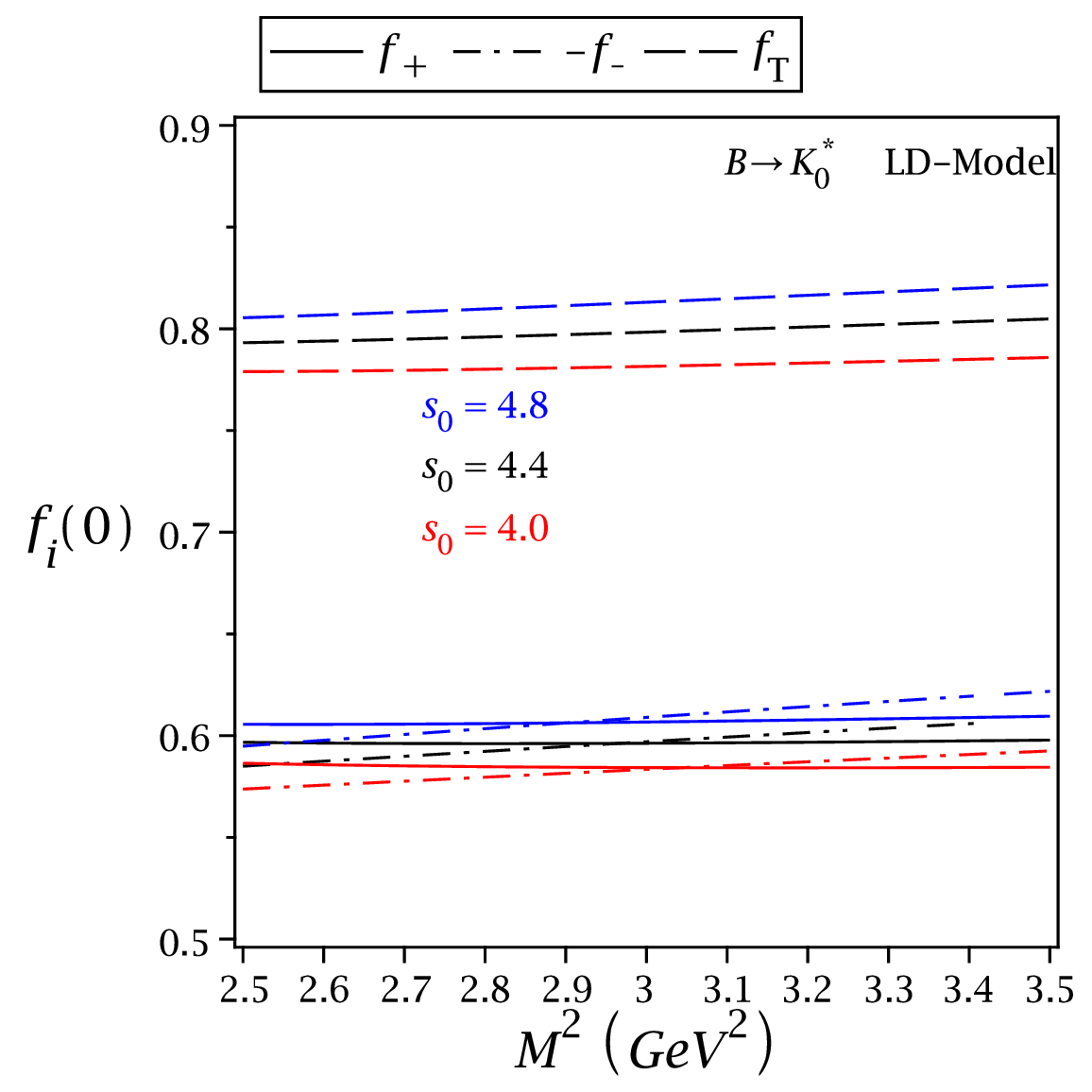}
\includegraphics[width=4cm,height=4cm]{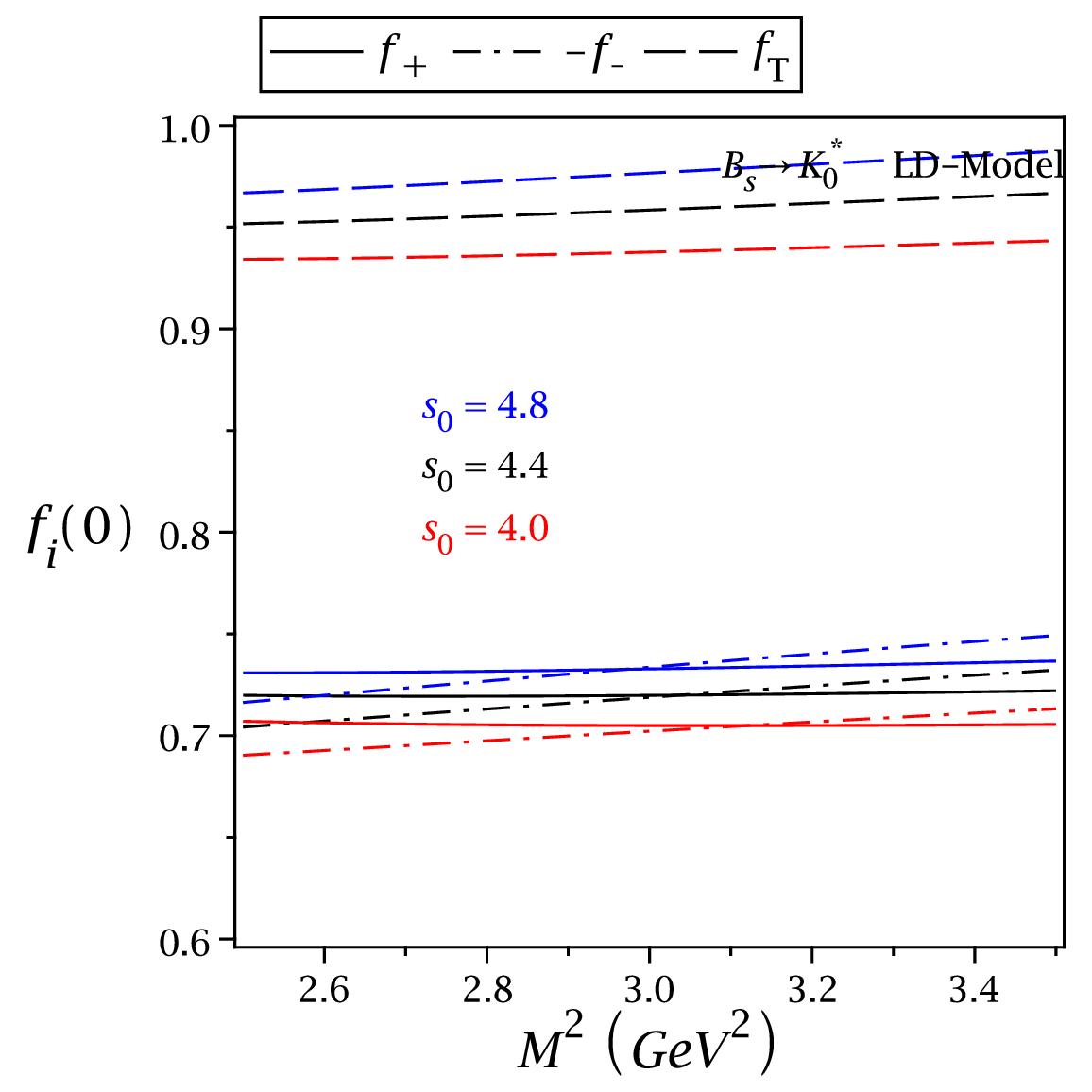}
\includegraphics[width=4cm,height=4cm]{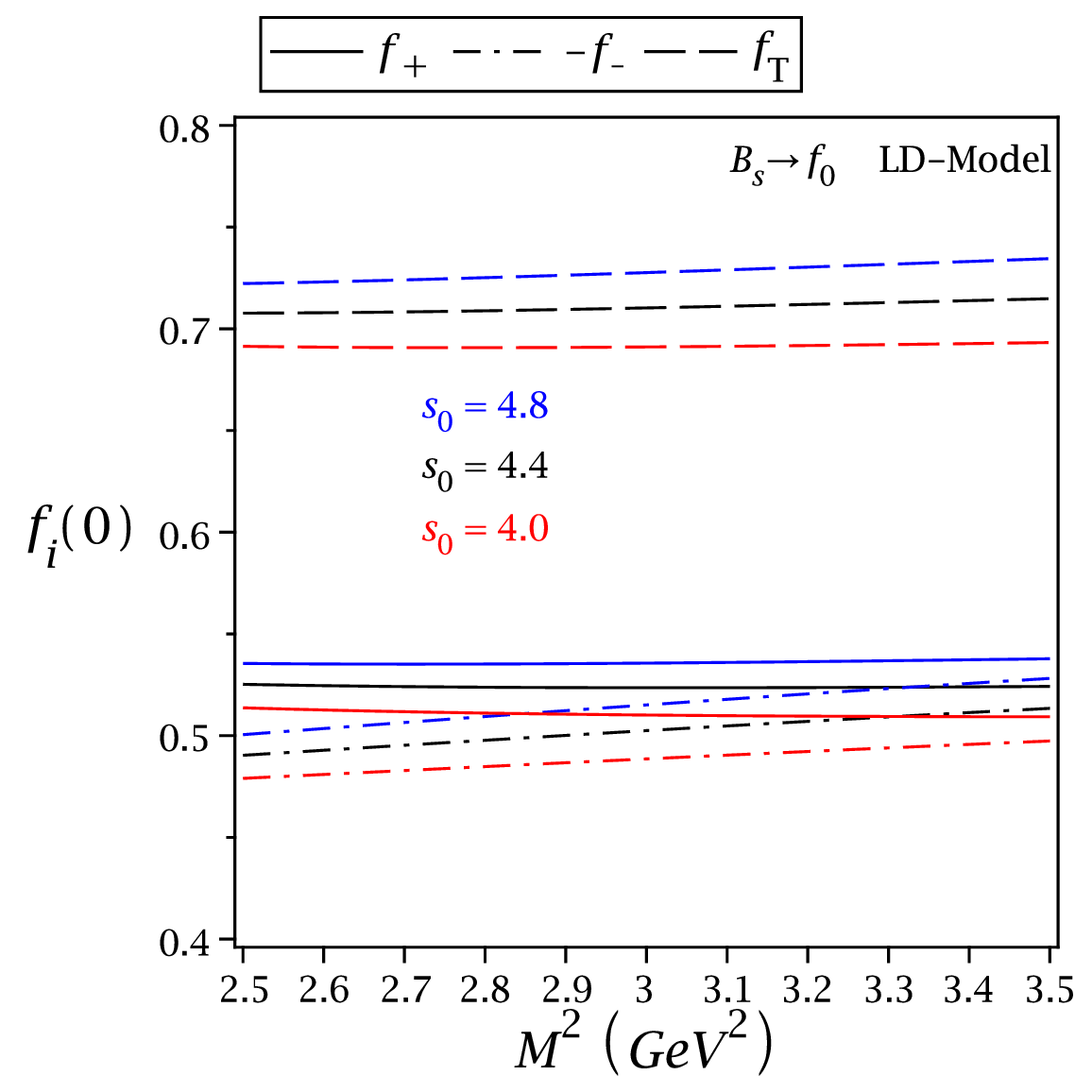}
\caption{The same as Fig. \ref{F304} but for the LD-model.}
\label{F305}
\end{center}
\end{figure}

After introducing the DA's of $B$-meson, we set the values of the
parameters. There are two auxiliary parameters in Eq. (\ref{eq211});
the Borel mass square $M^2$ and the continuum threshold $s_0$, that
the values of which must be determined before analyzing the form
factors of the semileptonic $B_{(s)}\to S$ decays. These parameters
are not physical quantities, so the form factors as physical
quantities should be independent of them. The continuum threshold
$s_0$ is not completely arbitrary and it is related to the energy of
the first exited state of the scalar meson. The working region for
the continuum threshold for the scalar mesons $K_0^*(1430),
a_0(1450)$ and $f_0(1500)$ is taken to be $s_0=(4.4 \pm 0.4)\,
\mbox{GeV}^2$ \cite{Dong}. The Borel parameter $M^2$ is chosen in
the region where (a) the contributions of the higher states and
continuum are effectively suppressed, which can ensure that the sum
rule does not sensitively depend on the approximation for the higher
states and continuum, and (b) the contributions of the condensates
should not be too large, which can ensure that the contributions of
the higher-dimensional operators are small and the truncated OPE is
effective. Considering the central value $4.4\, \mbox{GeV}^2$ for
$s_0$, a good stability of the form factors with respect to the
Borel parameter is obtained at $q^2=0$ in the interval $2.5\,
\mbox{GeV}^2\leq M^2\leq 3.5\,\mbox{GeV}^2$. The dependence of the
form factors $f_+$, $f_{-}$ and $f_{T}$ for the semileptonic decays
$B\to (a_0, K_0^*)$ and $B_{s}\to (K_0^*, f_0)$ on the Borel
parameter $M^2$ at three fixed values of the continuum threshold
i.e., $s_0=4.0, ~4.4$ and $4.8$, and $q^2=0$ is shown in Figs.
\ref{F304} and \ref{F305} through the two Exp and LD models,
respectively.
We take $M^2=3~ \mbox{GeV}^2$ in our calculations.

\subsection{Form factor analysis}
Inserting the values of the masses, the leptonic decay constants
$f_{S}$ and $f_{B_{(s)}}$, the continuum threshold $s_0$, the Borel
parameter $M^2$, the $B$-meson DA's and other quantities and
parameters related to them such as $\omega_0, \lambda^2_E$ and
$\lambda^2_H$, in addition considering all sources of uncertainties,
the central values of the form factors $f_{+}$, $f_{-}$, and $f_{T}$
and also their errors can be estimated for the semileptonic decays
$B_{(s)}\to (K_0^*, a_0, f_0)$ at $q^2=0$ via the $B$-meson LCSR
approach. Our results for the form factors at $q^2=0$ using the
$B$-meson DA's through the two Exp and LD models, as well as the
predictions of other approaches such as the pQCD \cite{LiLuWaWa},
CLF \cite{ChChuHwa}, QCDSR \cite{AliAzSav,MZYang,Ghahramany}, LFQM
\cite{CheGenLiLi}, MSSM \cite{AsluWang}, and the LCSR with the
light-meson DA's \cite{HanWuFu,WanAslLu,SunLiHu} are collected in
Table \ref{T302}.
\begin{table}[th]
\caption{The form factors of the semileptonic $B_{(s)} \to (K_0^*,
a_0, f_0)$ transitions at zero momentum transfer from different
approaches.}\label{T302}
\begin{ruledtabular}
\begin{tabular}{cccccc}
Decay Mode &Method &Ref. &$f_+(0)$ &$f_-(0)$ &$f_T(0)$  \\ \hline
$B\to a_0$
                                   &LCSR(Exp-model)&  Ours              &  $0.52^{+0.28}_{-0.23}$ &$-0.50^{+0.13}_{-0.35}$ &$0.71^{+0.37}_{-0.30}$ \\
                                   &LCSR(LD-model) &  Ours              &  $0.58^{+0.38}_{-0.23}$ &$-0.58^{+0.29}_{-0.37}$ &$0.78^{+0.44}_{-0.35}$ \\
                                   &LCSR     & \cite{HanWuFu}           &  $0.44$       &$-0.26$       &$0.43$       \\
                                   &LCSR(S2) & \cite{WanAslLu}          &  $0.52$       &$-0.44$       &$0.66$       \\
                                   &LCSR(S2) & \cite{SunLiHu}           &  $0.53$       &$-0.53$       &$--$         \\
                                   &pQCD(S2) & \cite{LiLuWaWa}          &  $0.68$       &$--$          &$0.92$       \\
                                   &LCSR(S1) & \cite{SunLiHu}           &  $0.26$       &$-0.26$       &$--$         \\
                                   &pQCD(S1) & \cite{LiLuWaWa}          &  $-0.31$      &$--$          &$-0.41$      \\
                                   &CLF      & \cite{ChChuHwa}          &  $0.26$       &$--$          &$--$         \\
\hline
$B \to K^*_0$                     &LCSR(Exp-model)&  Ours               &  $0.53^{+0.28}_{-0.22}$ &$-0.51^{+0.13}_{-0.36}$ &$0.72^{+0.39}_{-0.31}$ \\
                                  &LCSR(LD-model) &  Ours               &  $0.60^{+0.34}_{-0.28}$ &$-0.59^{+0.30}_{-0.41}$ &$0.79^{+0.45}_{-0.35}$ \\
                                  &LCSR      & \cite{HanWuFu}           &  $0.45$       &$-0.28$       &$0.46$       \\
                                  &LCSR(S2)  & \cite{WanAslLu}          &  $0.49$       &$-0.41$       &$0.60$       \\
                                  &LCSR(S2)  & \cite{SunLiHu}           &  $0.49$       &$-0.49$       &$0.69$       \\
                                  &pQCD(S2)  & \cite{LiLuWaWa}          &  $0.60$       &$--$          &$0.78$       \\
                                  &LCSR(S1)  & \cite{SunLiHu}           &  $0.17$       &$-0.17$       &$0.24$       \\
                                  &pQCD(S1)  & \cite{LiLuWaWa}          &  $-0.34$      &$--$          &$-0.44$      \\
                                  &CLF       & \cite{ChChuHwa}          &  $0.26$       &$--$          &$--$         \\
                                  &QCDSR     & \cite{AliAzSav}          &  $0.31$       &$-0.31$       &$-0.26$      \\
                                  &LFQM      & \cite{CheGenLiLi}        &  $-0.26$      &$0.21$        &$-0.34$      \\
                                  &MSSM      & \cite{AsluWang}          &  $0.49$       &$-0.41$       &$0.60$       \\
\hline
$B_s \to K^{*}_0$                 &LCSR(Exp-model) &  Ours              &  $0.51^{+0.38}_{-0.24}$ &$-0.48^{+0.24}_{-0.40}$ &$0.70^{+0.50}_{-0.32}$ \\
                                  &LCSR(LD-model)  &  Ours              &  $0.56^{+0.48}_{-0.27}$ &$-0.54^{+0.34}_{-0.50}$ &$0.75^{+0.59}_{-0.37}$ \\
                                  &LCSR      & \cite{HanWuFu}           &  $0.39$       &$-0.25$       &$0.41$       \\
                                  &LCSR(S2)  & \cite{WanAslLu}          &  $0.42$       &$-0.34$       &$0.52$       \\
                                  &LCSR(S2)  & \cite{SunLiHu}           &  $0.44$       &$-0.44$       &$--$         \\
                                  &pQCD(S2)  & \cite{LiLuWaWa}          &  $0.56$       &$--$          &$0.72$       \\
                                  &LCSR(S1)  & \cite{SunLiHu}           &  $0.10$       &$-0.10$       &$--$         \\
                                  &pQCD(S1)  & \cite{LiLuWaWa}          &  $-0.32$      &$--$          &$-0.41$      \\
                                  &QCDSR     & \cite{MZYang}            &  $0.24$       &$--$          &$--$         \\
                                  &QCDSR     & \cite{Ghahramany}        &  $0.25$       &$-0.17$       &$0.21$       \\
\hline
$B_s \to f_0$                     &LCSR(Exp-model) &  Ours              &  $0.47^{+0.36}_{-0.20}$ &$-0.45^{+0.22}_{-0.35}$ &$0.66^{+0.44}_{-0.31}$ \\
                                  &LCSR(LD-model)  &  Ours              &  $0.52^{+0.43}_{-0.24}$ &$-0.50^{+0.28}_{-0.47}$ &$0.71^{+0.51}_{-0.38}$ \\
                                  &LCSR      & \cite{HanWuFu}           &  $0.38$       &$-0.24$       &$0.40$       \\
                                  &LCSR(S2)  & \cite{WanAslLu}          &  $0.43$       &$-0.37$       &$0.56$       \\
                                  &LCSR(S2)  & \cite{SunLiHu}           &  $0.41$       &$-0.41$       &$0.59$       \\
                                  &pQCD(S2)  & \cite{LiLuWaWa}          &  $0.60$       &$--$          &$0.82$       \\
                                  &LCSR(S1)  & \cite{SunLiHu}           &  $0.14$       &$-0.14$       &$0.20$       \\
                                  &pQCD(S1)  & \cite{LiLuWaWa}          &  $-0.26$      &$--$          &$-0.34$      \\
\end{tabular}
\end{ruledtabular}
\end{table}
The most important sources of uncertainties in our calculations are
$\omega_0$, and then the decay constants of the light mesons. For
example, considering the form factor $f_{+}$ of the semileptonic
decay $B \to a_0$, and taking into account the variation of the
input values and parameters, we obtain the following results in the
Exp-model:
\begin{eqnarray}\label{eq319}
f^{B\to a_0}_{+}(0)={0.52}^{+0.21}_{-0.13}\mid_{\delta\omega_0}
{}^{+0.06}_{-0.05}\mid_{\delta f_{a_0}}
{}^{+0.01}_{-0.02}\mid_{\delta s_0}
{}^{+0.00}_{-0.01}\mid_{\delta\lambda^2_{H}}
{}^{+0.00}_{-0.01}\mid_{\delta M^2} {}^{+0.00}_{-0.01}\mid_{\delta
m_{a_0}} {}^{+0.00}_{-0.00}\mid_{\delta\lambda^2_{E}}.
\end{eqnarray}
As the calculations show, the most value of error in $f_{+}$ enters
through the variation of $\omega_0$.

Table \ref{T302} shows that considering the uncertainties, there is
a good agreement between our results in the Exp-model and
predictions of the conventional LCSR in S2 \cite{WanAslLu,SunLiHu}
for all cases. As a result, our calculations confirm scenario 2 for
describing the scalar mesons $K^*_{0}(1430), a_0(1450)$ and
$f_0(1500)$. This means that the scalar mesons $K^*_{0}(1430),
a_0(1450)$ and $f_0(1500)$ can be seen as the lowest lying states
with two quarks in the quark model.

Table \ref{T303} shows the individual contributions of the two- and
three-particle DA's to the semileptonic form factors $B_{(s)}\to
(K_0^*, a_0, f_0)$ at $q^2 = 0$. Note that the contributions of the
two-particle DA's are listed based on twist level $\varphi_{_+}$,
$\varphi_{_-}$, and $g_{_+}$, while $g_{_-}$ does not appear in the
results of the form factors. As can be seen, the two-particle
leading-twist DA of  $B$-meson, $\varphi_{_+}$ has the most
important contribution in calculation of the form factors.
\begin{table}[th]
\caption{Contributions of the two-particle DA's (2-P DA's) and
three-particle DA's (3-P DA's) to the form factor results at $q^2 =
0$ in the two Exp and LD models.} \label{T303}
\begin{ruledtabular}
\begin{tabular}{ccccc|cccc}
&&Exp-model&&&&LD-model&&\\ \cline{2-9}
\multirow{2}{*}{\vspace{2.1em}\rm{Form Factor}} & \multicolumn{3}{c}{2-P DA's} &\multirow{2}{*}{3-P DA's}& \multicolumn{3}{c}{2-P DA's} &\multirow{2}{*}{3-P DA's}    \\
\cline{2-4}\cline{6-8}&$\varphi_{_+}$ &$\varphi_{_-}$ & $g_{_+}$ & &$\varphi_{_+}$ &$\varphi_{_-}$ & $g_{_+}$ & \\
\hline\\[-2mm]
$f_+^{B \to a_0}$      &$0.49$&$0.07$&$-0.05$&$0.01$&$0.52$&$0.08$&$-0.04$&$0.02$        \\[2mm]
$f_-^{B \to a_0}$      &$-0.61$&$0.07$&$0.05$&$-0.01$&$-0.68$&$0.08$&$0.04$&$-0.02$       \\[2mm]
$f_T^{B \to a_0}$      &$0.70$&$0.00$&$0.00$&$0.01$&$0.77$&$0.00$&$0.00$&$0.01$       \\[2mm]
$f_+^{B \to K^*_0}$    &$0.50$&$0.07$&$-0.05$&$0.01$&$0.54$&$0.08$&$-0.04$&$0.02$       \\[2mm]
$f_-^{B \to K^*_0}$    &$-0.63$&$0.07$&$0.06$&$-0.01$&$-0.69$&$0.08$&$0.04$&$-0.02$       \\[2mm]
$f_T^{B \to K^*_0}$    &$0.71$&$0.00$&$0.00$&$0.01$&$0.78$&$0.00$&$0.00$&$0.01$       \\[2mm]
$f_+^{B_s \to K^*_0}$  &$0.48$&$0.08$&$-0.06$&$0.01$&$0.51$&$0.08$&$-0.05$&$0.02$       \\[2mm]
$f_-^{B_s \to K^*_0}$  &$-0.62$&$0.08$&$0.07$&$-0.01$&$-0.66$&$0.08$&$0.06$&$-0.02$       \\[2mm]
$f_T^{B_s \to K^*_0}$  &$0.69$&$0.00$&$0.00$&$0.01$&$0.74$&$0.00$&$0.00$&$0.01$       \\[2mm]
$f_+^{B_s \to f_0}$    &$0.44$&$0.07$&$-0.05$&$0.01$&$0.46$&$0.08$&$-0.04$&$0.02$       \\[2mm]
$f_-^{B_s \to f_0}$    &$-0.57$&$0.07$&$0.06$&$-0.01$&$-0.61$&$0.08$&$0.05$&$-0.02$       \\[2mm]
$f_T^{B_s \to f_0}$    &$0.65$&$0.00$&$0.00$&$0.01$&$0.70$&$0.00$&$0.00$&$0.01$       \\[2mm]
\end{tabular}
\end{ruledtabular}
\end{table}
In this table, the higher-twist contributions of the three-particle
DA's are not presented separately, because their contributions are
usually less than 0.01. For instance, the contributions of the eight
twist functions $\Psi_{_A}$, $\Psi_{_V}$, $X_{_A}$, $Y_{_A}$,
$\overline{X}_{_A}$, $\overline{Y}_{A}$, $W$, and $Z$ for the form
factor $f_+^{B \to a_0}$ at $q^2 = 0$ is reported in Table
\ref{T304} in the Exp-model.
\begin{table}[th]
\caption{Contributions of the eight twist functions $\Psi_{_A}$,
$\Psi_{_V}$, $X_{_A}$, $Y_{_A}$, $\overline{X}_{_A}$,
$\overline{Y}_{A}$, $W$, and $Z$ for the form factor $f_+^{B \to
a_0}$ at $q^2 = 0$ up to $\mathcal{O}(10^{-3})$ in the Exp-model.}
\label{T304}
\begin{ruledtabular}
\begin{tabular}{ccccccccc}
&$\Psi_{_A}$& $\Psi_{_V}$& $X_{_A}$& $Y_{_A}$&
$\overline{X}_{_A}$& $\overline{Y}_{A}$& $W$& $Z$ \\
\hline $f_+^{B \to a_0}$
&$-0.002$&$0.015$&$-0.002$&$0.001$&$0.000$&$0$&$-0.001$&$-0.001$
\end{tabular}
\end{ruledtabular}
\end{table}

Due to the cut-off in QCD theories, the form factors for each
aforementioned semileptonic decay can be estimated by the $B$-meson
LCSR method in half of the physical region $0\leq q^2 \leq
{(m_{B_{(s)}}- m_{S})}^2$, nearly. In order to extend our results to
the full physical region, we look for a parametrization of the form
factors in such a way that in the validity region of the LCSR, this
parametrization coincides with the LCSR predictions. We use the
following fit functions of the form factors with respect to $q^2$
as:
\begin{eqnarray}\label{eq320}
f^{\rm{I}}(q^{2})&=&\frac{f(0)}{1-\alpha\,s+\beta\,s^2},
\nonumber\\
f^{\rm{II}}(q^{2})&=&\frac{1}{1-s}\sum_{k=0}^{2}
b_{k}{[z(q^2)-z(0)]}^k\, ,
\end{eqnarray}
where $s=q^2/m_{B_{(s)}}^2$,
$z(t)=\frac{\sqrt{t_{+}-t}\,-\,\sqrt{t_{+}-t_0}}{\sqrt{t_{+}-t}\,+\,\sqrt{t_{+}-t_0}}$,
$t_0=t_+\,(1-\sqrt{1-t_-/t_+})$, and $t_{\pm}=(m_{B_{(s)}}\pm
m_{S})^2$. The parameters $(f(0),~ \alpha,~ \beta)$ and ($b_{0},~
b_{1},~ b_{2}$), related to the fit functions $f^{\rm{I}}(q^2)$ and
$f^{\rm{II}}(q^2)$ respectively, are determined from the fitting
procedure. Table \ref{T305} shows the values of these parameters for
the form factors of the semileptonic decays $B_{(s)} \to (K_0^*,
a_0, f_0)$ in the Exp-model. Table \ref{T306} shows the same values
as Table \ref{T305} but for the LD-model.
\begin{table}[th]
\caption{Values of parameters $(f(0),~ \alpha,~ \beta)$ and
($b_{0},~ b_{1},~ b_{2}$) connected to the fit functions
$f^{\rm{I}}(q^2)$ and $f^{\rm{II}}(q^2)$ respectively, for the
fitted form factors of $B_{(s)}\to (K_0^*, a_0, f_0)$ transitions in
the Exp-model.}\label{T305}
\begin{ruledtabular}
\begin{tabular}{cccc|ccc}
\rm{Form Factor} & $f(0)$ & $\alpha$ & $\beta$ & $b_{0}$ & $b_{1}$ & $b_{2}$ \\
\hline\\[-3mm]
$f_{+}^{B \to a_0}$ & $0.52$ & $-0.49$ & $1.68$ & $0.52$ & $-0.23$ & $-6.83$ \\[2mm]
$f_{-}^{B \to a_0}$ & $-0.50$ & $0.44$ & $-1.51$ & $-0.50$ & $0.54$ & $5.41$ \\[2mm]
$f_{T}^{B \to a_0}$ & $0.71$ & $-0.59$ & $2.53$ & $0.71$ & $-0.95$ & $-3.79$ \\[2mm]
$f_{+}^{B \to K^*_0}$ & $0.53$ & $-0.50$ & $1.71$ & $0.53$ & $-0.23$ & $-6.79$ \\[2mm]
$f_{-}^{B \to K^*_0}$ & $-0.51$ & $0.44$ & $-1.55$ & $-0.51$ & $0.55$ & $5.39$ \\[2mm]
$f_{T}^{B \to K^*_0}$ & $0.72$ & $-0.60$ & $2.57$ & $0.72$ & $-0.95$ & $-3.77$ \\[2mm]
$f_{+}^{B_s \to K^{*}_0}$ & $0.51$ & $-0.43$ & $1.51$ & $0.51$ & $-0.65$ & $-4.46$ \\[2mm]
$f_{-}^{B_s \to K^{*}_0}$ & $-0.48$ & $0.38$ & $-1.37$ & $-0.48$ & $0.89$ & $2.32$ \\[2mm]
$f_{T}^{B_s \to K^{*}_0}$ & $0.70$ & $-0.51$ & $2.25$ & $0.70$ & $-1.65$ & $2.99$ \\[2mm]
$f_{+}^{B_s \to f_0}$ & $0.47$ & $-0.39$ & $1.10$ & $0.47$ & $-0.69$ & $-6.15$ \\[2mm]
$f_{-}^{B_s \to f_0}$ & $-0.45$ & $0.36$ & $-1.34$ & $-0.45$ & $0.75$ & $2.53$ \\[2mm]
$f_{T}^{B_s \to f_0}$ & $0.66$ & $-0.48$ & $2.11$ & $0.66$ & $-1.58$ & $3.05$ \\[2mm]
\end{tabular}
\end{ruledtabular}
\end{table}
\begin{table}[th]
\caption{The same as Table \ref{T305} but for the
LD-model.}\label{T306}
\begin{ruledtabular}
\begin{tabular}{cccc|ccc}
\rm{Form Factor} & $f(0)$ & $\alpha$ & $\beta$ & $b_{0}$ & $b_{1}$ & $b_{2}$ \\
\hline\\[-3mm]
$f_{+}^{B \to a_0}$ & $0.58$ & $-0.48$ & $0.95$ & $0.58$ & $-0.86$ & $-13.82$ \\[2mm]
$f_{-}^{B \to a_0}$ & $-0.58$ & $0.42$ & $-0.83$ & $-0.58$ & $1.49$ & $11.43$ \\[2mm]
$f_{T}^{B \to a_0}$ & $0.78$ & $-0.58$ & $1.15$ & $0.78$ & $-1.82$ & $-16.30$ \\[2mm]
$f_{+}^{B \to K^*_0}$ & $0.60$ & $-0.49$ & $0.97$ & $0.60$ & $-0.87$ & $-13.77$ \\[2mm]
$f_{-}^{B \to K^*_0}$ & $-0.59$ & $0.43$ & $-0.85$ & $-0.59$ & $1.50$ & $11.41$ \\[2mm]
$f_{T}^{B \to K^*_0}$ & $0.79$ & $-0.59$ & $1.16$ & $0.79$ & $-1.83$ & $-16.15$ \\[2mm]
$f_{+}^{B_s \to K^{*}_0}$ & $0.56$ & $-0.41$ & $1.07$ & $0.56$ & $-1.44$ & $-4.24$ \\[2mm]
$f_{-}^{B_s \to K^{*}_0}$ & $-0.54$ & $0.35$ & $-0.93$ & $-0.54$ & $1.94$ & $-2.27$ \\[2mm]
$f_{T}^{B_s \to K^{*}_0}$ & $0.75$ & $-0.49$ & $1.26$ & $0.75$ & $-2.67$ & $1.49$ \\[2mm]
$f_{+}^{B_s \to f_0}$ & $0.52$ & $-0.38$ & $1.00$ & $0.52$ & $-1.36$ & $-4.07$ \\[2mm]
$f_{-}^{B_s \to f_0}$ & $-0.50$ & $0.32$ & $-0.86$ & $-0.50$ & $1.85$ & $-2.33$ \\[2mm]
$f_{T}^{B_s \to f_0}$ & $0.71$ & $-0.46$ & $1.19$ & $0.71$ & $-2.56$ & $1.62$ \\[2mm]
\end{tabular}
\end{ruledtabular}
\end{table}

\begin{figure}[th]
\begin{center}
\includegraphics[width=4cm,height=4cm]{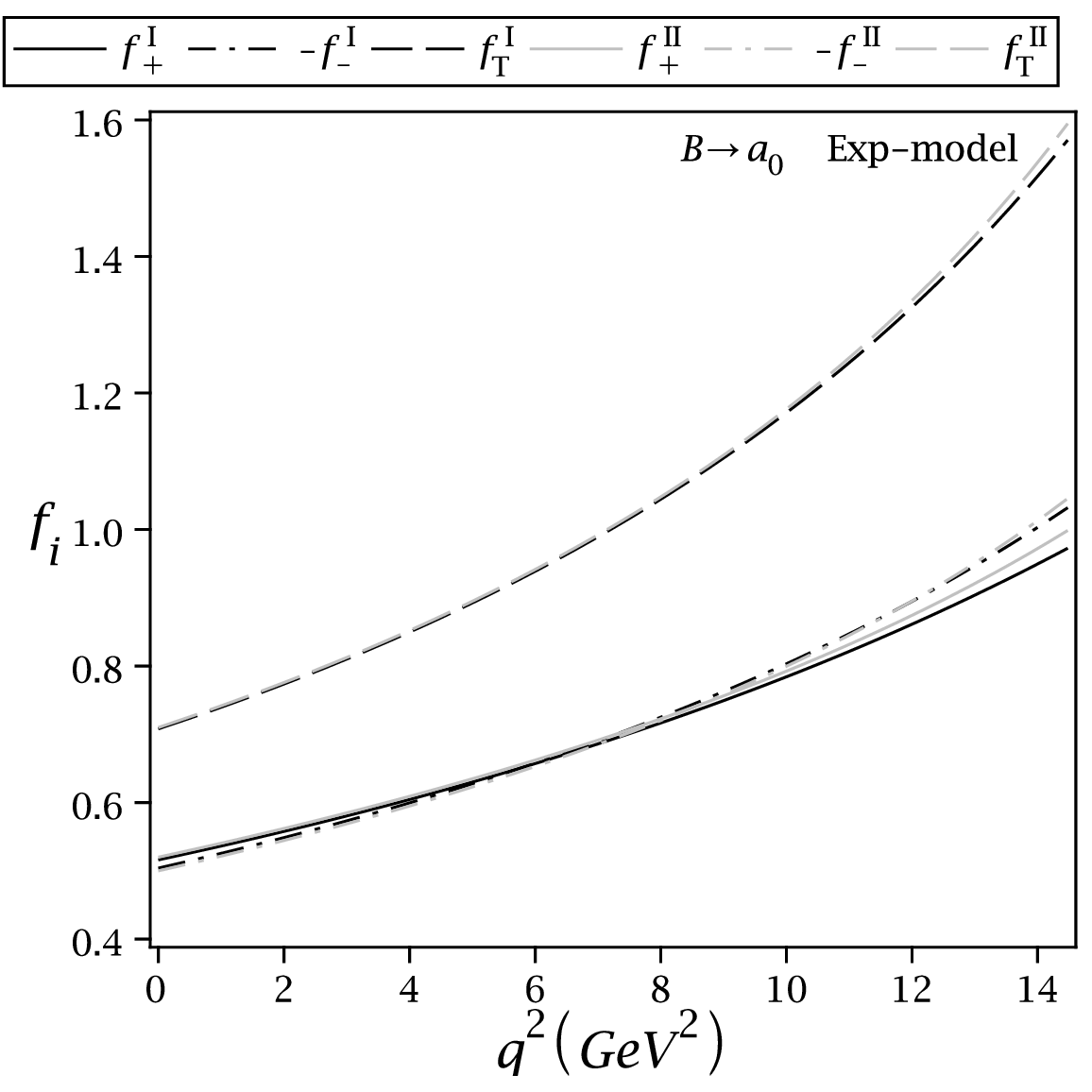}
\includegraphics[width=4cm,height=4cm]{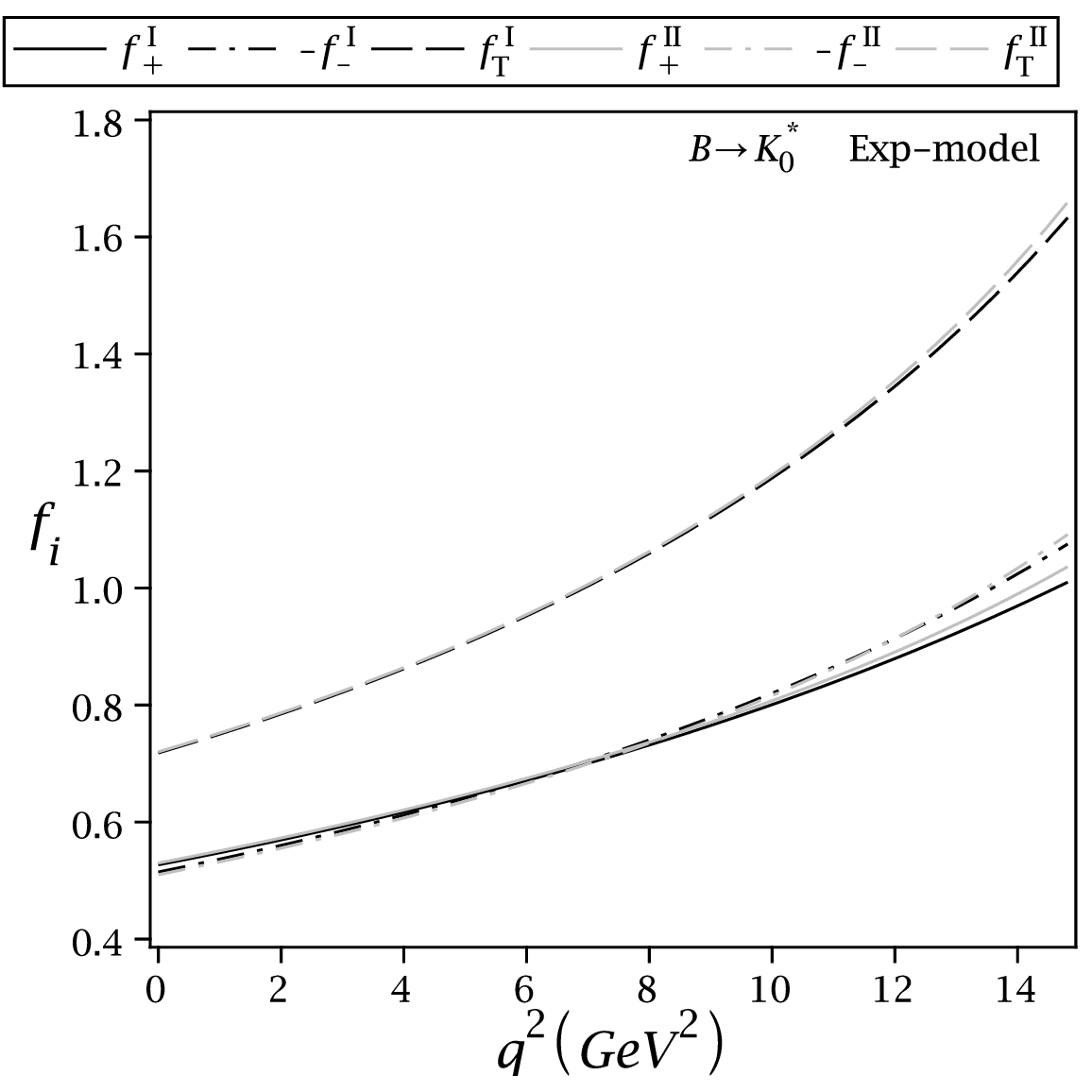}
\includegraphics[width=4cm,height=4cm]{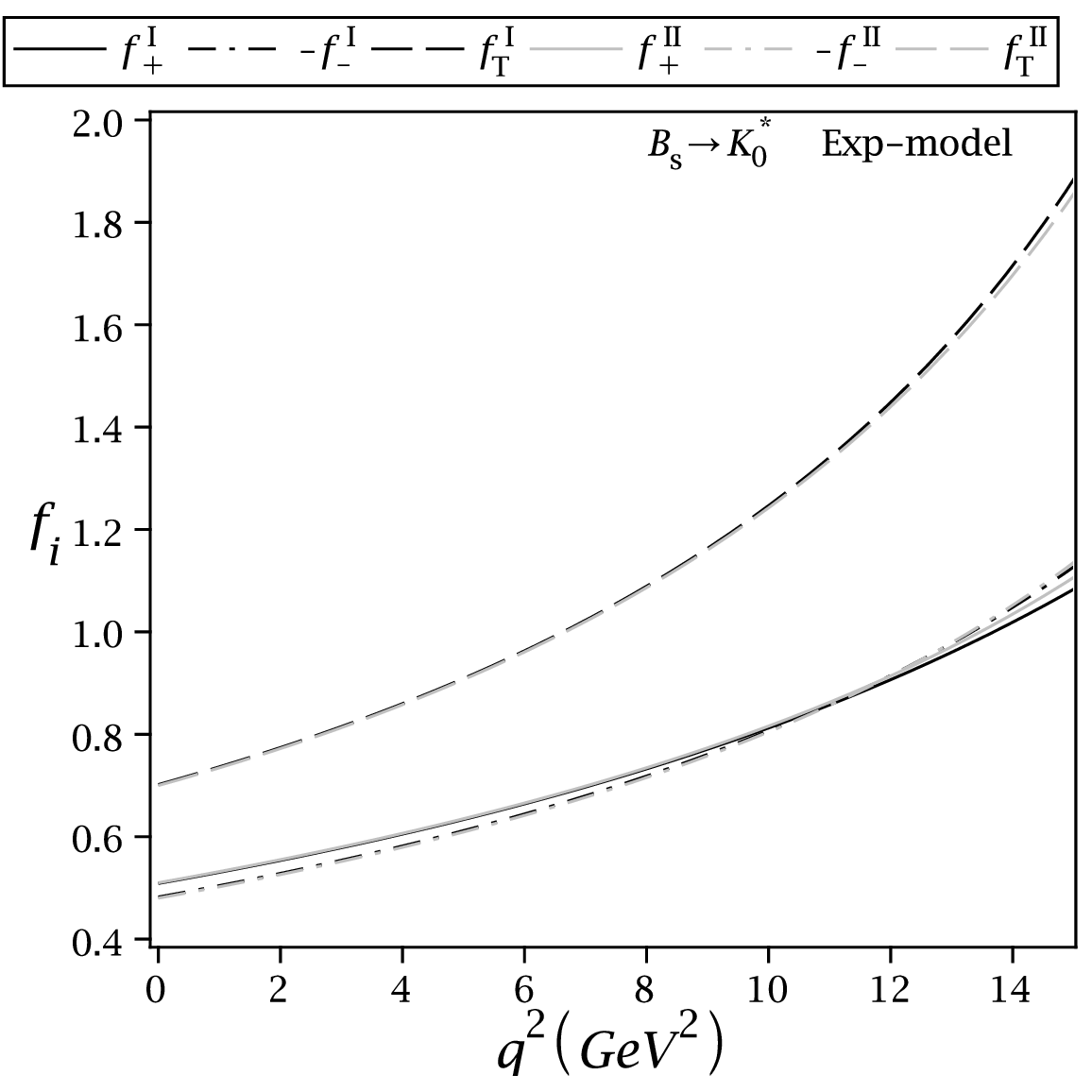}
\includegraphics[width=4cm,height=4cm]{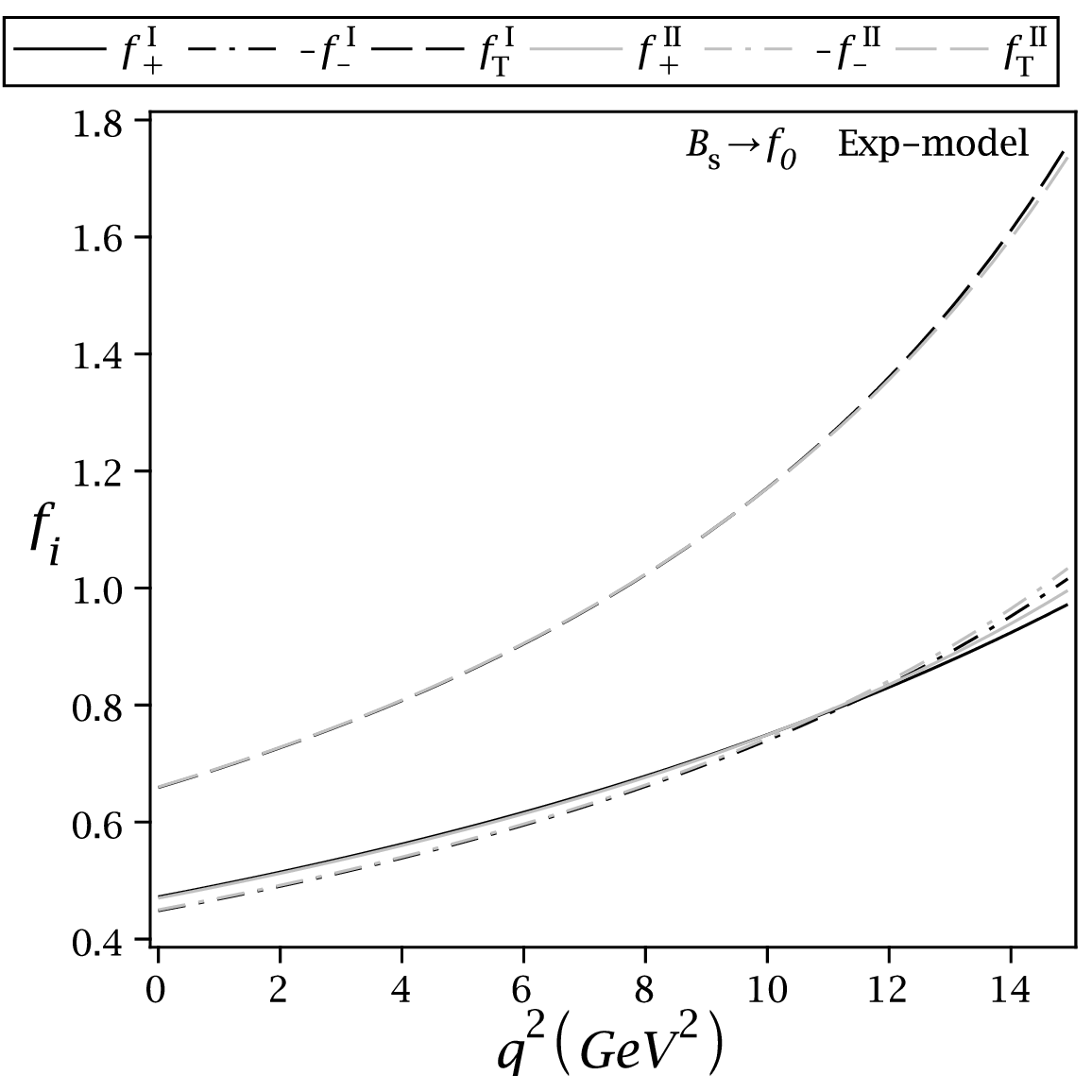}
\caption{Black and gray lines show the fitted form factors $f_{+}$,
$-f_{-} $, and $f_{T}$ of the $B_{(s)}\to S$ transitions by using
the fit functions $f^{\rm{I}}(q^2)$ and $f^{\rm{II}}(q^2)$
respectively, with respect to $q^2$ in the Exp-model.} \label{F306}
\end{center}
\end{figure}

\begin{figure}[th]
\begin{center}
\includegraphics[width=4cm,height=4cm]{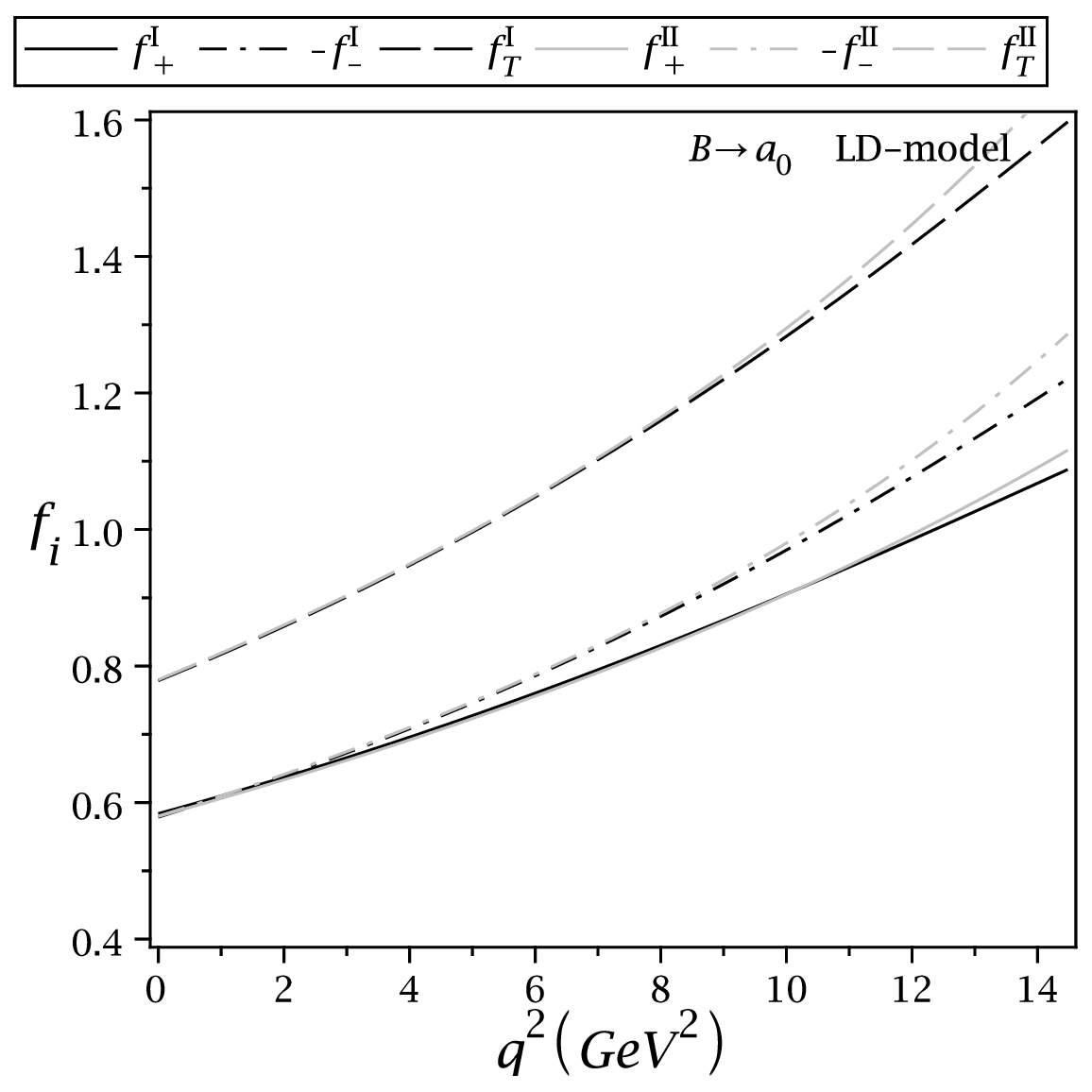}
\includegraphics[width=4cm,height=4cm]{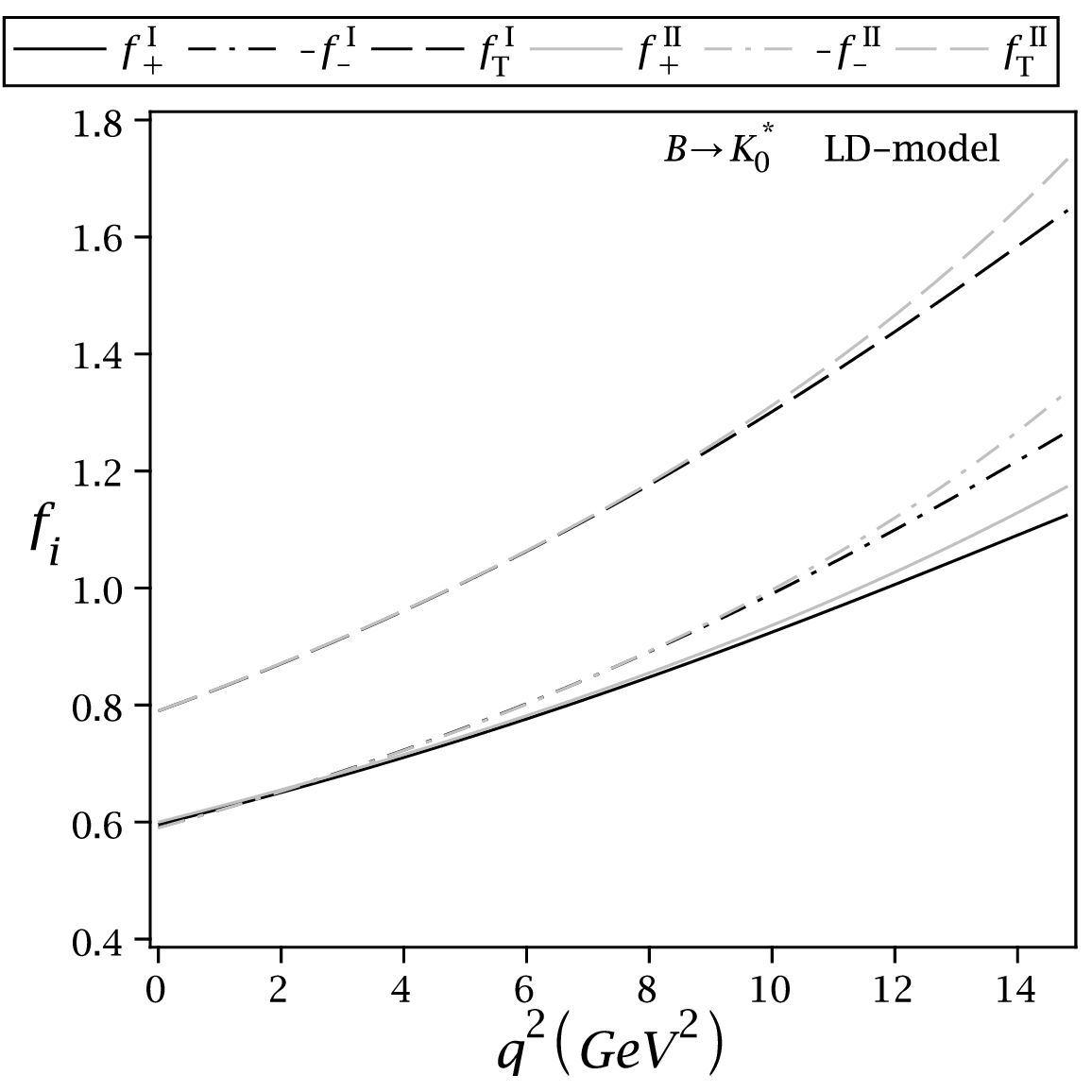}
\includegraphics[width=4cm,height=4cm]{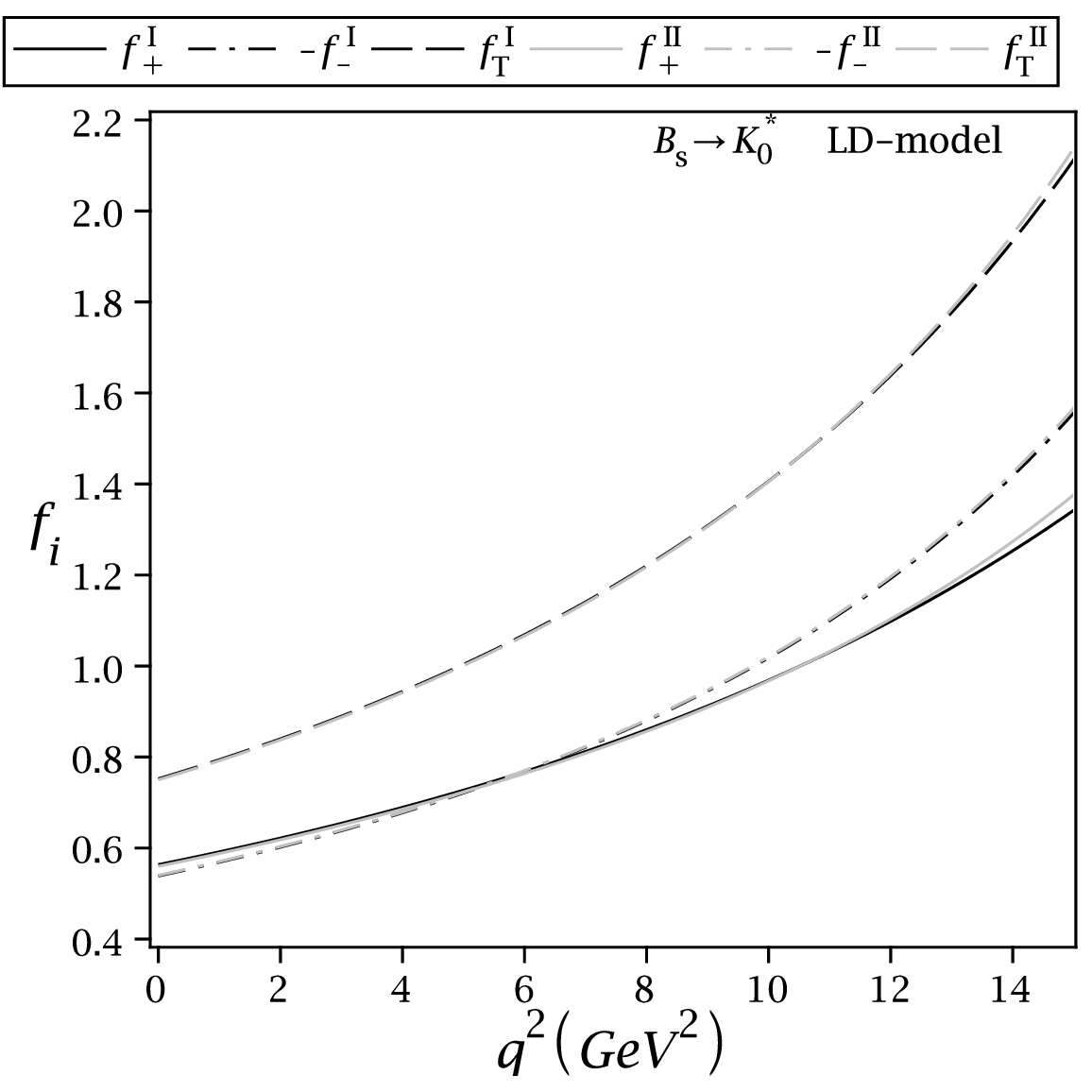}
\includegraphics[width=4cm,height=4cm]{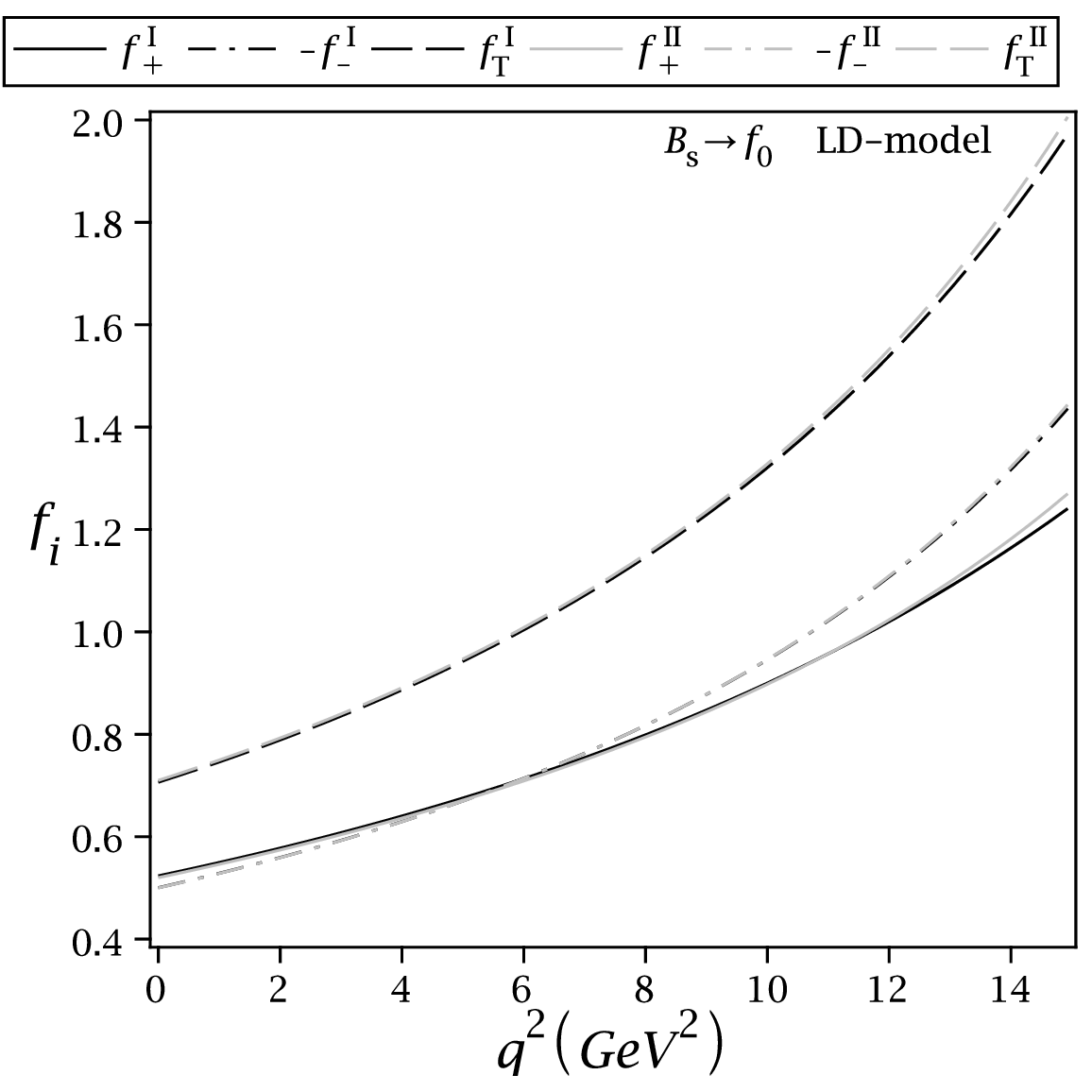}
\caption{The same as Fig. \ref{F306} but for the LD-model.}
\label{F307}
\end{center}
\end{figure}

The dependence of the fitted form factors $f_{i}\, (i=+,-,T)$ on
$q^2$ is given in Fig. \ref{F306} for $B_{(s)}\to S\, (S=K_0^*, a_0,
f_0)$ transitions. These form factors are related to the Exp-model.
Fig. \ref{F307} depicts the same results as Fig. \ref{F306}, but for
the LD-model.
In these figures, the black and gray lines show the results for
$f^{\rm{I}}(q^2)$ and $f^{\rm{II}}(q^2)$ fit functions,
respectively. According to Figs. \ref{F306} and \ref{F307}, the
fitted form factors obtained for the two fit functions are
consistent in each case.

The form factors at large recoil should satisfy the following
relations \cite{ColFazWan}:
\begin{eqnarray}\label{eq321}
f_{-}(q^2)&=&-\frac{m^2_{B_{(s)}}-m^2_{S}}{m_{b}\,m_{B_{(s)}}}f_{+}(q^2)\,,\nonumber\\
f_{T}(q^2)&=&\frac{m_{B_{(s)}}+m_{S}}{m_{B_{(s)}}}f_{+}(q^2).
\end{eqnarray}
Figs. \ref{F306} and \ref{F307} show that the computed form factors
from the LCSR with the $B$-meson DA's for the two Exp and LD models
satisfy the relations in Eq. (\ref{eq321}), by considering the
errors.

\begin{figure}[th]
\begin{center}
\includegraphics[width=4cm,height=4cm]{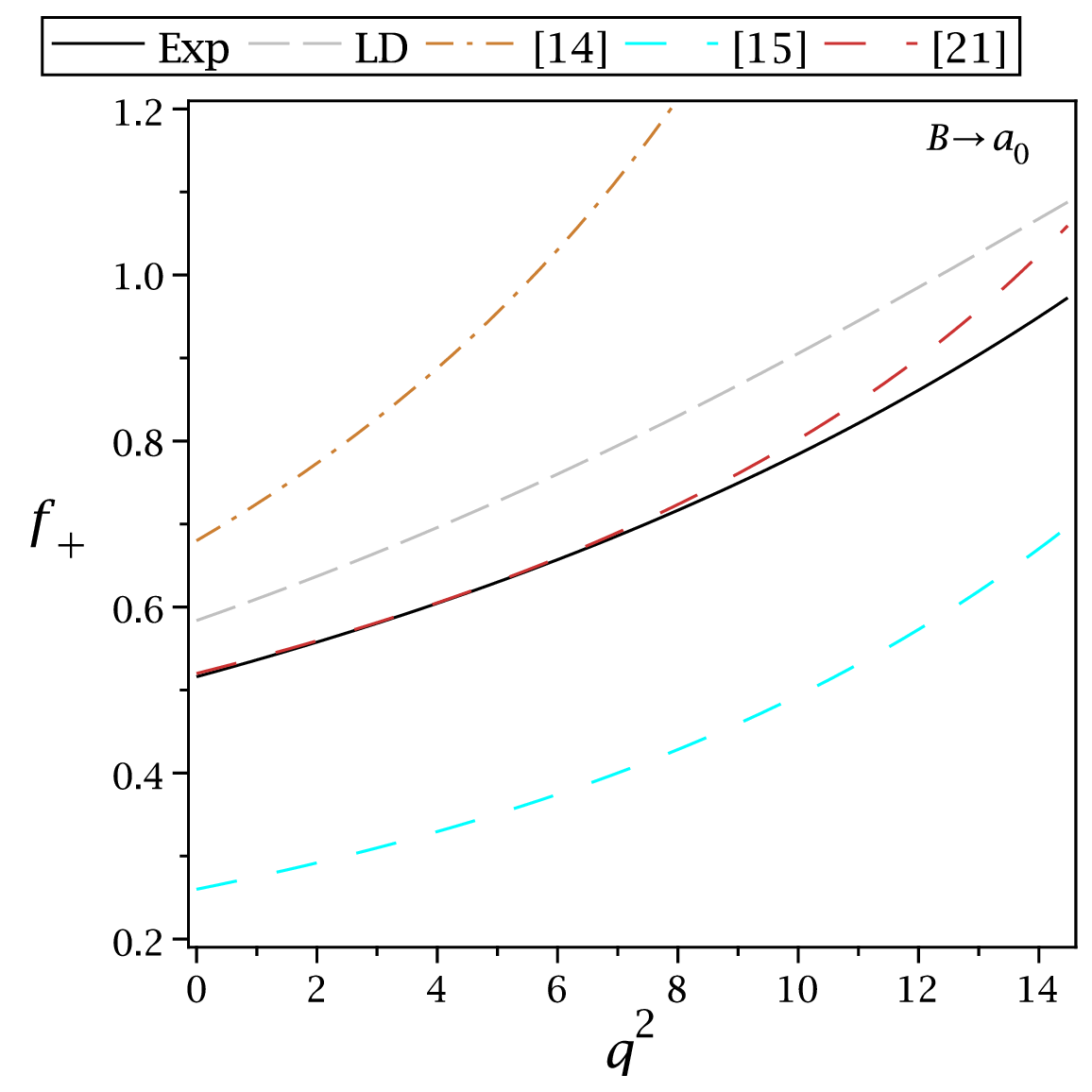}
\includegraphics[width=4cm,height=4cm]{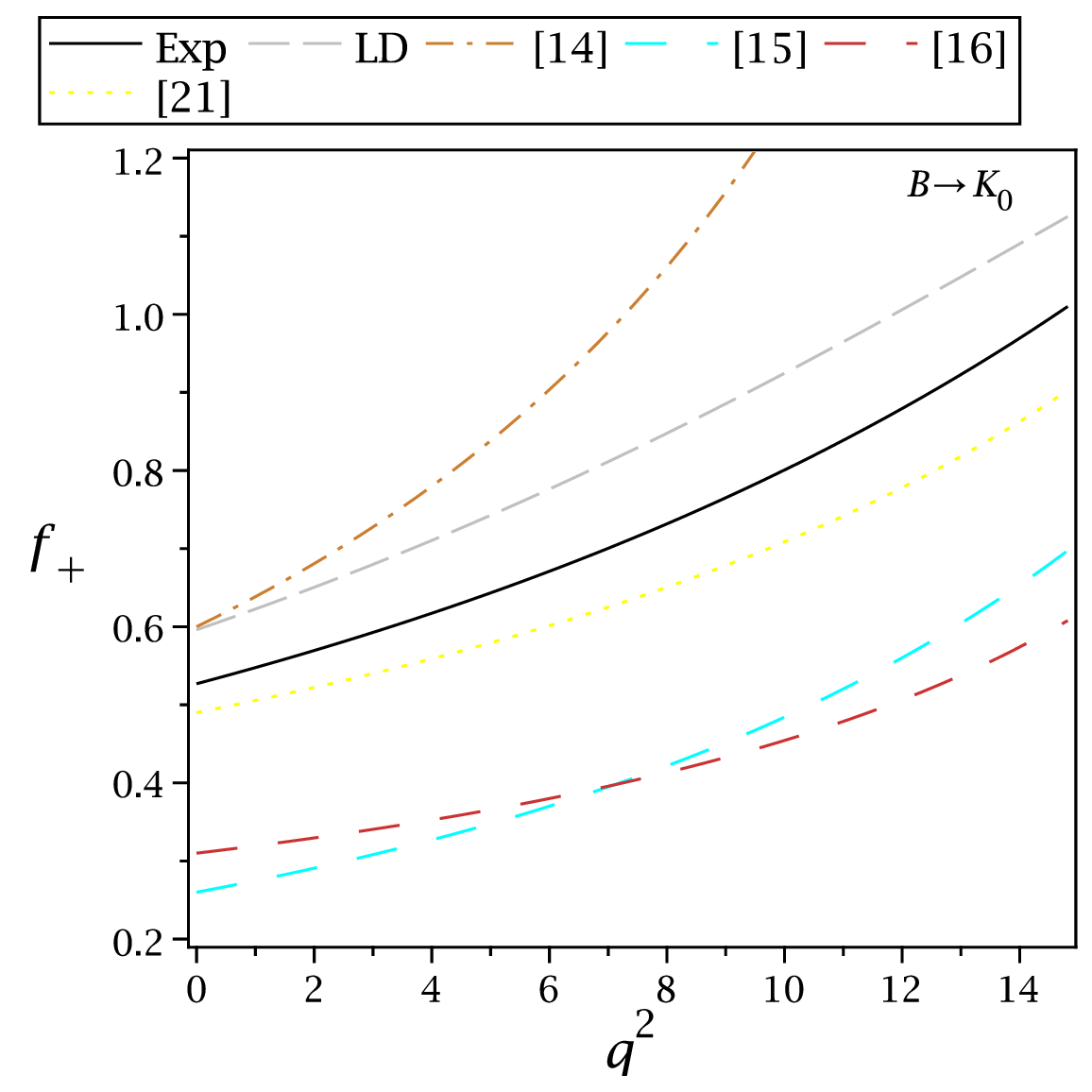}
\includegraphics[width=4cm,height=4cm]{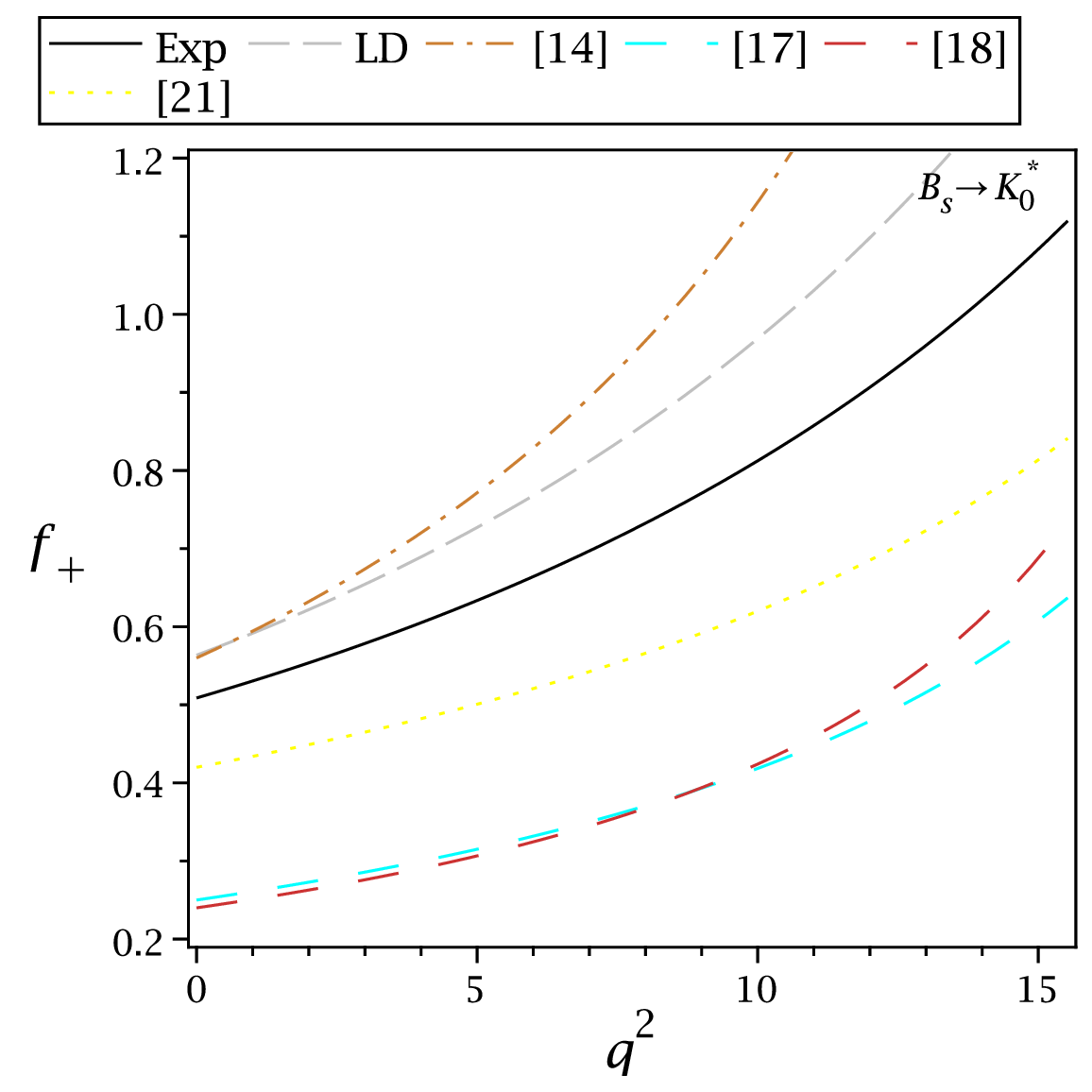}
\includegraphics[width=4cm,height=4cm]{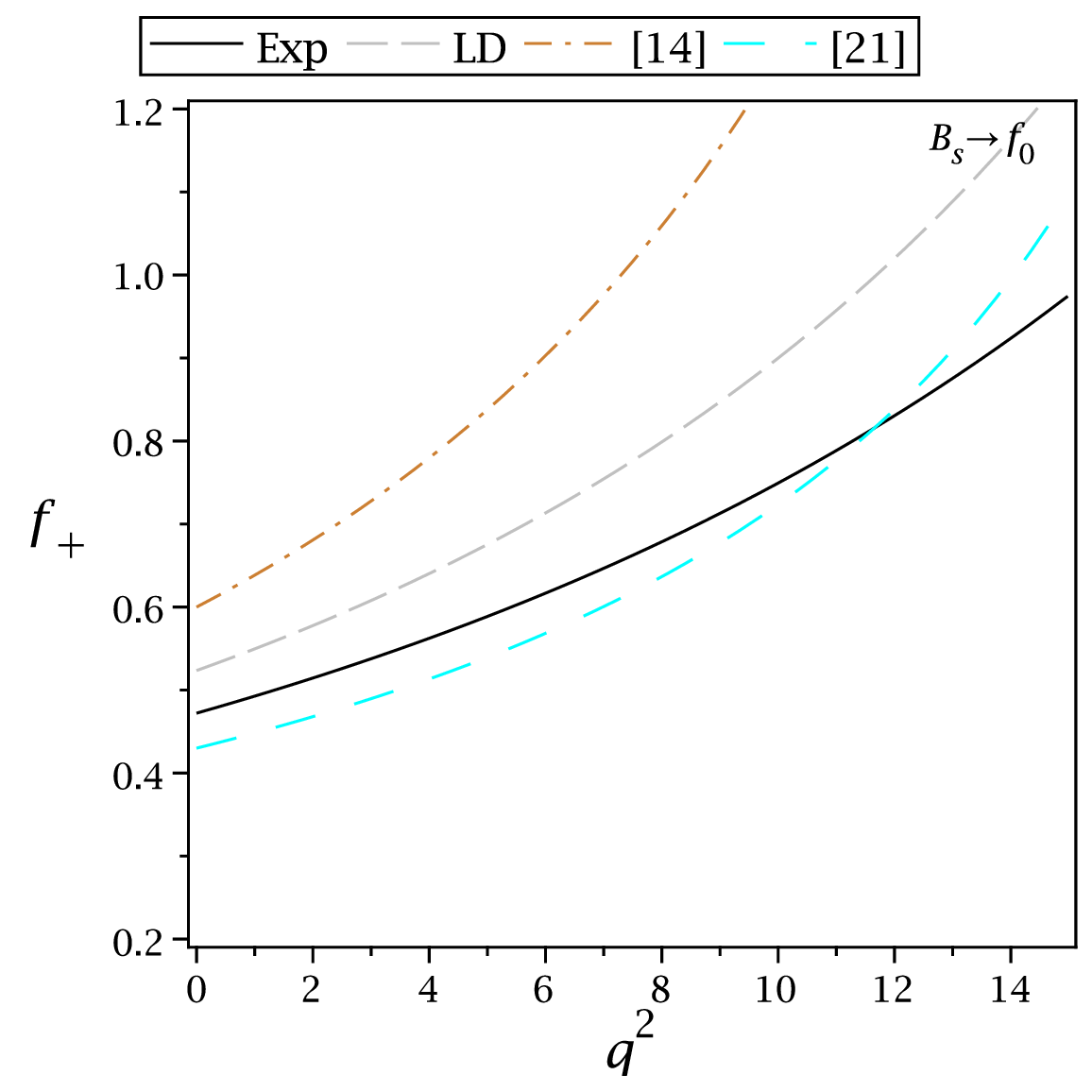}
\caption{Form factor $f_{+}$ for $B \to (a_0, K_0^*)$ and $B_s \to
(K_0^*, f_0)$ decays in different models such as the pQCD
\cite{LiLuWaWa}, CLF \cite{ChChuHwa}, QCDSR
\cite{AliAzSav,MZYang,Ghahramany}, the LCSR with the light-meson
DA's \cite{WanAslLu} and our results via $B$-meson LCSR in two Exp
and LD models.}\label{F308}
\end{center}
\end{figure}

The results of the form factor $f_+(q^2)$ for the aforementioned
decays from different models are compared with our results in the
two Exp and LD models in Fig. \ref{F308}.

\subsection{Semileptonic $B^{0}_{s} \to K_0^{*+}\, l^- \bar{\nu}_{l}$ and $B^0 \to a^{+}_0\, l^- \bar{\nu}_{l}$ decays}

\begin{figure}[th]
\begin{center}
\includegraphics[width=4cm,height=4cm]{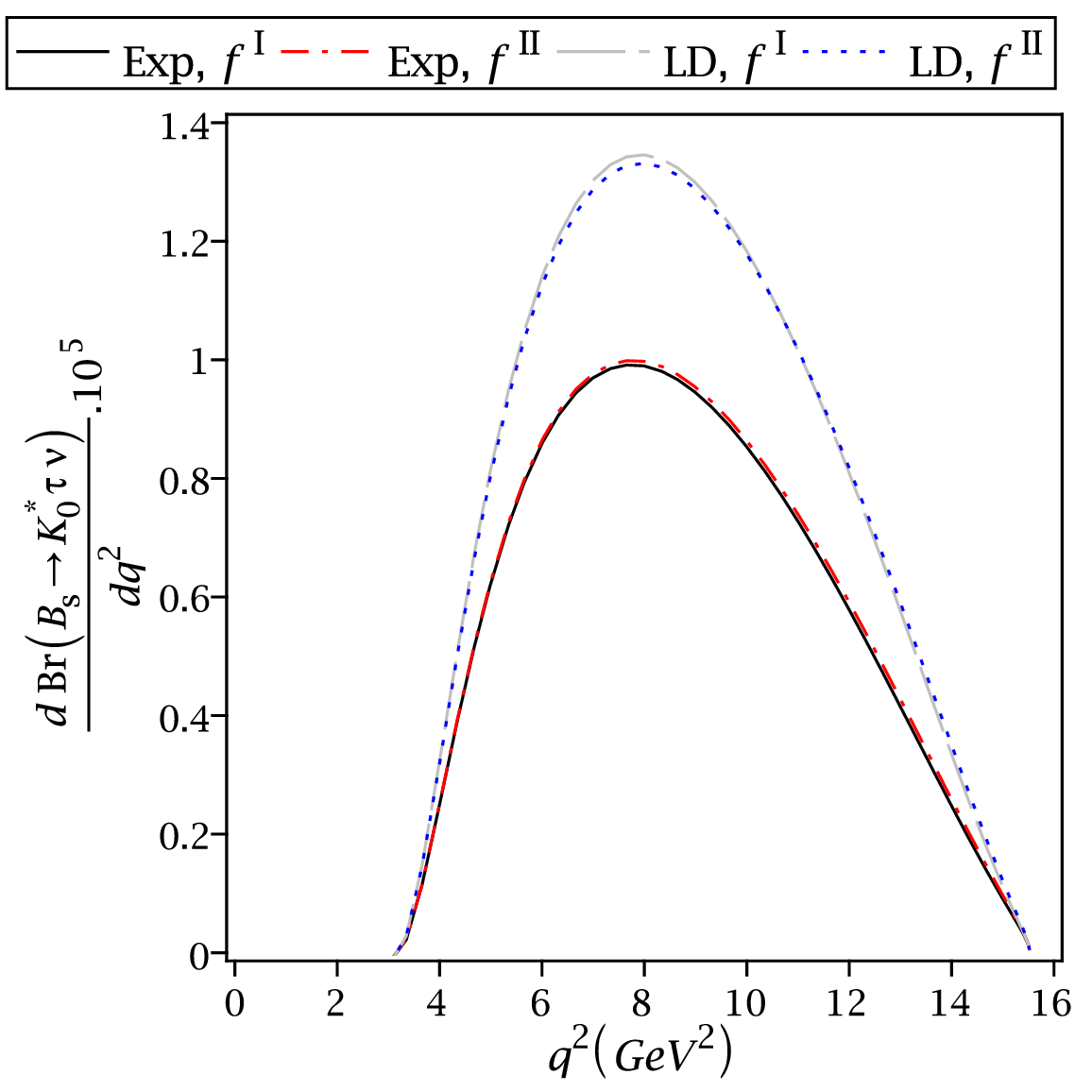}
\includegraphics[width=4cm,height=4cm]{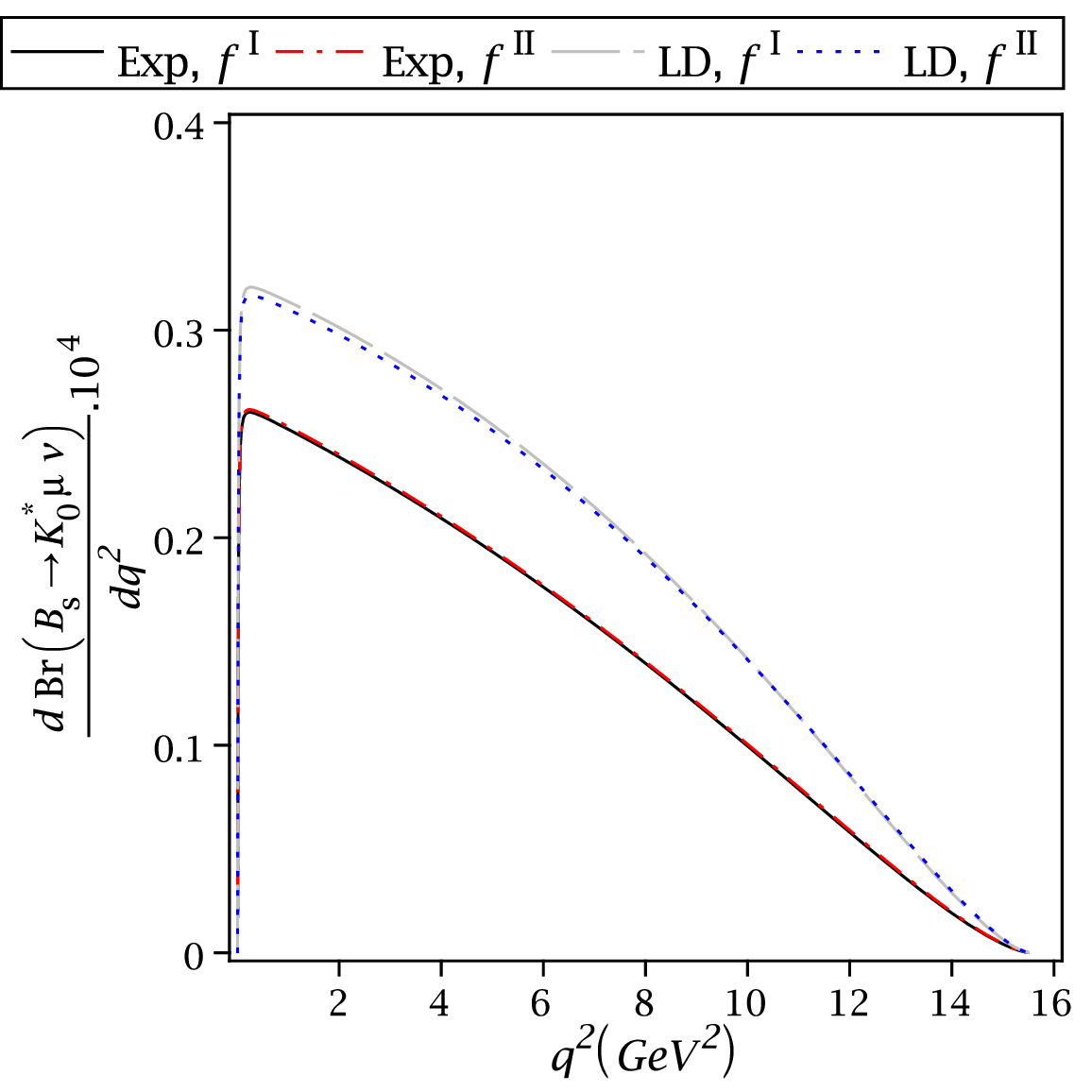}
\includegraphics[width=4cm,height=4cm]{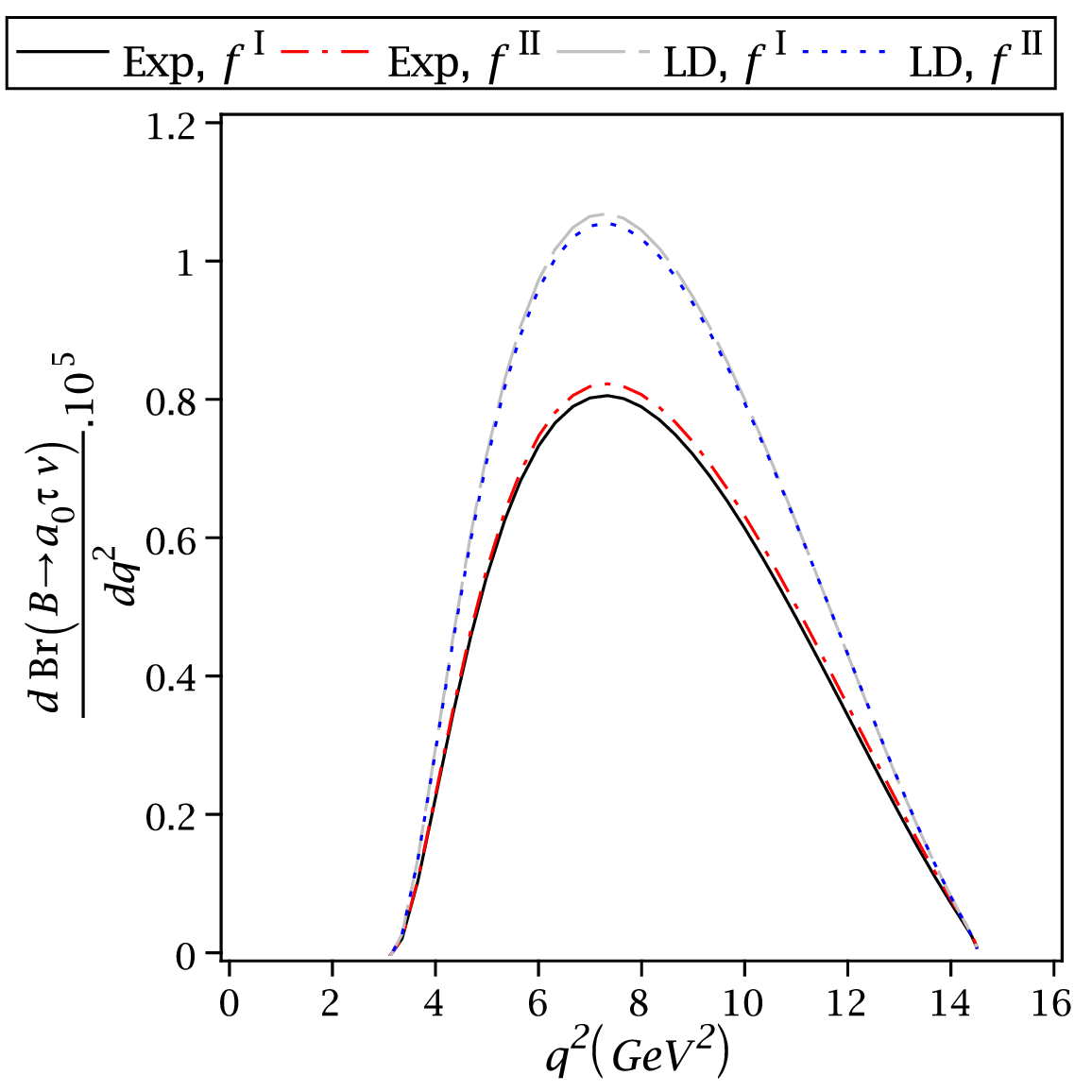}
\includegraphics[width=4cm,height=4cm]{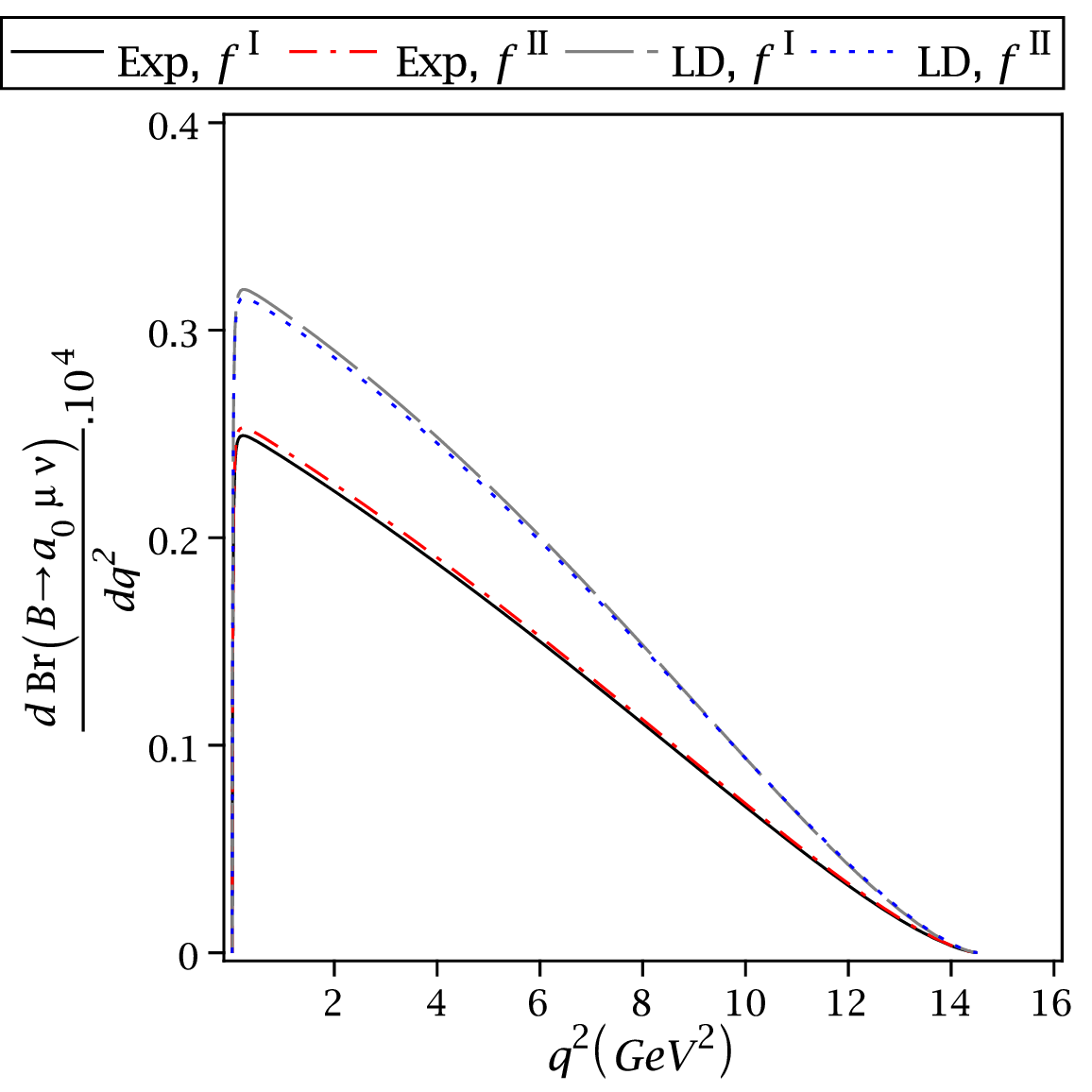}
\caption{Differential branching ratios of the semileptonic $B_{(s)}
\to (K_0^*, a_0)\, l \nu\, (l=\mu, \tau)$ transitions on $q^2$ for
the Exp-model and fit function $f^{\rm{I}}(q^2)$ [Exp,
$f^{\rm{I}}$], the Exp-model and fit function $f^{\rm{II}}(q^2)$
[Exp, $f^{\rm{II}}$], the LD-model and fit function
$f^{\rm{I}}(q^2)$ [LD, $f^{\rm{I}}$], and the LD-model and fit
function $f^{\rm{II}}(q^2)$ [LD, $f^{\rm{II}}$].} \label{F309}
\end{center}
\end{figure}

At the quark level, the tree-level $b \to u$ transition is
responsible for the $B_{s} \to K_0^*\, l \bar{\nu}_{l}$ and $B \to
a_0\, l \bar{\nu}_{l}$ decay modes. The Hamiltonian for this
transition is written as
\begin{eqnarray}\label{eq322}
\mathcal{H}_{\rm{eff}}(b \to u\, l
\bar{\nu}_{l})=\frac{G_{F}}{\sqrt{2}} V_{ub}\, \bar{u}\gamma_{\mu}
(1-\gamma_{5}) b\,\, \bar{l} \gamma^{\mu} (1-\gamma_{5}) \nu_{l}\,,
\end{eqnarray}
where $G_F$ is the Fermi constant, $V_{ub}=(3.82\pm0.24)\times
10^{-3}$. With this Hamiltonian, the differential decay width
$\frac{d\Gamma }{dq^2}$ for the processes $B_{(s)} \to S\, l
\bar{\nu}_{l}\, (S=K_0^*, a_0)$ in terms of the form factors can be
expressed as \cite{YJSun}
\begin{eqnarray}\label{eq323}
&&\frac{d\Gamma}{dq^2}(B_{(s)}\to S\, l\,
\bar{\nu}_l)=\frac{G_F^2|V_{ub}|^2}{384\, \pi^3\,
m_{B_{(s)}}^3}\frac{(q^2-m_l^2)^2}{(q^2)^3} \sqrt{(m_{B_{(s)}}^2-m_{
S}^2-q^2)^2 -4\, q^2\, m_{S}^2}
\,\,\,\Big\{(m_l^2+2q^2) \nonumber\\
&& \times \sqrt{(m_{B_{(s)}}^2-m_{S}^2-q^2)^2 -4 q^2 m_{
S}^2}\,\,f_+^2(q^2)
+3\,m_l^2\,(m_{B_{(s)}}^2-m_{S}^2)^2\,\Big[f_+(q^2)+\frac{q^2}{m_{B_{(s)}}^2-m_{
S}^2}f_-(q^2)\Big]^2\Big\},
\end{eqnarray}
where $m_{l}$ is the mass of the lepton. The dependency of the
differential branching ratios of $B_{s}\to K_0^*\, l\, \bar{\nu}_l$
and $B \to a_0\, l\, \bar{\nu}_l$\, ($l=\mu, \tau$) decays on $q^2$
is shown in Fig. \ref{F309} for both the Exp and LD models as well
as the two fit functions $f^{\rm{I}}(q^2)$ and $f^{\rm{II}}(q^2)$.

Integrating Eq. (\ref{eq323}) over $q^2$ in the whole physical
region $m_l^2 \leq q^2 \leq {(m_{B_{(s)}}- m_{S})}^2,$ and using the
total mean lifetimes $\tau_{B_{s}}= (1.515\pm0.004)~ps$ and
$\tau_{B^{0}}= (1.519\pm0.004)~ps$ \cite{PDG}, we present the
branching ratio values of the semileptonic decays $B_{(s)}\to
(K_0^*, a_0)\, l\, \bar{\nu}_l$\, ($l=\mu, \tau$) in Table
\ref{T307}, for both the Exp and LD models, in addition the two fit
functions.
\begin{table}[th]
\caption{The branching ratio values of $B_{(s)}\to (K_0^*, a_0)\,
l\, \bar{\nu}_l$ for both the Exp and LD models as well as the two
fit functions in addition different approaches.} \label{T307}
\begin{ruledtabular}
\begin{tabular}{cccccccc}
\multirow{3}{*}{\vspace{1.1em}\rm{Mode}} & \multicolumn{4}{c}{This work} &\multirow{3}{*}{\vspace{1.1em} \rm{LCSR(S2)} \cite{WanAslLu}} &\multirow{3}{*}{\vspace{1.1em} \rm{pQCD(S2)} \cite{LiLuWaWa}} & \multirow{3}{*}{\vspace{1.1em}\rm{QCDSR} \cite{MZYang}}   \\
\cline{2-5}                                      &
Exp,\,$f^{\rm{I}}$ & Exp,\,$f^{\rm{II}}$ & LD,\,$f^{\rm{I}}$ &
LD,\,$f^{\rm{II}}$ &
\\
\hline\\[-2mm]
$\mbox{Br}(B_{s} \to K_0^*\, \mu \,\nu_{\mu})\times 10^{4}$  &$1.98^{+1.51}_{-0.65}$&$2.00^{+1.52}_{-0.66}$&$2.63^{+1.99}_{-0.87}$&$2.60^{+1.98}_{-0.86}$&$1.30^{+1.30}_{-0.40}$ & $2.45^{+1.77}_{-1.05}$ & $0.36^{+0.38}_{-0.24}$ \\[2mm]
$\mbox{Br}(B_{s} \to K_0^*\, \tau\,\nu_{\tau})\times 10^{4}$ &$0.70^{+0.57}_{-0.26}$&$0.71^{+0.58}_{-0.27}$&$0.95^{+0.77}_{-0.35}$&$0.95^{+0.77}_{-0.35}$&$0.52^{+0.57}_{-0.18}$ & $1.09^{+0.82}_{-0.47}$ & $--$                   \\[2mm]
$\mbox{Br}(B \to a_0\, \mu \,\nu_{\mu})\times 10^{4}$  &$1.67^{+1.27}_{-0.53}$&$1.70^{+1.29}_{-0.54}$&$2.20^{+1.67}_{-0.70}$&$2.18^{+1.66}_{-0.69}$&$1.80^{+0.90}_{-0.70}$ & $3.25^{+2.36}_{-1.36}$ & $--$ \\[2mm]
$\mbox{Br}(B \to a_0\, \tau\,\nu_{\tau})\times 10^{4}$ &$0.51^{+0.41}_{-0.19}$&$0.53^{+0.42}_{-0.20}$&$0.67^{+0.54}_{-0.25}$&$0.66^{+0.53}_{-0.24}$&$0.63^{+0.34}_{-0.25}$ & $1.32^{+0.97}_{-0.57}$ & $--$                   \\[2mm]
\end{tabular}
\end{ruledtabular}
\end{table}
The results obtained for the electron are very close to the results
of the muon. Therefore, the branching ratios for muon are only
presented in this table. Table \ref{T307} shows that the difference
between the calculations through the two fit functions can be
completely ignored. This table also contains the results estimated
via the conventional LCSR \cite{WanAslLu} and pQCD \cite{LiLuWaWa}
through S$2$ as well as the QCDSR \cite{MZYang} approach. In
general, the values obtained in this work are in a logical agreement
with the two models; the conventional LCSR and pQCD. Especially, the
obtained values of the Exp-model are in a good agreement with the
conventional LCSR. As can be seen in this table, uncertainties in
the values obtained for the branching ratios of the semileptonic
decays are very large. The main source of errors comes from the form
factor $f_{+}(q^2)$.

\subsection{Semileptonic $B^0 \to K_0^{*0}\, l^+ l^- /\nu \bar{\nu}$ and $B^0_s \to f^0_0\, l^+ l^- /\nu \bar{\nu}$ decays}

The semileptonic decays
$B_{(s)} \to (K_0^*, f_0)\, l^+
l^-/\nu\bar{\nu}$
are conducted by the FCNC $b \to s$ loop
transition. In the SM, the weak effective Hamiltonian responsible
for these rare decays, neglecting the CKM-suppressed contributions
proportional to $V_{ub}V^*_{us}$, and also considering the
approximation $|V_{tb} V_{ts}|\simeq |V_{cb}V_{cs}|$, is described
at the energy scale $\mu= m_b$ as \cite{BuBuLa,Buras0,KhMaWa}:
\begin{eqnarray}\label{eq324}
\mathcal{H}_{\rm{eff}}(b \to s\, l^+ l^-)= - \frac{4\,G_F}{\sqrt{2}}
V_{tb} V^*_{ts} \left(C_1 O^c_1 +C_2 O^c_2+\sum_{i=3}^{10} C_i O_i
\right),
\end{eqnarray}
where $C_i (\mu)$ are the Wilson coefficients. $O^c_{1,2}$ are
current-current operators, $O_{3-6}$ are QCD penguin operators,
$O_{7,8}$ are magnetic penguin operators, and $O_{9,10}$ are
semileptonic electroweak penguin operators. The contributions of the
operators $O_7$ and $O_{9,10}$ in the decay amplitudes $B_{(s)} \to
(K_0^*, f_0)$ are factorized in the form factors $f_{\pm}$ and
$f_{T}$. The effect of other operators appears as the factorizable
and nonfactorizable contributions.

The factorizable contributions have the same form factor dependence
as $C_9$ which can be absorbed into an effective Wilson coefficient
$C^{\rm eff}_9$. The dominant factorizable contribution is generated
by the tree-level four quark operators $O^c_{1,2}$ with large Wilson
coefficients $|V_{cb}V_{cs}|$. This contribution includes
intermediate vector charmonium states in the upper part of the decay
kinematical region as long-distance effect.

The nonfactorizable contributions arise from electromagnetic
corrections to the matrix elements of purely hadronic operators in
the weak effective Hamiltonian. The weak annihilation and quark-loop
diagrams with soft and hard gluon create the nonfactorizable
corrections  \cite{KhMaWa,KhodjRusov}. These contributions for the
FCNC $B_{(s)} \to (K_0^*, f_0) $ decays are highly suppressed due to
the large current uncertainties of the form factors, and also the
small Wilson coefficients of the penguin operators.

According to the effective weak Hamiltonian of the $b \to s~ l^+
l^-$ transition in Eq. (\ref{eq324}), the matrix element for this
FCNC decay by considering the contributions of the operators $O_7$
and $O_{9,10}$ as well as the factorizable contributions of the
operators through $C_9^{\rm eff}$, and ignoring the nonfactorizable
contributions, can be written as
\begin{eqnarray*}\label{eq325}
{\cal M} (b \to s\, l^+ l^-)= \frac{G_{F}\alpha}{2\sqrt{2}\pi}
V_{tb}V_{ts}^{*}\Bigg[ C_9^{\rm eff} \,\, \bar {s} \gamma_\mu
(1-\gamma_5) b~ \overline {l }\gamma_\mu  l + C_{10}\,\, \bar {s}
\gamma_\mu (1-\gamma_5) b~ \overline {l } \gamma_\mu \gamma_5 \l -
2\, C_7^{\rm eff}\,\,\frac{m_b}{q^2}\, \bar {s} ~i\sigma_{\mu\nu}
q^\nu (1+\gamma_5) b~  \overline {l}  \gamma_\mu l \Bigg],
\end{eqnarray*}
where $\alpha$  is the fine structure constant at $Z$ mass scale,
the CKM matrix elements $|V_{tb}V^*_{ts}|=0.041$ \cite{KinKirLen},
and the Wilson coefficients $C^{\rm{eff}}_7=-0.313$ and
$C_{10}=-4.669$ \cite{BuBuLa}. The effective Wilson coefficient
$C_{9}^{\rm{eff}}$ includes both the short-distance and
long-distance effects as
\begin{eqnarray}\label{326}
C^{\rm eff}_9 = C_9 + Y_{S}(q^2)+Y_{L}(q^2),
\end{eqnarray}
where $Y_{S}(q^2)$ describes the short-distance contributions from
four-quark operators far away from the resonance regions, which can
be calculated reliably in perturbative theory as \cite{Buras0}:
\begin{eqnarray}\label{327}
Y_{S}(q^2)&=&0.124~ \omega(s)+h(\hat{m_c},s)C_0
-\frac{1}{2}h(1,s)(4C_3+4C_4+3C_5+C_6)\nonumber\\
&-&\frac{1}{2}h(0,s)(C_3+3C_4) +\frac{2}{9}(3C_3+C_4+3C_5+C_6),
\end{eqnarray}
where $s=q^2/m_b^2$, $\hat{m_c}=m_c/m_b$,
$C_0=3C_1+C_2+3C_3+C_4+3C_5+C_6$,  and
\begin{eqnarray}\label{328}
\omega(s)= -\frac{2}{9} \pi^2 -\frac{4}{3} {\rm Li}_2(s)-
\frac{2}{3} \ln (s) \ln(1-s) - \frac{5+4s}{3(1+2s)} \ln(1-s) -
\frac{2 s(1+s)(1-2s)}{3(1-s)^2(1+2s)} \ln (s)+ \frac{5+9 s-6
s^2}{6(1-s)(1+2s)}.
\end{eqnarray}
The functional form of the $h(\hat{m_c}, s)$ and $h(0, s)$ are as:
\begin{eqnarray}\label{329}
h(\hat m_c,  s)=- \frac{8}{9}\ln \frac{m_b}{\mu} -
\frac{8}{9}\ln\hat m_c + \frac{8}{27} + \frac{4}{9} x -\frac{2}{9}
(2+x) |1-x|^{1/2} \left\{
\begin{array}{ll}
\left( \ln\left| \frac{\sqrt{1-x} + 1}{\sqrt{1-x} - 1}\right| - i\pi
\right), &
\mbox{for } x \equiv \frac{4 \hat m_c^2}{ s} < 1 \nonumber \\
& \\
2 \arctan \frac{1}{\sqrt{x-1}}, & \mbox{for } x \equiv \frac {4 \hat
m_c^2}{ s} > 1
\end{array}
\right. \nonumber\\
\end{eqnarray}
and
\begin{eqnarray}\label{330}
h(0,  s)  =  \frac{8}{27} -\frac{8}{9} \ln\frac{m_b}{\mu} -
\frac{4}{9} \ln\ s + \frac{4}{9} i\pi\,.
\end{eqnarray}

The long-distance contributions, $Y_{L}(q^2)$ from four-quark
operators near the $c\bar{c}$ resonances can not be calculated from
the first principles of QCD and are usually parametrized in the form
of a phenomenological Breit-Wigner formula as \cite{Buras0}:
\begin{eqnarray}\label{331}
Y_{L}(q^2)&=&\frac{3\pi}{\alpha^2} \sum_{V_i=J/\psi,
\psi(2S)}\frac{\Gamma(V_i\to l^+ l^-)m_{V_i}}{m_{V_i}^2-q^2-i
m_{V_i} \Gamma_{V_i}}.
\end{eqnarray}
In the range of $4m_l^2\leq q^2\leq(m_{B_{(s)}}-m_{S})^2$, there are
two charm-resonances $J/\psi(3.097)$ and $\psi(3.686)$. To avoid the
background of charmonium resonances, it is common to delete the
experimental measurements around the resonance regions. For this
reason, the long-distance contributions are ignored in our
calculations.

Using the parametrization of the aforementioned decays in terms of
the form factors, the differential decay width in the rest frame of
$B_{(s)}$-meson can be written as:
\begin{eqnarray}\label{eq332}
\frac{d{\Gamma}}{dq^2}(B_{(s)}\rightarrow S {\nu} \bar \nu) &=& \frac{%
G_{F}^2{\mid}V_{tb}V^*_{ts}{\mid}^2
m_{B_{(s)}}^3\alpha^2} {2^8 \pi ^5}~\frac{{\mid}D_{\nu}(x_t){\mid}^2}{\rm{sin}^4\theta_W}~\phi%
^{3/2}(1,\hat{r},\hat{s}){\mid}f_+ (q^2){\mid}%
^2~,\nonumber\\
\frac{d\Gamma  }{dq^2}\left( B_{(s)} \rightarrow S l^+l^-\right) &=&
\frac{G_{F}^2{\mid}V_{tb}V^*_{ts}{\mid}^2 m_{B_{(s)}}^3\alpha^2}
{3\times 2^9\pi ^5}v \,\phi^{1/2}(1,\hat{r},\hat{s})\left[
\left(1+\frac{2\hat{l}} {\hat{s}}\right)
\phi(1,\hat{r},\hat{s})\alpha _1+12~\hat{l}\beta_1\right],
\end{eqnarray}
where $\hat{r} = \frac{m_{S}^2}{m_{B_{(s)}}^2}$, $\hat{s}
=\frac{q^2}{m_{B_{(s)}}^2}$, $\hat{l} = \frac{m_{l}^2}{
m_{B_{(s)}}^2}$, $x_t = \frac{m_t^2}{m_W^2}$, ${\hat m}_b =
\frac{m_b}{m_{B_{(s)}}}$, $v =\sqrt{1-\frac{4\hat{l}}{\hat{s}}}$,
${\phi}(1,\hat{r},\hat{s}) = 1+\hat{r}^2
+\hat{s}^2-2\hat{r}-2\hat{s}-2\hat{r}\hat{s}$, and the functions
$D_{\nu}(x_t)$, $\alpha_1$ and $\beta_1$ are defined as:
\begin{eqnarray}\label{eq333}
&&D_{\nu}(x_t)=\frac{x_t}{8}\Bigg(\frac{2+x_t}{x_t-1}+\frac{3x_t-6}{(x_t-1)^2}\ln
x_t\Bigg)~,\nonumber\\
&&\alpha_1 = \biggl|C_{9}^{\rm eff}\,f_{+}(q^2) +\frac{2\,{\hat
m}_b\,C_{7}^{\rm eff}\,f_{T}(q^2)}{1+\sqrt{\hat{r}}}\biggr|^{2}
+|C_{10}f_{+}(q^2)|^{2} ~,\nonumber\\
&&\beta_1 = |C_{10}|^{2}\biggl[ \left( 1+\hat{r}-{\hat{s}\over
2}\right) |f_{+}(q^2)|^{2}+\left( 1-\hat{r}\right) {\rm
Re}\left[f_{+}(q^2)f_{-}^{*}(q^2)\right]+\frac{1}{2}\hat{s}\,|f_{-}(q^2)|^{2}\biggr].~
\end{eqnarray}

\begin{figure}[th]
\begin{center}
\includegraphics[width=5cm,height=5cm]{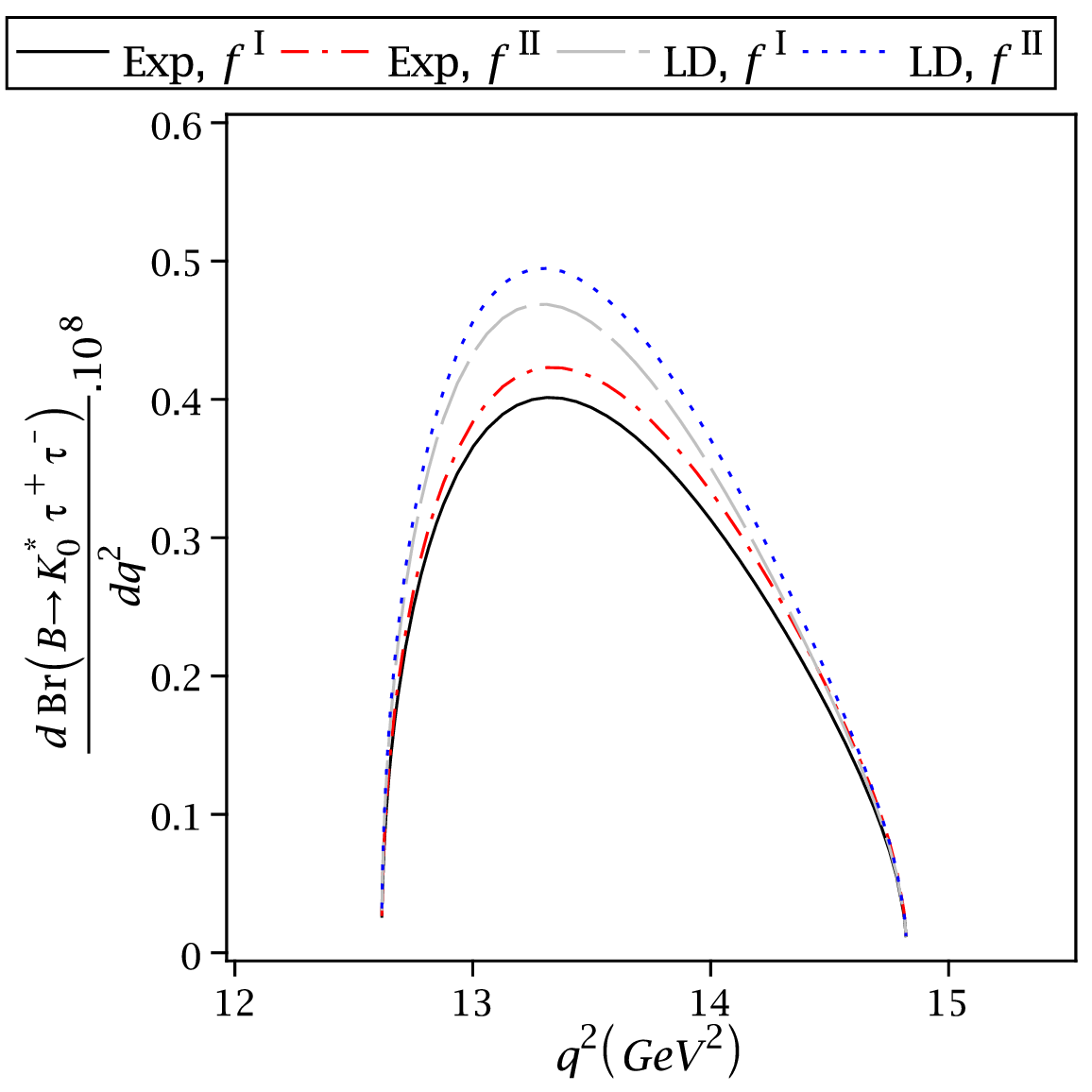}
\includegraphics[width=5cm,height=5cm]{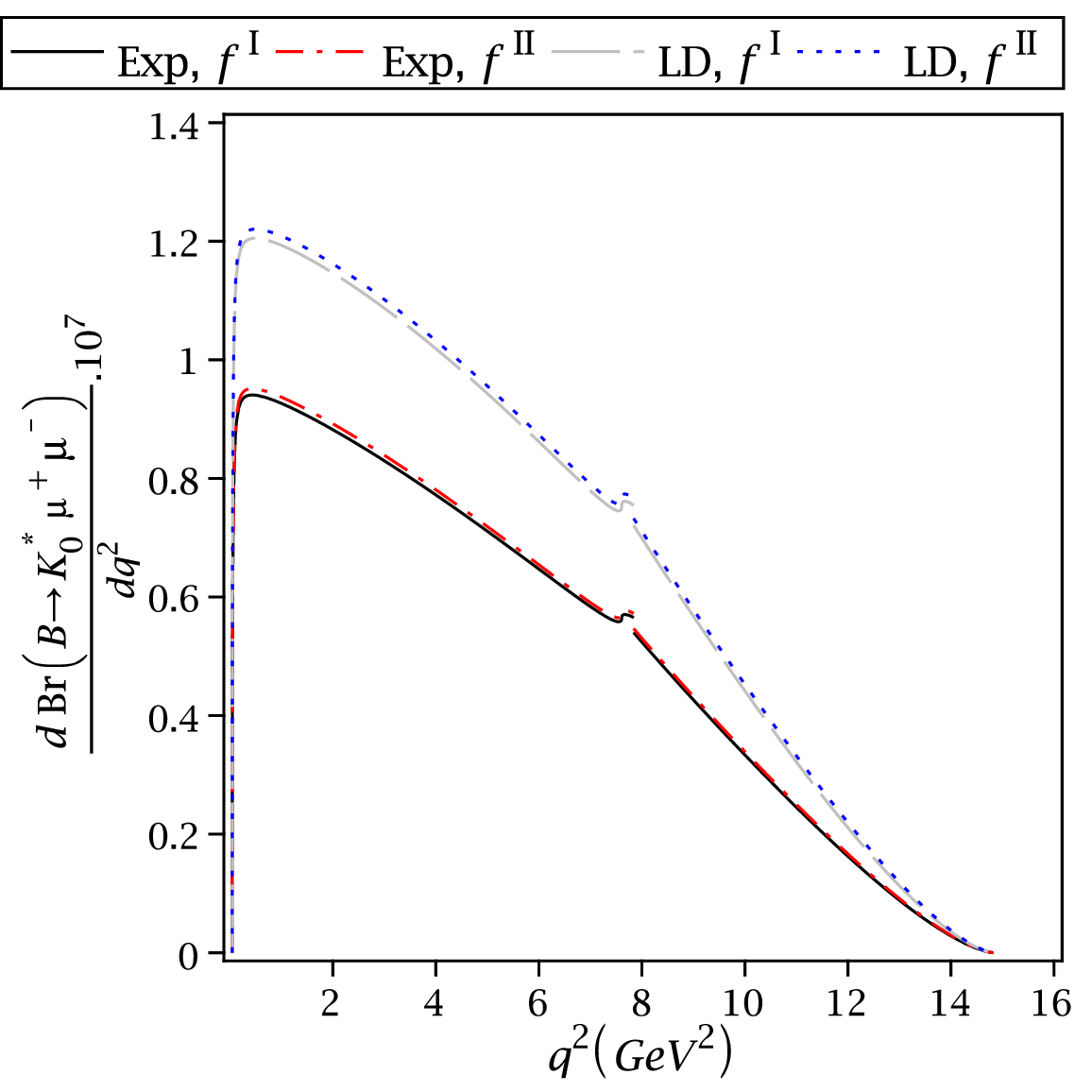}
\includegraphics[width=5cm,height=5cm]{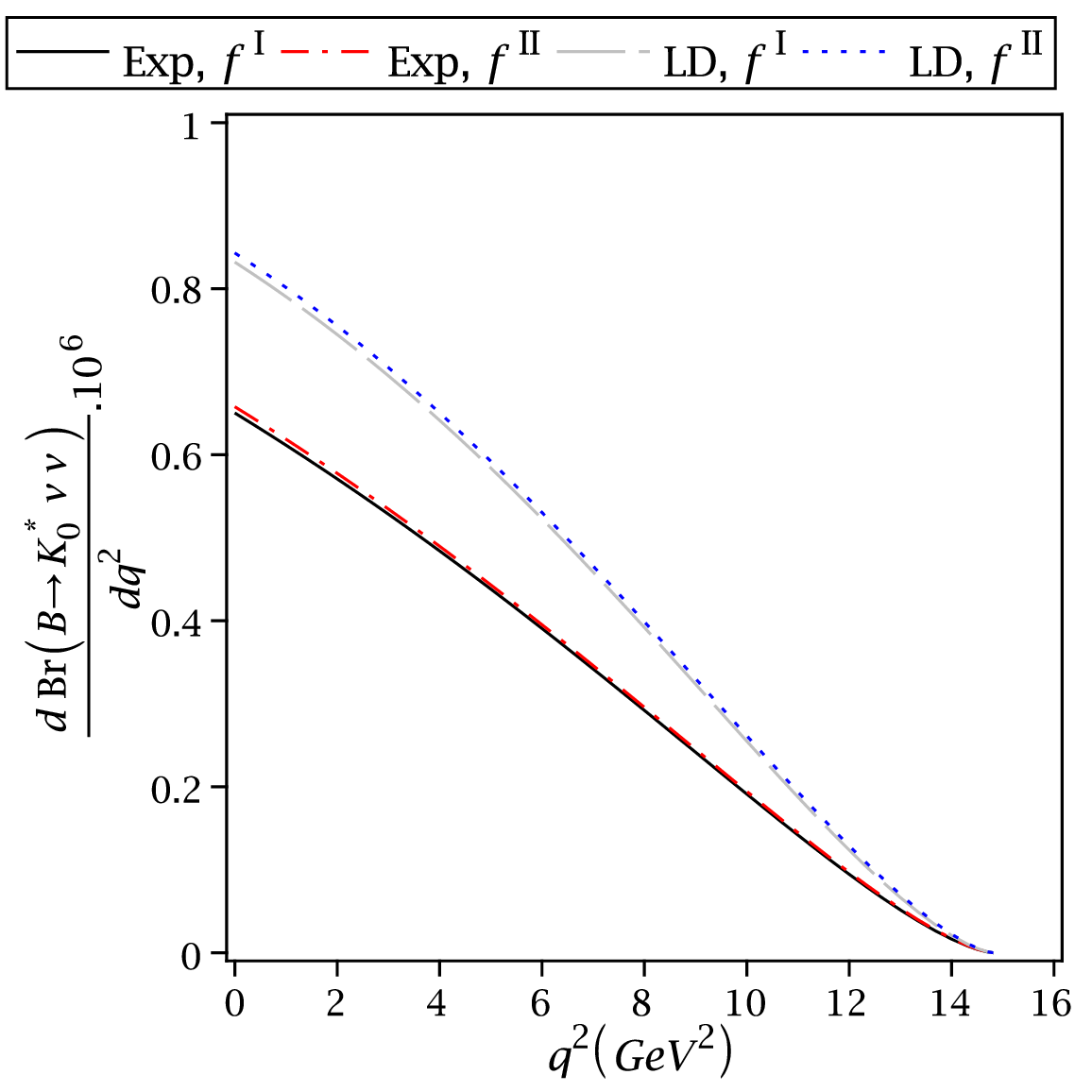}
\caption{The differential branching ratios of the semileptonic $B
\to K_0^* l^{+}l^{-}/\nu\bar{\nu}$ decays ($ l=\mu,\tau$)  on $q^2$
for both the Exp and LD models as well as the two different fit
functions.} \label{F310}
\end{center}
\end{figure}

\begin{figure}[th]
\begin{center}
\includegraphics[width=5cm,height=5cm]{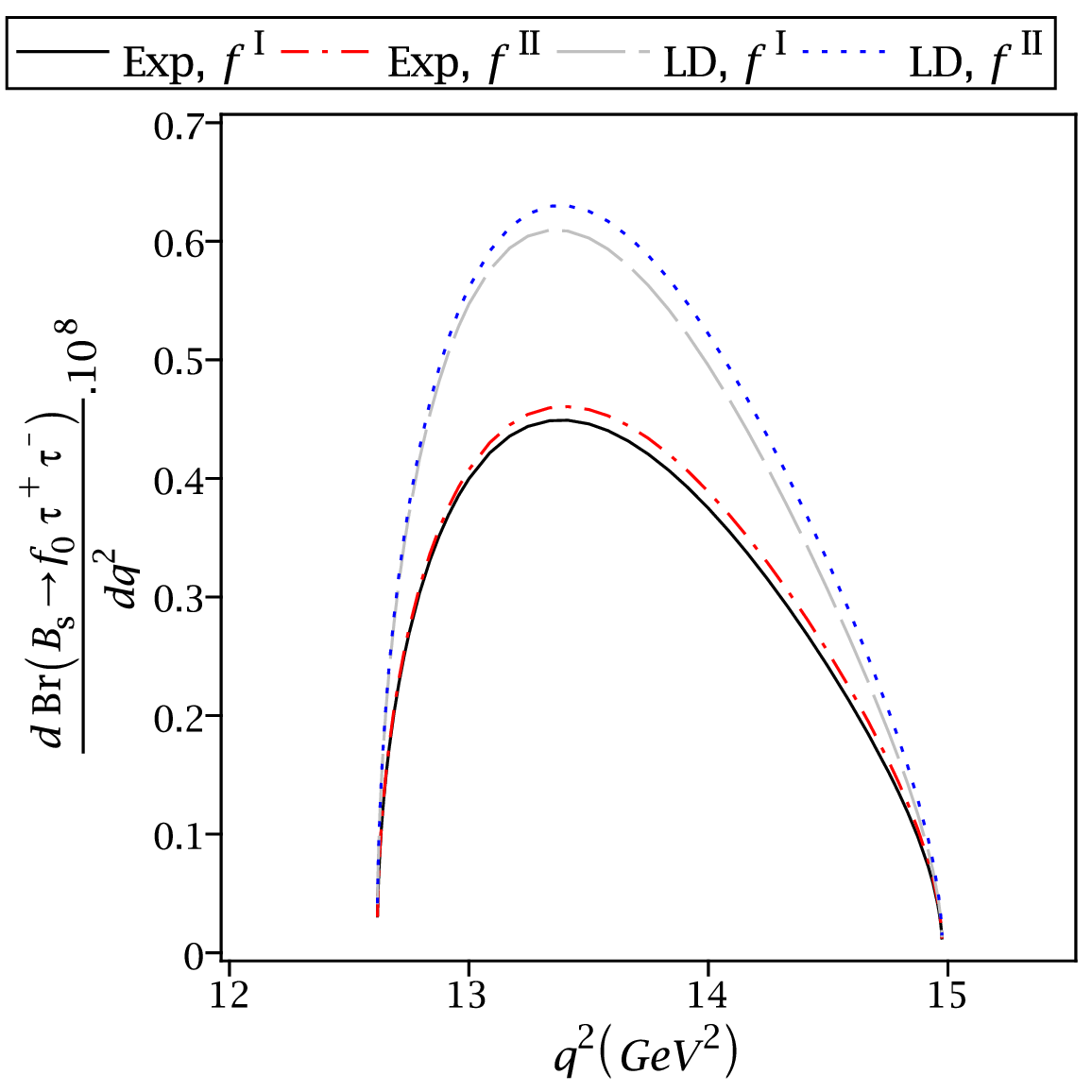}
\includegraphics[width=5cm,height=5cm]{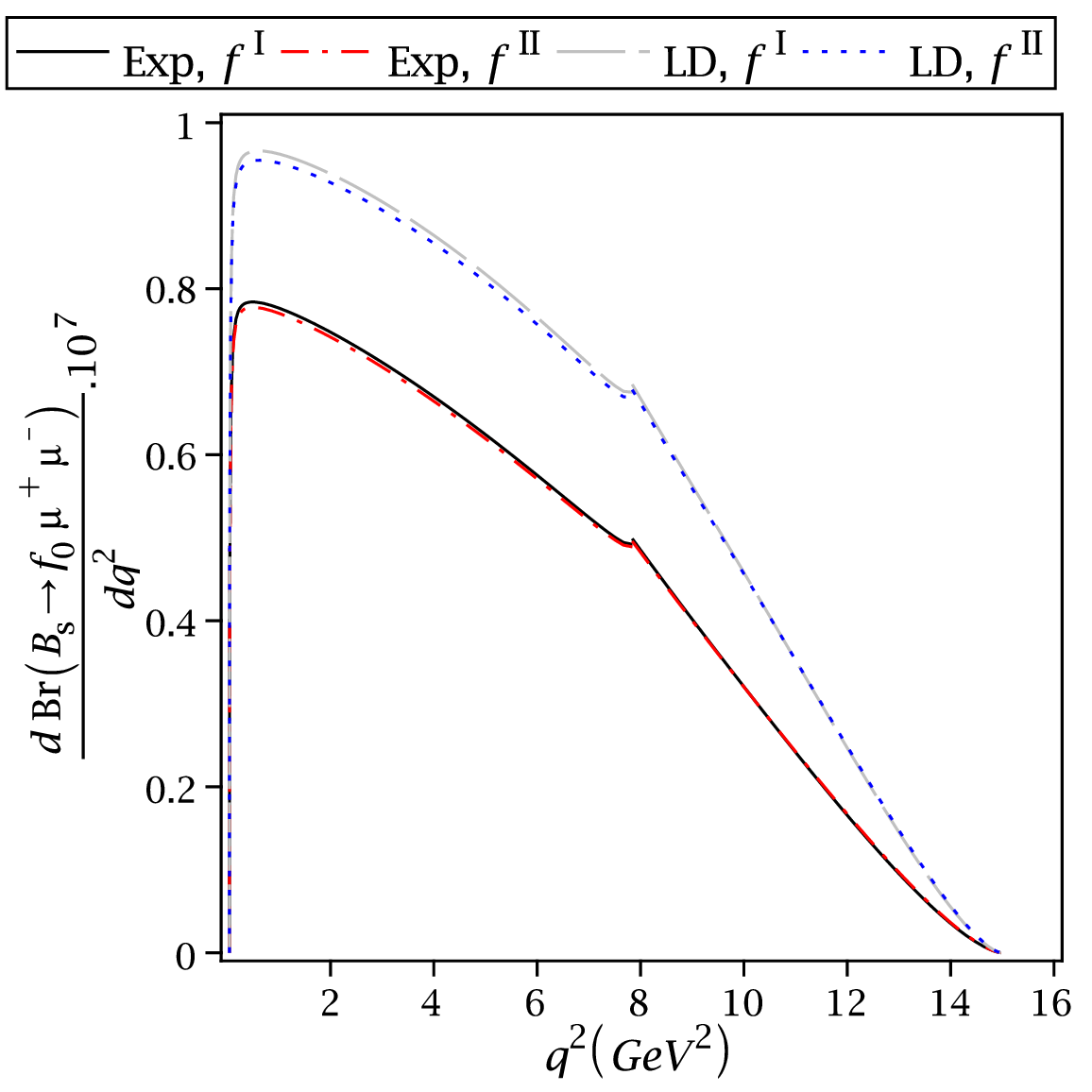}
\includegraphics[width=5cm,height=5cm]{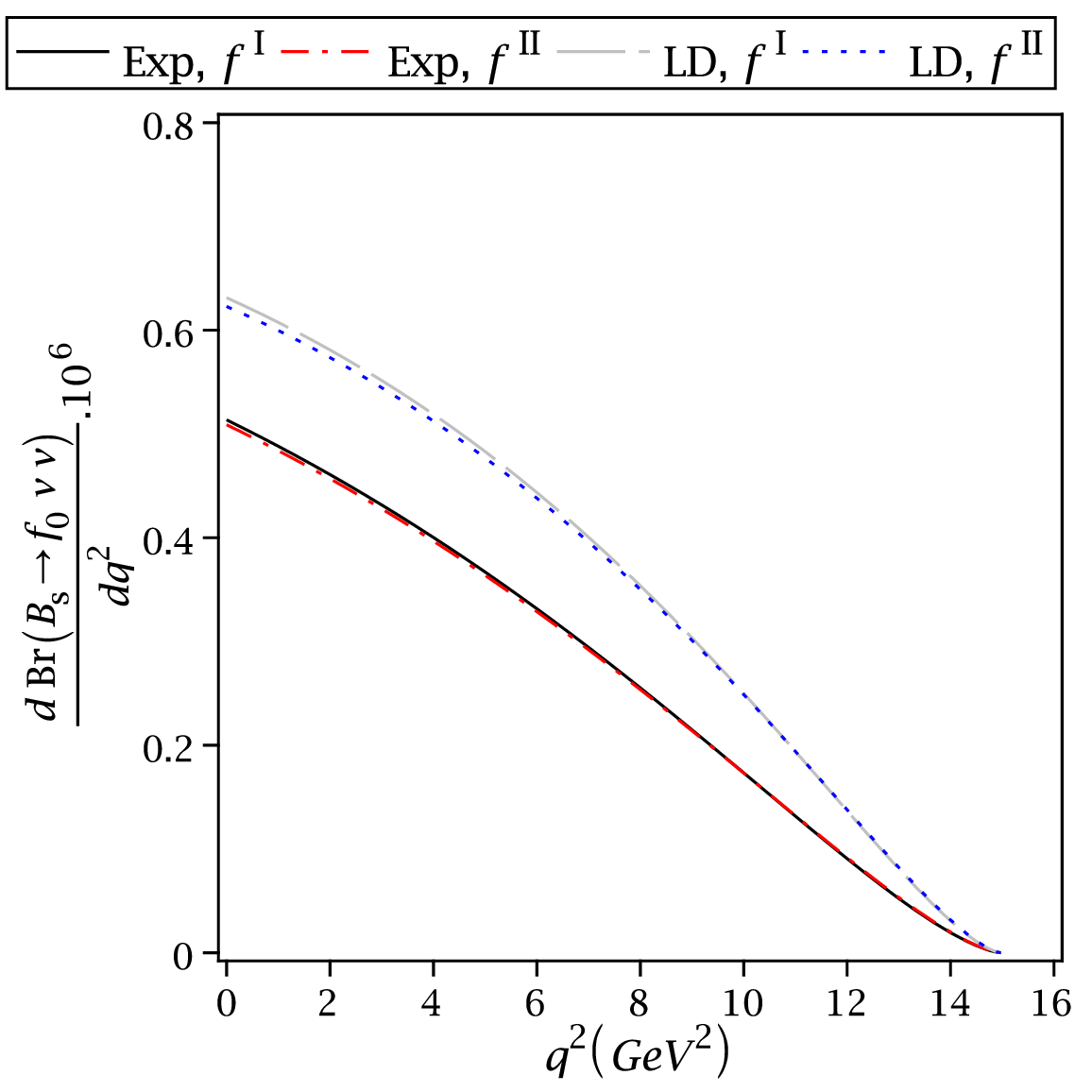}
\caption{The same as Fig. \ref{F310} but for  $B_s \to f_0
l^{+}l^{-}/\nu\bar{\nu}$ decays ($ l=\mu,\tau$).} \label{F311}
\end{center}
\end{figure}

The dependency of the differential branching ratios for $B_{(s)} \to
(K^*_0, f_0) l^{+}l^{-}/\nu \bar{\nu}$ on $q^2$ for both the Exp and
LD models as well as the two different fit functions is shown in
Figs. \ref{F310} and \ref{F311}.

\newpage
Integrating Eq. (\ref{eq332}) over $q^2$ in the physical region $4
m_l^2 \leq q^2 \leq (m_{B_{(s)}}- m_{S})^2,$ and using
$\tau_{B_{(s)}}$, the branching ratio results of the $B_{(s)}
\rightarrow S l^{+}l^{-}/\nu\bar{\nu}$ are obtained. Table
\ref{T308} shows the branching ratios of the aforementioned decays
for both the Exp and LD models as well as the two different fit
functions, in addition predictions by the conventional LCSR(S2)
\cite{WanAslLu}, pQCD(S2) \cite{LiLuWaWa}, and LFQM(S2)
\cite{CheGenLiLi}.
\begin{table}[th]
\caption{The branching ratio values of $B_{(s)}\to (K_0^*, f_0)\,
l^+l^-/\nu \bar{\nu}$ for both the Exp and LD models as well as the
two different fit functions in addition the LCSR, pQCD, and LFQM
approaches.} \label{T308}
\begin{ruledtabular}
\begin{tabular}{ccccccccc}
\multirow{3}{*}{\vspace{1.1em}\rm{Mode}} & \multicolumn{4}{c}{This work} &\multirow{4}{*}{\vspace{1.1em} \rm{LCSR(S2)}\cite{WanAslLu}}&\multirow{4}{*}{\vspace{1.1em} \rm{pQCD(S2)}\cite{LiLuWaWa}}&\multirow{4}{*}{\vspace{1.1em} \rm{LFQM(S2)}\cite{CheGenLiLi}}  \\
\cline{2-5}                                      &Exp, $f^{\rm{I}}$
&Exp, $f^{\rm{II}}$ & LD, $f^{\rm{I}}$ & LD, $f^{\rm{II}}$ &
\\
\hline\\[-2mm]
$\mbox{Br}(B     \to K_0^* \nu\bar{\nu})~~~~\times 10^{6}$ & $4.49^{+1.35}_{-0.90} $&$4.78^{+1.43}_{-0.95}  $&$6.22^{+1.87}_{-1.24}$&$6.32^{+1.90}_{-1.26}$&$-- $&$-- $&$-- $\\[2mm]
$\mbox{Br}(B     \to K_0^* \mu^+ \mu^-)\times 10^{7}$      & $5.66^{+2.77}_{-1.98} $&$6.03^{+2.95}_{-2.11}  $&$7.89^{+3.86}_{-2.76}$&$8.03^{+3.93}_{-2.81}$&$5.6^{+3.1}_{-2.3} $&$9.78^{+7.66}_{-4.40} $&$1.62 $\\[2mm]
$\mbox{Br}(B     \to K_0^* \tau^+ \tau^-)\times 10^{8}$    & $0.55^{+0.36}_{-0.28} $&$0.61^{+0.40}_{-0.31}  $&$0.65^{+0.42}_{-0.33}$&$0.69^{+0.44}_{-0.35}$&$0.98^{+1.24}_{-0.55} $&$0.63^{+0.57}_{-0.30} $&$0.29 $\\[2mm]
$\mbox{Br}(B_{s} \to f_0 \nu\bar{\nu})~~~~\times 10^{6}$   & $3.97^{+1.07}_{-0.71} $&$4.14^{+1.12}_{-0.75}  $&$5.52^{+1.49}_{-0.99}$&$5.46^{+1.47}_{-0.98}$&$-- $&$-- $&$-- $\\[2mm]
$\mbox{Br}(B_{s} \to f_0 \mu^+ \mu^-)\times 10^{7}$        & $5.00^{+1.95}_{-1.35} $&$5.22^{+2.04}_{-1.41}  $&$7.00^{+2.73}_{-1.89}$&$6.92^{+2.70}_{-1.87}$&$5.2^{+2.3}_{-1.7} $&$10.0^{+8.5}_{-3.8} $&$-- $\\[2mm]
$\mbox{Br}(B_{s} \to f_0 \tau^+ \tau^-)\times 10^{8}$      & $0.65^{+0.38}_{-0.21} $&$0.71^{+0.41}_{-0.23}  $&$0.90^{+0.53}_{-0.30}$&$0.95^{+0.56}_{-0.31}$&$1.2^{+0.8}_{-0.5}$&$1.3^{+1.2}_{-0.6} $&$-- $\\[2mm]
\end{tabular}
\end{ruledtabular}
\end{table}

\begin{figure}[th]
\begin{center}
\includegraphics[width=4cm,height=4cm]{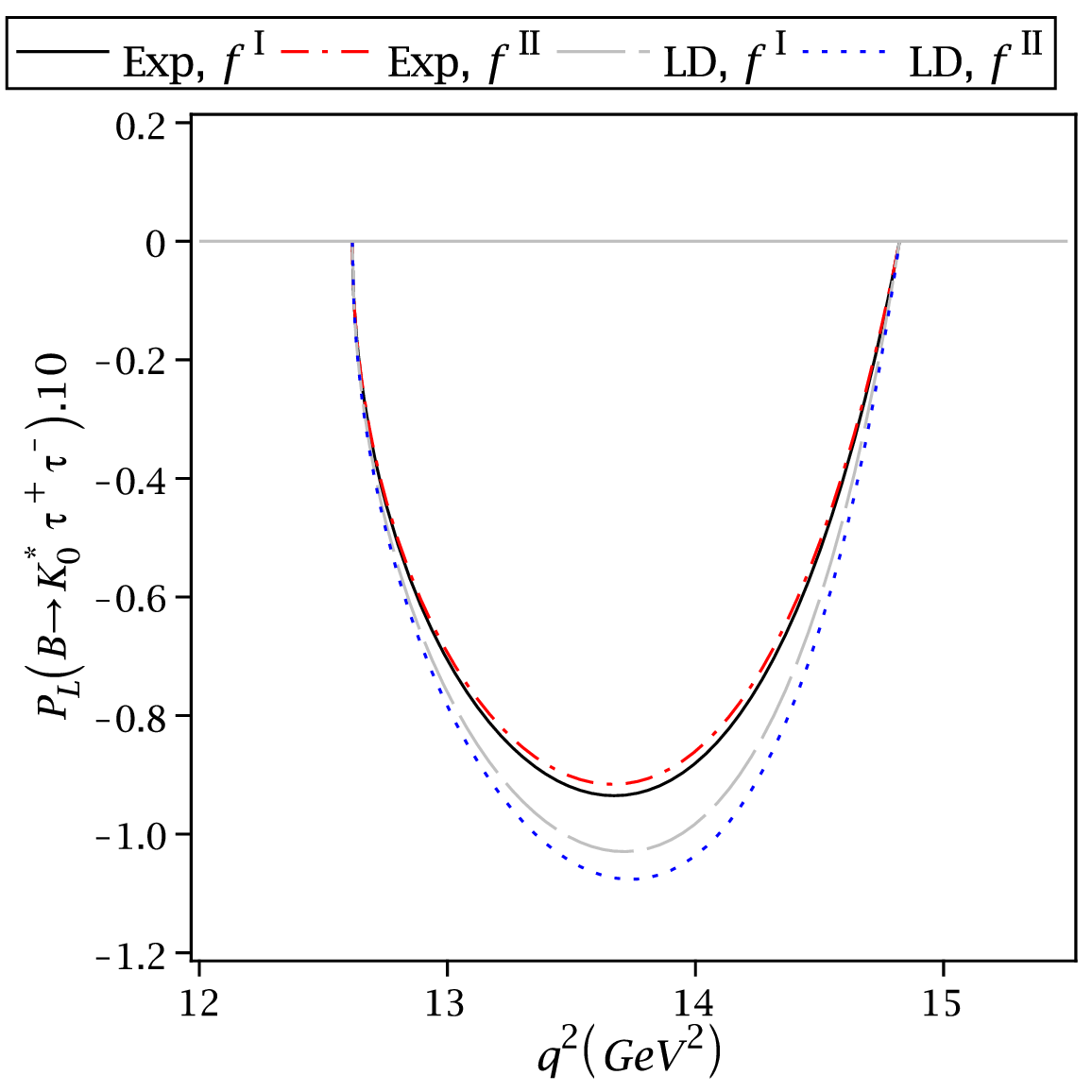}
\includegraphics[width=4cm,height=4cm]{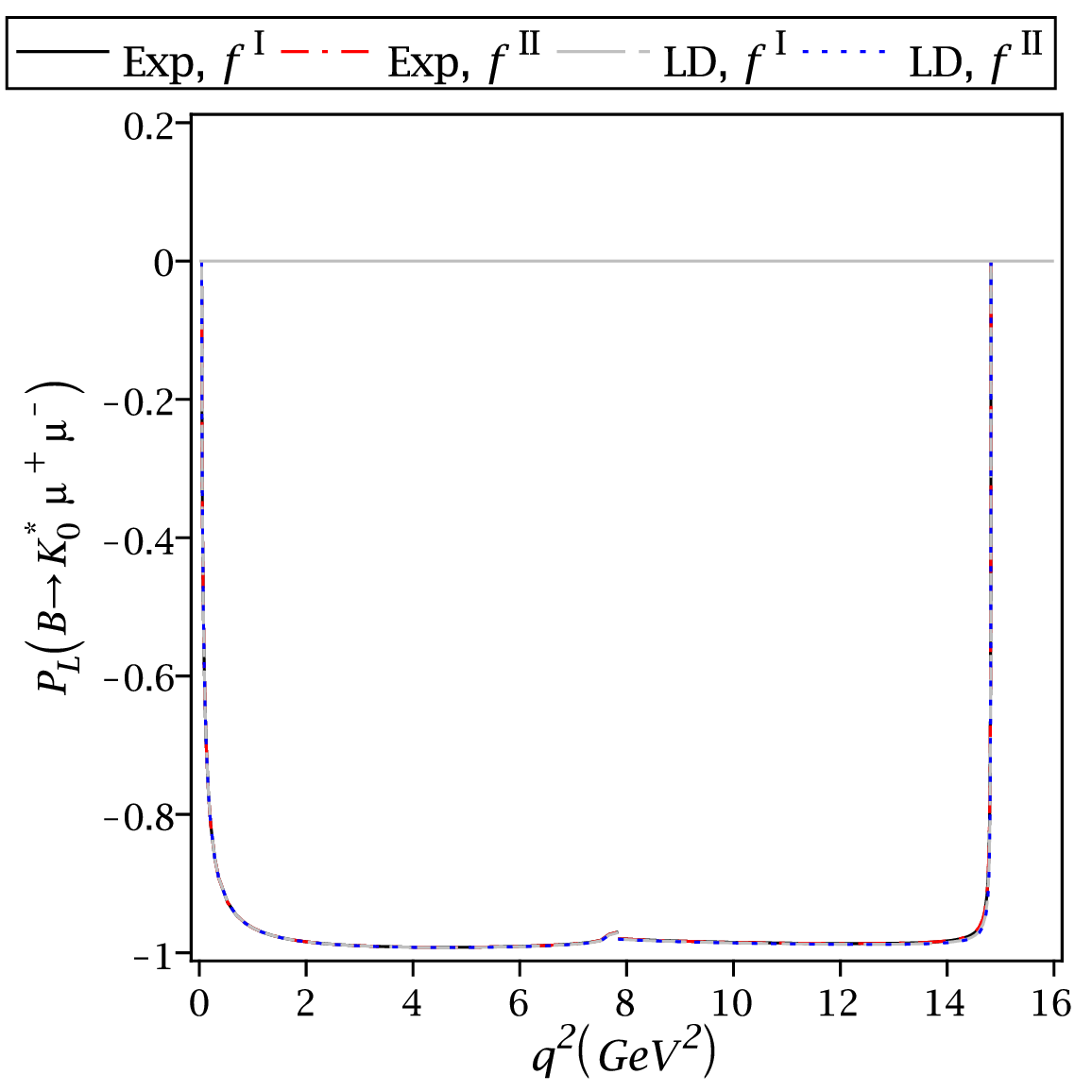}
\includegraphics[width=4cm,height=4cm]{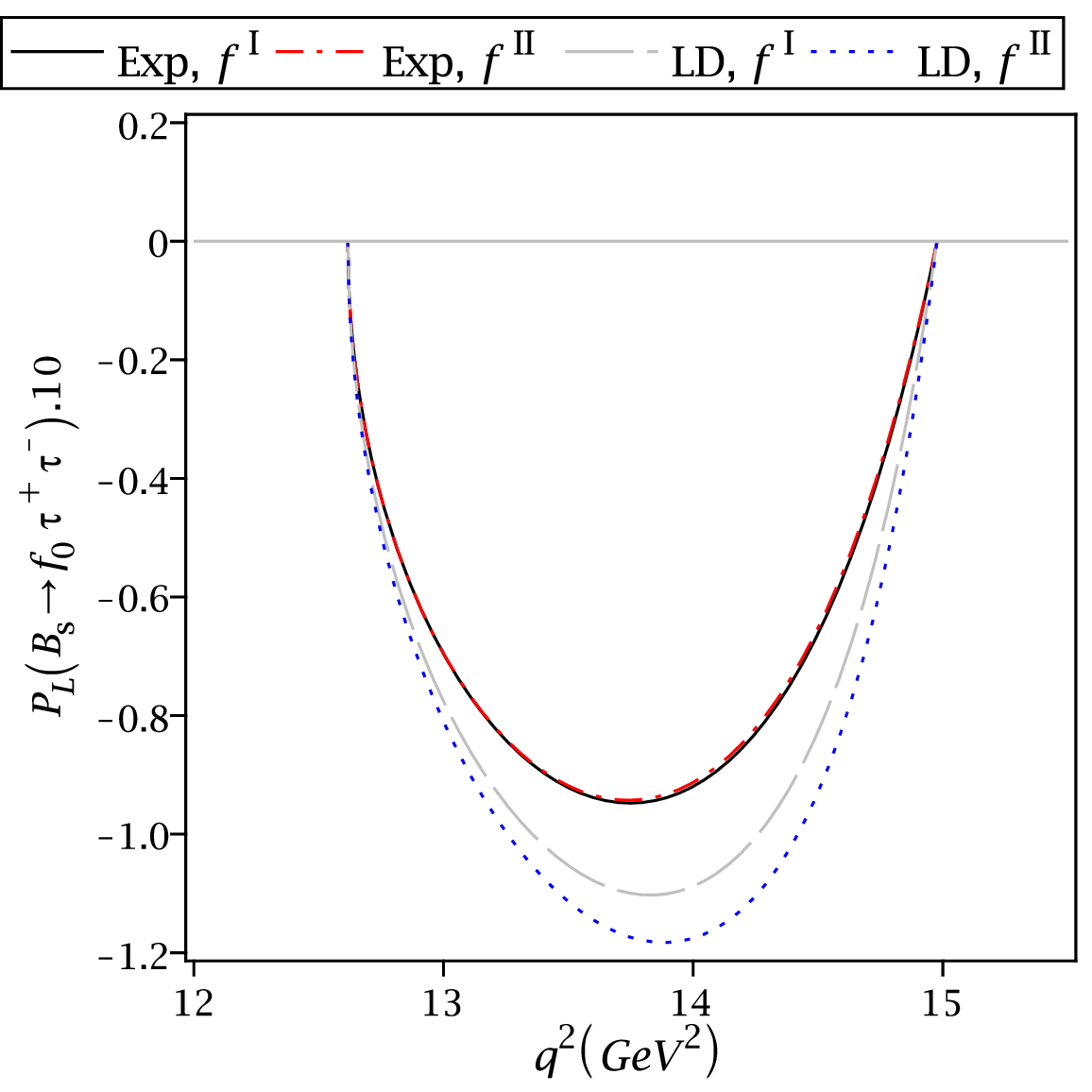}
\includegraphics[width=4cm,height=4cm]{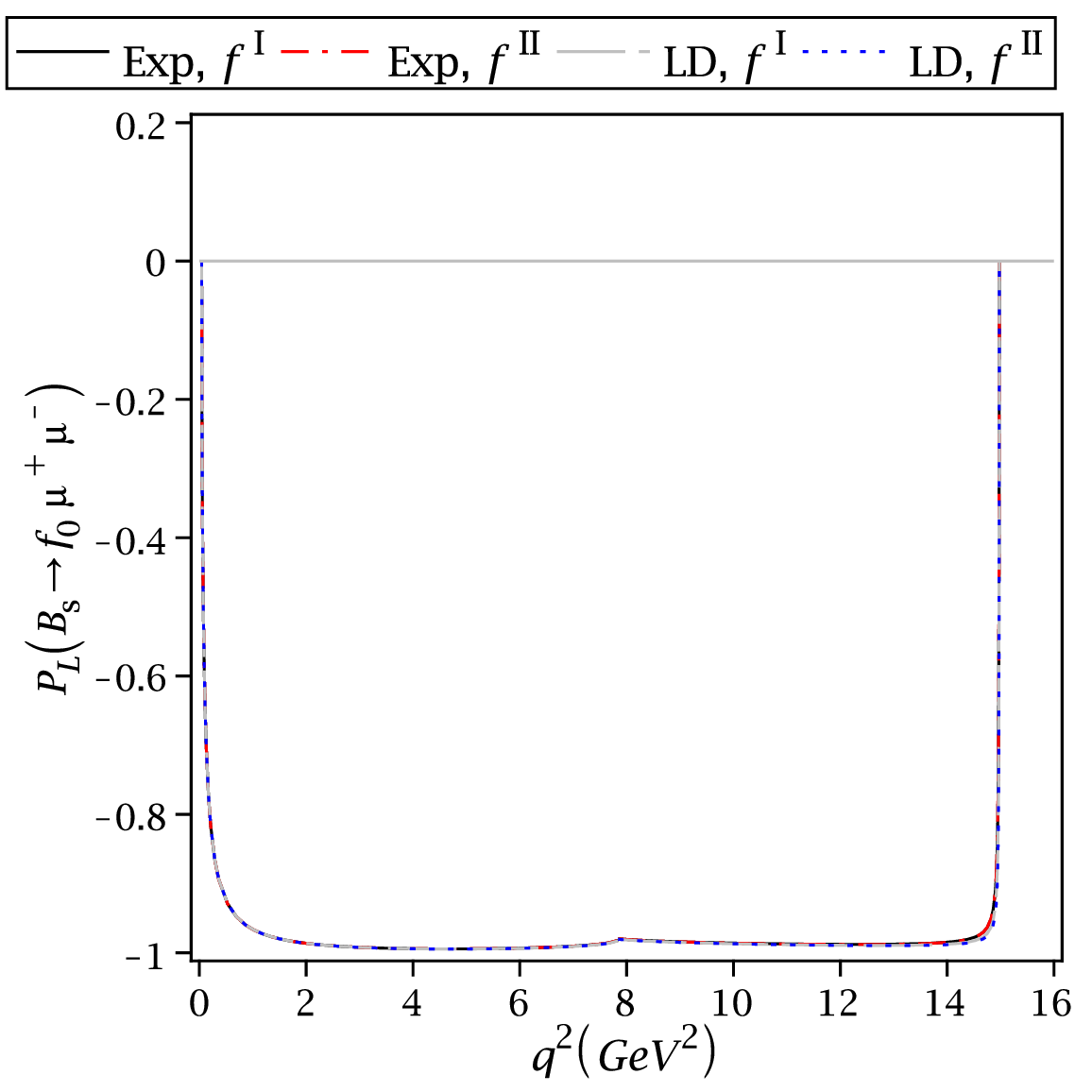}
\caption{The dependence of the longitudinal lepton polarization
asymmetries on $q^2$ for both the Exp and LD models as well as the
two different fit functions.} \label{F312}
\end{center}
\end{figure}

The polarization asymmetries provide valuable information on the
flavor changing loop effects in the SM. The longitudinal lepton
polarization asymmetry formula for $B_{(s)} \to S l^{+}l^{-}$ is
given as:
\begin{eqnarray}\label{eq334}
P_L=\frac{2v}{(1+\frac{2\hat{l}}{\hat{s}})\phi(1,\hat{r},\hat{s})\alpha_1+12
\hat{l}\beta_1}{\rm{Re}}\left[\phi(1,\hat{r},\hat{s})\left(C_9^{eff}f_+(q^2)-\frac{2C_7
f_T(q^2)}{1+\sqrt{\hat{r}}} \right)(C_{10}f_+(q^2))^*\right],
\end{eqnarray}
where $v, ~\hat{l}, ~\hat{r}, ~\hat{s}, ~\phi(1,\hat{r},\hat{s}),
\alpha_1$ and $\beta_1$ were defined before. The dependence of the
longitudinal lepton polarization asymmetries for the $B_{(s)} \to
(K_0^*, f_0)\, l^{+}l^{-}\,(l=\mu, \tau)$ decays on the transferred
momentum square $q^2$ for both the Exp and LD models as well as the
two different fit functions is plotted in Fig. \ref{F312}.

It should be noted that the forward-backward asymmetry for the decay
modes $B_{(s)} \to (K_0^*, f_0) l^+ l^-$ is exactly equal to zero in
the SM \cite{BelGen,GengKao}, due to the absence of scalar-type
coupling between the lepton pair.

In summary, our main goal was to calculate the form factors of the
semileptonic decays $B_{(s)} \to S$ ($S=K_0^*(1430), a_0(1450),
f_0(1500)$) in the frame work of the LCSR with the $B$-meson DA's.

$\bullet$ Two different phenomenological models including
exponential and local duality models were used for the shapes of the
$B$-meson DA's.

$\bullet$ The $B$-meson DA's were also applied for $B_s$ meson in
the $\rm SU(3)_F$ symmetry limit.

$\bullet$ The form factors of the aforementioned decays were
estimated at $q^2=0$ through the two exponential and local duality
models, and compared with the predictions of other approaches.

$\bullet$ It was shown that the two-particle leading-twist DA of
$B$-meson, $\varphi_{_+}$ has the most important contribution in
calculation of the form factors.

$\bullet$ In addition, it was shown that the main sources of the
uncertainties in estimation of the form factors were the shape
parameter $\omega_{0}$ and the decay constants of the scalar mesons.

$\bullet$ Considering the uncertainties, there was a good agreement
between our results in the exponential-model and predictions of the
conventional LCSR in scenario 2. As a result, our calculations
confirmed that the scalar mesons $K^*_{0}(1430), a_0(1450)$ and
$f_0(1500)$ can be viewed as the lowest lying states with two quarks
in the quark model.

$\bullet$ For a better analysis, the results obtained for the form
factors via the $B$-meson LCSR method were parameterized to the two
different fit functions. The form factors obtained by the both fit
functions were consistent very well in each case.

$\bullet$ Using the form factors $f_{+}(q^2)$, $f_{-}(q^2)$ and
$f_{T}(q^2)$, the branching ratio values for the semileptonic $B_s
\to K_0^* \,l \bar{\nu_{l}}$ and $B \to a_0 \,l \bar{\nu_{l}}$
decays, and also the FCNC semileptonic transitions $B \to K^*_0$ and
$B_s \to f_0$ were calculated.

$\bullet$ The dependence of the differential branching ratios as
well as the longitudinal lepton polarization asymmetries for the
aforementioned decays were plotted with respect to $q^2$.

$\bullet$ Future experimental measurement can give valuable
information about these aforesaid decays and the nature of the
scalar mesons.

\end{document}